\documentclass[twocolumn]{aastex63}

\definecolor{DarkOrange}{RGB}{204, 85, 0}
\definecolor{LincolnGreen}{RGB}{17, 102, 0}
\definecolor{Rust}{HTML}{9B4F0F}
\definecolor{DarkCyan}{HTML}{008B8B}
\definecolor{MediumAquaMarine}{HTML}{66CDAA}

\def\ion#1#2{#1$\;${\footnotesize\rm{#2}}\relax}

\newcommand{\rztf}{$r_\mathrm{ZTF}$}
\newcommand{\gztf}{$g_\mathrm{ZTF}$}
\newcommand{\iztf}{$i_\mathrm{ZTF}$}
\newcommand{\tfl}{$t_\mathrm{fl}$}

\newcommand{\tbmax}{$T_{B,\mathrm{max}}$}
\newcommand{\kms}{km\,s$^{-1}$}
\newcommand{\RSiII}{$\mathcal{R}($\ion{Si}{II}$)$}
\newcommand{\radni}{$^{56}$Ni}

\newcommand{\sn}{SN\,2019yvq}

\usepackage{lineno}
\usepackage{pifont}
\newcommand{\cmark}{\ding{51}}%
\newcommand{\xmark}{\ding{55}}%

\received{2020 May 14}
\revised{2020 June 03}
\accepted{2020 June 04}
\submitjournal{ApJ}

\shorttitle{Spectacular UV Flash of \sn}
\shortauthors{Miller et al.}
\graphicspath{{./}{figures/}}

\begin{document}

\title{The Spectacular Ultraviolet Flash From the \\ Peculiar Type Ia Supernova 2019yvq}

\author[0000-0001-9515-478X]{A.~A.~Miller}
\affiliation{Center for Interdisciplinary Exploration and Research in 
             Astrophysics (CIERA) and Department of Physics and Astronomy, 
             Northwestern University, 
             1800 Sherman Road, Evanston, IL 60201, USA}
\affiliation{The Adler Planetarium, Chicago, IL 60605, USA}
\email{amiller@northwestern.edu}

\author[0000-0002-0629-8931]{M.~R.~Magee}
\affiliation{School of Physics,
             Trinity College Dublin,
             The University of Dublin,
             Dublin 2, Ireland}

\author[0000-0002-1633-6495]{A.~Polin}
\affiliation{Departments of Physics and Astronomy,
             University of California, Berkeley,
             Berkley, CA 94720, USA}

\author[0000-0002-9770-3508]{K.~Maguire}
\affiliation{School of Physics,
             Trinity College Dublin,
             The University of Dublin,
             Dublin 2, Ireland}

\author[0000-0001-8985-2493]{E.~Zimmerman}
\affiliation{Department of Particle Physics and Astrophysics,
             Weizmann Institute of Science,
             234 Herzl St, 76100 Rehovot, Israel}

\author[0000-0001-6747-8509]{Y.~Yao}
\affiliation{Cahill Center for Astrophysics,
             California Institute of Technology,
             1200 E.~California Boulevard, Pasadena, CA 91125, USA}

\author[0000-0003-1546-6615]{J.~Sollerman}
\affiliation{Department of Astronomy, 
             The Oskar Klein Center, Stockholm University, 
             AlbaNova, SE-10691 Stockholm, Sweden}

\author[0000-0001-6797-1889]{S.~Schulze}
\affiliation{Department of Particle Physics and Astrophysics, 
             Weizmann Institute of Science, 
             234 Herzl St, 76100 Rehovot, Israel}

\author[0000-0001-8472-1996]{D.~A.~Perley}
\affiliation{Astrophysics Research Institute, 
             Liverpool John Moores University, 
             IC2, Liverpool Science Park, 146 Brownlow Hill, 
             Liverpool L3 5RF, UK}

\author[0000-0003-4380-7536]{M.~Kromer}
\affiliation{Heidelberger Institut f\"{u}r Theoretische Studien,
             Schloss-Wolfsbrunnenweg 35, D-69118 Heidelberg, Germany}

\author[0000-0002-2376-6979]{S.~Dhawan}
\affiliation{The Oskar Klein Centre, 
             Department of Physics, 
             Stockholm University, AlbaNova, SE-10691 Stockholm, Sweden}

\author[0000-0002-8255-5127]{M.~Bulla}
\affiliation{Nordita,
             KTH Royal Institute of Technology and Stockholm University,
             Roslagstullsbacken 23, SE-106 91 Stockholm, Sweden}

\author[0000-0002-8977-1498]{I.~Andreoni}
\affiliation{Cahill Center for Astrophysics,
             California Institute of Technology,
             1200 E.~California Boulevard, Pasadena, CA 91125, USA}

\author[0000-0001-8018-5348]{E.~C.~Bellm}
\affiliation{DIRAC Institute, 
             Department of Astronomy, University of Washington, 
             3910 15th Avenue NE, Seattle, WA 98195, USA} 

\author[0000-0002-8989-0542]{K.~De}
\affiliation{Cahill Center for Astrophysics,
             California Institute of Technology,
             1200 E.~California Boulevard, Pasadena, CA 91125, USA}

\author[0000-0002-5884-7867]{R.~Dekany}
\affiliation{Caltech Optical Observatories, 
             California Institute of Technology, Pasadena, CA 91125, USA}

\author{A.~Delacroix}
\affiliation{Caltech Optical Observatories, 
             California Institute of Technology, Pasadena, CA 91125, USA}

\author[0000-0002-4223-103X]{C.~Fremling}
\affiliation{Cahill Center for Astrophysics,
             California Institute of Technology,
             1200 E.~California Boulevard, Pasadena, CA 91125, USA}

\author[0000-0002-3653-5598]{A.~Gal-Yam}
\affiliation{Department of Particle Physics and Astrophysics,
             Weizmann Institute of Science,
             234 Herzl St, 76100 Rehovot, Israel}

\author[0000-0003-3461-8661]{D.~A.~Goldstein}
\altaffiliation{Hubble Fellow}
\affiliation{Cahill Center for Astrophysics,
             California Institute of Technology,
             1200 E.~California Boulevard, Pasadena, CA 91125, USA}

\author[0000-0001-8205-2506]{V.~Z.~Golkhou}
\affiliation{DIRAC Institute, 
             Department of Astronomy, University of Washington, 
             3910 15th Avenue NE, Seattle, WA 98195, USA} 
\affiliation{The eScience Institute, 
             University of Washington, Seattle, WA 98195, USA}

\author[0000-0002-4163-4996]{A.~Goobar}
\affiliation{The Oskar Klein Centre, 
             Department of Physics, 
             Stockholm University, AlbaNova, SE-10691 Stockholm, Sweden}

\author[0000-0002-3168-0139]{M.~J.~Graham}
\affiliation{Cahill Center for Astrophysics,
             California Institute of Technology,
             1200 E.~California Boulevard, Pasadena, CA 91125, USA}

\author[0000-0002-7996-8780]{I.~Irani}
\affiliation{Department of Particle Physics and Astrophysics,
             Weizmann Institute of Science,
             234 Herzl St, 76100 Rehovot, Israel}

\author[0000-0002-5619-4938]{M.~M.~Kasliwal}
\affiliation{Cahill Center for Astrophysics,
             California Institute of Technology,
             1200 E.~California Boulevard, Pasadena, CA 91125, USA}

\author{S.~Kaye}
\affiliation{Caltech Optical Observatories, 
             California Institute of Technology, Pasadena, CA 91125, USA}

\author[0000-0002-1031-0796]{Y.-L.~Kim}
\affiliation{Universit\'e de Lyon, 
             Universit\'e Claude Bernard Lyon 1, CNRS/IN2P3, 
             IP2I Lyon, F-69622, Villeurbanne, France}

\author[0000-0003-2451-5482]{R.~R.~Laher}
\affiliation{IPAC, California Institute of Technology, 
             1200 E. California Blvd, 
             Pasadena, CA 91125, USA}

\author[0000-0003-2242-0244]{A.~A.~Mahabal}
\affiliation{Cahill Center for Astrophysics,
             California Institute of Technology,
             1200 E.~California Boulevard, Pasadena, CA 91125, USA}
\affiliation{Center for Data Driven Discovery, 
             California Institute of Technology, Pasadena, CA 91125, USA}

\author[0000-0002-8532-9395]{F.~J.~Masci}
\affiliation{IPAC, California Institute of Technology, 
             1200 E. California Blvd, 
             Pasadena, CA 91125, USA}

\author[0000-0002-3389-0586]{P.~E.~Nugent}
\affiliation{Computational Cosmology Center, 
             Lawrence Berkeley National Laboratory, 
             1 Cyclotron Road, Berkeley, CA 94720, USA}
\affiliation{Departments of Physics and Astronomy,
             University of California, Berkeley,
             Berkley, CA 94720, USA}

\author[0000-0002-6786-8774]{E.~Ofek}
\affiliation{Department of Particle Physics and Astrophysics,
             Weizmann Institute of Science,
             234 Herzl St, 76100 Rehovot, Israel}

\author[0000-0002-9656-4032]{E.~S.~Phinney}
\affiliation{Cahill Center for Astrophysics,
             California Institute of Technology,
             1200 E.~California Boulevard, Pasadena, CA 91125, USA}

\author[0000-0003-0486-6242]{S.~J.~Prentice}
\affiliation{School of Physics,
             Trinity College Dublin,
             The University of Dublin,
             Dublin 2, Ireland}

\author[0000-0002-0387-370X]{R.~Riddle}
\affiliation{Caltech Optical Observatories,
             California Institute of Technology, Pasadena, CA 91125, USA}

\author[0000-0002-8121-2560]{M.~Rigault}
\affiliation{Universit\'{e} Clermont Auvergne,
             CNRS/IN2P3, Laboratoire de Physique de Clermont,
             F-63000 Clermont-Ferrand, France}

\author[0000-0001-7648-4142]{B.~Rusholme}
\affiliation{IPAC, California Institute of Technology,
             1200 E. California Blvd,
             Pasadena, CA 91125, USA}

\author{T.~Schweyer}
\affiliation{Department of Astronomy,
             The Oskar Klein Center, Stockholm University,
             AlbaNova, SE-10691 Stockholm, Sweden}

\author[0000-0003-4401-0430]{D.~L.~Shupe}
\affiliation{IPAC, California Institute of Technology,
             1200 E. California Blvd,
             Pasadena, CA 91125, USA}

\author[0000-0001-6753-1488]{M.~T.~Soumagnac}
\affiliation{Lawrence Berkeley National Laboratory,
             1 Cyclotron Road, Berkeley, CA 94720, USA}
\affiliation{Department of Particle Physics and Astrophysics,
             Weizmann Institute of Science,
             234 Herzl St, 76100 Rehovot, Israel}

\author[0000-0003-0794-5982]{G.~Terreran}
\affiliation{Center for Interdisciplinary Exploration and Research in 
             Astrophysics (CIERA) and Department of Physics and Astronomy, 
             Northwestern University, 
             1800 Sherman Road, Evanston, IL 60201, USA}

\author{R.~Walters}
\affiliation{Cahill Center for Astrophysics,
             California Institute of Technology,
             1200 E.~California Boulevard, Pasadena, CA 91125, USA}
\affiliation{Caltech Optical Observatories,
             California Institute of Technology, Pasadena, CA 91125, USA}

\author[0000-0003-1710-9339]{L.~Yan}
\affiliation{Cahill Center for Astrophysics,
             California Institute of Technology,
             1200 E.~California Boulevard, Pasadena, CA 91125, USA}

\author{J.~Zolkower}
\affiliation{Caltech Optical Observatories,
             California Institute of Technology, Pasadena, CA 91125, USA}

\author[0000-0001-5390-8563]{S.~R.~Kulkarni}
\affiliation{Cahill Center for Astrophysics,
             California Institute of Technology,
             1200 E.~California Boulevard, Pasadena, CA 91125, USA}


\begin{abstract}

Early observations of Type Ia supernovae (SNe\,Ia) provide essential clues for
understanding the progenitor system that gave rise to the terminal
thermonuclear explosion. We present exquisite observations of \sn, the second
observed SN\,Ia, after iPTF\,14atg, to display an early flash of emission in
the ultraviolet (UV) and optical. Our analysis finds that \sn\ was unusual,
even when ignoring the initial flash, in that it was moderately underluminous
for an SN\,Ia ($M_g \approx -18.5$\,mag at peak) yet featured very high
absorption velocities ($v \approx 15,000$\,\kms\ for \ion{Si}{II}
$\lambda$6355 at peak). We find that many of the observational features of
\sn, aside from the flash, can be explained if the explosive yield of
radioactive \radni\ is relatively low (we measure $M_{^{56}\mathrm{Ni}} = 0.31
\pm 0.05\,M_\odot$) and it and other iron-group elements are concentrated in
the innermost layers of the ejecta. To explain both the UV/optical flash and
peak properties of \sn\ we consider four different models: interaction between
the SN ejecta and a nondegenerate companion, extended clumps of \radni\ in the
outer ejecta, a double-detonation explosion, and the violent merger of two
white dwarfs. Each of these models has shortcomings when compared to the
observations; it is clear additional tuning is required to better match \sn.
In closing, we predict that the nebular spectra of \sn\ will feature either H
or He emission, if the ejecta collided with a companion, strong [\ion{Ca}{II}]
emission, if it was a double detonation, or narrow [\ion{O}{I}] emission, if
it was due to a violent merger.

\end{abstract}


\section{Introduction}

There is now no doubt that Type Ia supernovae (SNe\,Ia) are the result of
thermonuclear explosions in C/O white dwarfs (WDs) in multiple star systems
\citep[see, e.g.,][and references therein]{Maoz14}. Despite this certainty,
the nature of the binary companion, which plays an essential role in driving
the primary WD toward explosion, remains highly uncertain.

Historically, most studies have focused on whether or not the companion is
also a WD, the double degenerate (DD) scenario \citep[e.g.,][]{Webbink84}, or
some other nondegenerate star, the single degenerate (SD) scenario
\citep[e.g.,][]{Whelan73}. In addition to this fundamental question, recent
efforts have also focused on whether or not sub-Chandrasekhar mass WDs can
explode \citep[e.g.,][]{Fink10,Scalzo14a,Shen14,Polin19,Gronow20} and the
specific scenario in which the WD explodes \citep[see][and references
therein]{Hillebrandt13,Ropke18}.

Unfortunately, maximum-light observations of SNe\,Ia have not provided the
discriminatory power necessary to answer these questions and infer the
progenitor system \citep[e.g.,][]{Ropke12}.\footnote{Indeed, SNe Ia are
standardizable candles precisely because they are so uniform at this phase.}
It has recently been recognized that extremely early observations, in the
hours to days after explosion, may help to constrain which progenitor
scenarios are viable and which are not. In particular, \citet{Kasen10a} showed
that for favorable configurations in the SD scenario, the SN ejecta will
collide with the nondegenerate companion producing a shock that gives rise to
an ultraviolet (UV)/optical flash in excess of the typical emission from an
SN\,Ia.

The findings in \citet{Kasen10a} launched a bevy of studies to search for such
a signal \citep[e.g.,][]{Hayden10,Bianco11,Ganeshalingam11,Nugent11,Olling15},
including several claims of a detection of the interaction with a
nondegenerate companion (e.g.,
\citealt{Cao15,Marion16,Hosseinzadeh17,Dimitriadis19}; though see also
\citealt{Kromer16,Jiang18,Shappee18,Shappee19} for alternative explanations).
In the meantime, it has been found that an early optical bump, or flash, in
the light curves of SNe\,Ia is not uniquely limited to the SD scenario
\citep[e.g.,][]{Raskin13,Piro16,Jiang17,Levanon17,Noebauer17,Maeda18,De19,
Polin19, Magee20a}.

Despite some observational degeneracies, early observations have and will
continue to play a critical role in understanding the progenitors of SNe\,Ia
\citep[e.g., early photometry of SN\,2011fe constrained the size of the
exploding star to be $\lesssim 0.02$\,$R_\odot$, providing the most direct
evidence to date that SNe\,Ia come from WDs;][]{Bloom12a}.

Here we present X-ray, UV, and optical observations of the spectacular \sn,
only the second observed SN\,Ia, after iPTF\,14atg \citep{Cao15}, to exhibit
an early UV flash.\footnote{``Excess'' emission or early optical bumps have
been observed and claimed in many other SNe\,Ia
\citep[e.g.,][]{Goobar15,Marion16,Hosseinzadeh17,Jiang17,Dimitriadis19,
Shappee19}. These events lack a distinct early decline in the UV, however,
which distinguishes iPTF\,14atg and \sn.} \sn\ declined by $\sim$2.5\,mag in
the UV in the $\sim$3\,d after discovery followed by a more gradual rise and
fall, typical of SNe\,Ia, in the ensuing weeks. Our observations and analysis
show that, even if the early flash had been observationally missed, we would
conclude that \sn\ is unusual relative to normal SNe\,Ia. We consider several
distinct models to explain the origin of \sn\ and find that they all have
considerable shortcomings. Spectroscopic observations of \sn\ obtained during
the nebular phase will narrow the range of potential explanations for this
highly unusual explosion.

Along with this paper, we have released our open-source analysis and the data
utilized in this study. These are available online at
\href{https://github.com/adamamiller/SN19yvq}
{\url{https://github.com/adamamiller/SN19yvq}}; a version of these materials
is archived on Zenodo
(doi:\href{https://doi.org/10.5281/zenodo.3897419}{10.5281/zenodo.3897419}).

\section{Discovery and Observations}\label{sec:obs}

\sn\ was discovered by \citet{Itagaki19}, and detected at an unfiltered
magnitude of 16.7\,mag, in an image obtained on 2019 December 28.74
UT.\footnote{UT times are used throughout this paper.} The transient candidate
was announced $\sim$2\,hr later on the Transient Name Server, and given the
designation AT\,2019yvq \citep{Itagaki19}. Subsequent spectroscopic
observations confirmed the SN nature of the transient, with an initial report
that the event was an SN\,Ib/c, and subsequent spectra confirming the event as
an SN\,Ia.\footnote{The initial classification is from \citet{Kawabata20},
while the SN\,Ia classifications are from Prentice, Mazzali, Teffs \& Medler
and Dahiwale \& Fremling (see
\href{https://wis-tns.weizmann.ac.il/search?&name=SN2019yvq}
{\url{{https://wis-tns.weizmann.ac.il/search?&name=SN2019yvq}}}).} These
spectroscopic observations also showed \sn\ to be at the same redshift as
NGC\,4441, its host galaxy.

\subsection{Zwicky Transient Facility (ZTF) Photometric Observations}

\begin{deluxetable}{lrrrrrcccccccccc}
\tabletypesize{\footnotesize}
\tablewidth{0pt}
\tablecaption{ZTF P48 Photometry of \sn\label{tab:p48}}
\tablehead{
\colhead{MJD}
&&&&& \colhead{Flux}
&&&&& \colhead{$\sigma_\mathrm{flux}$}
&&&&& \colhead{Filter} \\
\colhead{}
&&&&& \colhead{($\mu\mathrm{Jy}$)}
&&&&& \colhead{($\mu\mathrm{Jy}$)}
&&&&& \colhead{}
}
\startdata
58,846.4699 &&&&& 504.81 &&&&& 7.28 &&&&& \rztf \\
58,846.5385 &&&&& 374.33 &&&&& 4.99 &&&&& \iztf \\
58,846.5583 &&&&& 595.33 &&&&& 5.56 &&&&& \gztf \\
58,849.4489 &&&&& 487.54 &&&&& 7.75 &&&&& \rztf \\
58,849.5078 &&&&& 379.06 &&&&& 5.54 &&&&& \gztf \\
\enddata
\tablecomments{
Observed fluxes in the ZTF passbands, no correction for reddening has been applied.
Due to poor observing conditions, \sn\ is not detected in 
one \gztf\ and one \iztf\ image from 2020 March 9, and we therefore 
do not provide a flux measurement for those epochs. \\
(This table is available in its entirety in a machine-readable 
form in the online journal.)
}
\end{deluxetable}

\begin{figure*}
    \centering
    \includegraphics[width=6in]{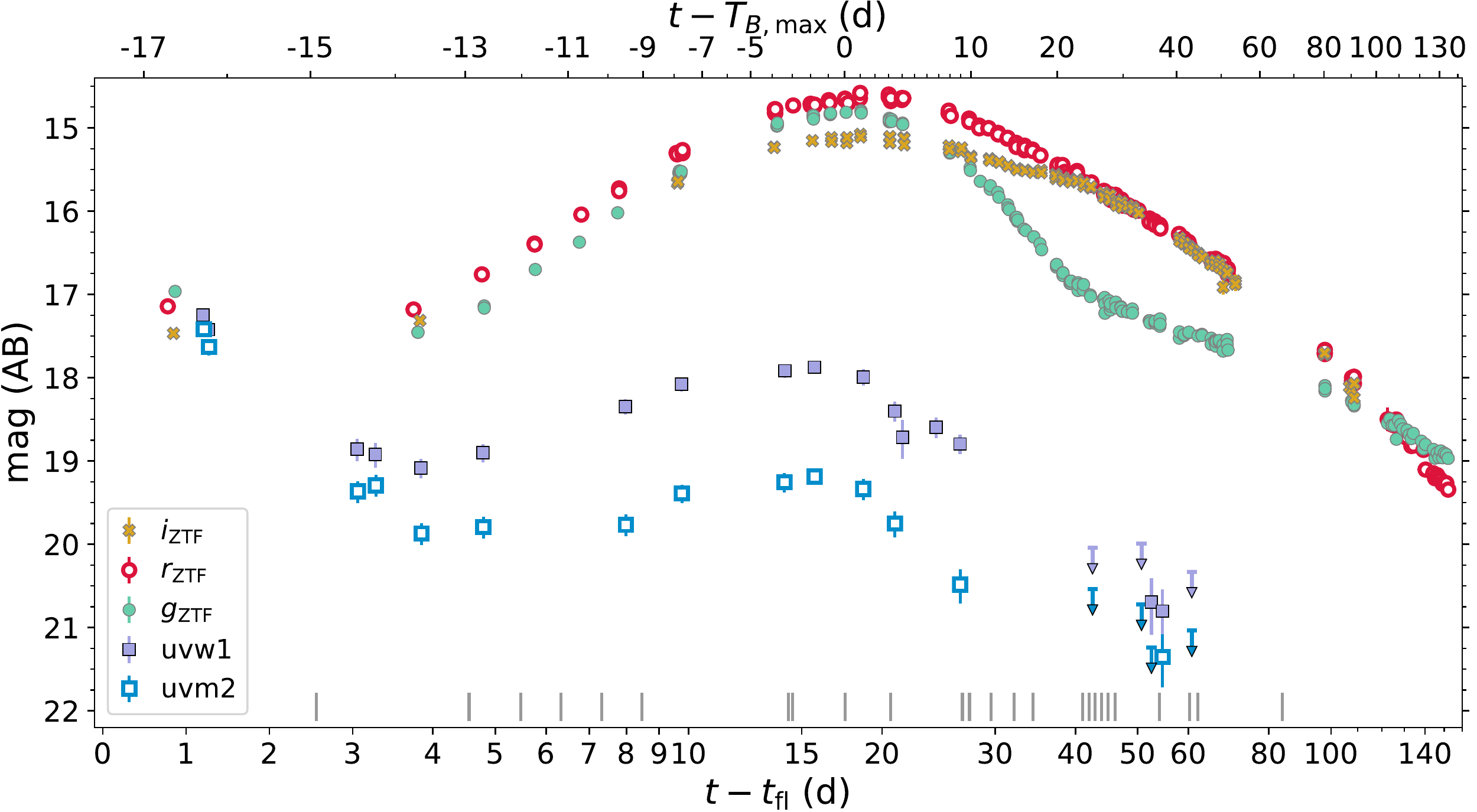}
    \caption{Photometric evolution of \sn, highlighting the initial decline
    observed in the light curve. \gztf, \rztf, and \iztf\ observations are
    shown as filled green circles, open red circles, and filled golden
    crosses, respectively. UVOT $uvw1$ and $uvm2$ are shown as filled and open
    squares, respectively. Upper limits are shown as downward pointing arrows.
    The lower axis shows time measured in rest-frame days relative to the time
    of first light, \tfl\ (see \S\ref{sec:phot}), while the upper axis shows
    time relative to the time of $B$-band maximum, \tbmax. Note that the
    horizontal axis is shown with a linear scale from $0\,\mathrm{d} \le t -
    t_\mathrm{fl} \le 3$\,d and a log scale for $t - t_\mathrm{fl} > 3$\,d.
    Vertical gray ticks show the epochs of the spectroscopic observations.}
    \label{fig:p48}
\end{figure*}

ZTF \citep{Bellm19,Graham19,Dekany20}) simultaneously conducts multiple
time-domain surveys using the ZTF camera on the the Palomar Oschin Schmidt 48
inch (P48) telescope. \sn\ was first detected by ZTF on 2019 December 29.46,
as part of the ZTF public survey (see \citealt{Bellm19a}). The automated ZTF
pipeline \citep{Masci19} detected \sn\ using the image-differencing technique
of \citet{Zackay16}. The candidate passed internal thresholds (e.g.,
\citealt{Mahabal19}), leading to the production and dissemination of a
real-time alert \citep{Patterson19} and the internal designation ZTF19adcecwu.
The public alert included the position, $\alpha =
12^{\mathrm{h}}27\arcmin21\farcs836$, $\delta = +64\degr47\arcmin59\farcs87$
(J2000), and brightness, \rztf$ = 17.14\pm0.05$\,mag, which, together with the
\citet{Itagaki19} discovery report suggested the SN was fading. There was an
$\sim$8\,d gap in ZTF observations prior to its initial detection of \sn,
meaning ZTF nondetections cannot directly constrain the time of explosion,
$t_\mathrm{exp}$. Continued monitoring with ZTF, and follow-up with other
telescopes, confirmed a spectacular decline in the early emission from \sn\
(Figure~\ref{fig:p48}).

The field of \sn\ was additionally observed by ZTF with nearly a nightly
cadence as part of the ZTF partnership Uniform Depth Survey (ZUDS;
D.~Goldstein et al.~2020, in preparation). Using images obtained as part of
the ZUDS program, we perform forced point-spread function (PSF) photometry at
the location of \sn\ following the procedure described in
\citet{Yao19}.\footnote{Images of \sn obtained as part of the ZTF public
survey have not been released, to either the public or members of the ZTF
collaboration, preventing us from applying forced-PSF measurements. We
therefore only include the ZTF partnership ZUDS images in the analysis
described herein. Our measurements are largely consistent with those provided
in the public ZTF alerts.} The evolution of \sn\ in the \gztf, \rztf, and
\iztf\ filters is shown in Figure~\ref{fig:p48}, and the ZTF flux measurements
are summarized in Table~\ref{tab:p48}.

\subsection{\textit{Swift} Ultraviolet/Optical Telescope (UVOT) and X-ray Telescope (XRT) Observations}\label{sec:swift}

\begin{deluxetable}{lrrrrrrrrrrccccc}
\tabletypesize{\footnotesize}
\tablewidth{0pt}
\tablecaption{UVOT Photometry of \sn\label{tab:uvot}}
\tablehead{
\colhead{MJD}
&&&&& \colhead{Flux}
&&&&& \colhead{$\sigma_\mathrm{flux}$}
&&&&& \colhead{Filter} \\
\colhead{}
&&&&& \colhead{($\mu\mathrm{Jy}$)}
&&&&& \colhead{($\mu\mathrm{Jy}$)}
&&&&& \colhead{}
}
\startdata
58,846.8969 &&&&& 457.90 &&&&& 30.80 &&&&& $uvw1$ \\
58,846.9017 &&&&& 314.30 &&&&& 22.96 &&&&& $uvw2$ \\
58,846.9066 &&&&& 392.00 &&&&& 24.90 &&&&& $uvm2$ \\
58,846.9607 &&&&& 390.40 &&&&& 27.64 &&&&& $uvw1$ \\
58,846.9655 &&&&& 307.40 &&&&& 22.56 &&&&& $uvw2$ \\
\enddata
\tablecomments{ 
Host-subtracted fluxes in the UVOT passbands, 
no correction for reddening has been applied. 
Epochs with a signal-to-noise ratio $\mathrm{(S/N)} < 3$ are shown as upper limits in Figure~\ref{fig:p48}. \\
(This table is available in its entirety in a machine-readable 
form in the online journal.)
}
\end{deluxetable}

UV observations of \sn\ were obtained with the UVOT \citep{Roming05}) onboard
the Neil Gehrels Swift Observatory (hereafter \textit{Swift};
\citealt{Gehrels04}) following a time-of-opportunity
request.\footnote{\textit{Swift} ToO requests for \sn\ (\textit{Swift} Target
ID: 13037) were submitted by D.~Hiramatsu, J.~Burke, and S.~Schulze.} Pre-SN
UVOT reference images are limited to the $uvw1$, $uvm2$, and $uvw2$ filters.
Therefore, accurate estimates of the SN flux in the \textit{Swift} $u$, $b$,
and $v$ filters are not possible.

We estimate the flux in the $uvw1$, $uvm2$, and $uvw2$ filters using a
circular aperture with a $3\arcsec$ radius at the SN position, and subtract
the flux measured using an identical procedure in the pre-SN images, as
summarized in Table~\ref{tab:uvot}. For clarity, we only show the
\textit{Swift} $uvw1$ and $uvm2$ light curves in
Figure~\ref{fig:p48}.\footnote{The $uvw2$ evolution is nearly identical to
$uvm2$. Furthermore, the red leak associated with the $uvw2$ filter (see e.g.,
\citealt{Breeveld11}), in combination with the relatively red spectral energy
distribution of SNe\,Ia, make it very difficult to interpret $uvw2$ light
curves of SNe\,Ia (see \citealt{Brown17} and references therein). Therefore,
unless otherwise noted, we exclude $uvw2$ measurements from the analysis
below.} \textit{Swift}/UVOT observations show that the initial decline seen in
the optical is even more dramatic in the UV.

While absolute flux measurements in the UVOT $u$, $b$, and $v$ filters are not
available, if we assume the underlying flux from the host is constant in time
we can estimate the time of $B$-band maximum, \tbmax, from the relative
$b$-band light curve. Using a second-order polynomial, we model the $b$-band
light curve near peak (including observations between MJD$\,> \,$58,855 and
MJD$\,<\,$58,871). From this fit we measure $T_{B,\mathrm{max}} = 58$,863.33$
\,\pm \,0.21$\,MJD. The UVOT $b$ filter is slightly different from the
Johnson $B$ filter, with the latter typically being used to estimate \tbmax.
Using nine SNe with \tbmax\ estimates from Johnson $B$-band observations
\citep{Krisciunas17}, we repeat the above procedure on \textit{Swift} $b$-band
observations \citep[data from][]{Brown14}. We find that most of these SNe have
\tbmax\ measurements consistent to within the uncertainties. On average,
\tbmax\ estimates from \textit{Swift} $b$-band observations occur later than
those in the Johnson $B$-band, with a median offset of $\sim$0.26\,d.

In parallel with the \textit{Swift}/UVOT observations, \textit{Swift} observed
\sn\ with the XRT \citep{Burrows05} between 0.3 and 10\,keV in the photon
counting mode from 2019 December 29 through 2020 February 27. We analyzed the
data with the online tools of the UK \textit{Swift}
team\footnote{\href{https://www.swift.ac.uk/user_objects/}
{\url{https://www.swift.ac.uk/user_objects/}}}, which uses the methods
described in \citet{Evans07,Evans09} and the software package
\texttt{HEASOFT}\footnote{
\href{https://heasarc.gsfc.nasa.gov/docs/software/heasoft/}
{\url{https://heasarc.gsfc.nasa.gov/docs/software/heasoft/}}} version 6.26.1
\citep{Heasarc}.

To build the light curve of \sn\ and test whether transient X-ray emission is
present at the SN position, we stack the data of each \textit{Swift} observing
segment. In the pre-SN observations from 2012, we detect X-ray emission at the
position of \sn. The average count rate in the 2012 observations is
$0.0026\pm0.0008\,\mathrm{ct\,s}^{-1}$ (0.3--10\,keV). The detected count rate
during observations of \sn\ is marginally higher than in 2012, however,
spectra of the two epochs show no differences to within the uncertainties.
Therefore, the same source from 2012 dominates the ongoing emission at the
position of \sn.

In the first epoch of XRT observations of \sn, corresponding to the time we
would expect the X-ray flux to be largest if the UV/optical flash is due to
the collision of the ejecta with either circumstellar material or a
nondegenerate companion, we marginally detect emission at the position of \sn\
with a count rate of $0.0031^{+0.0017}_{-0.0013}$\,ct\,s$^{-1}$. This flux is
identical to that measured in 2012 to within the uncertainties. To estimate an
upper limit on the SN flux, we take the difference between the 2019 and 2012
flux measurements and arrive at a $3\sigma$ upper limit on the SN count rate
of $< 0.0057$\,ct\,s$^{-1}$. The upper limits in future epochs of XRT
observations are less constraining than this first epoch.
 
To convert the count rate to flux, we extracted a spectrum of the 2019--2020
data set. The spectrum is adequately described with an absorbed power law
where the two absorption components represent absorption in the Milky Way and
the host galaxy. The Galactic equivalent neutral-hydrogen column density was
fixed to $2.03\times10^{20}~{\rm cm}^{-2}$ \citep{HI4PI2016a}. The best fit
suggests negligible host absorption, though we note that the data do not
constrain this parameter, and a photon index\footnote{The photon index is
defined as $A(E) \propto E^{-\Gamma}$.} of $\Gamma = 1.9^{+1.0}_{-0.5}$ (all
uncertainties at 90\% confidence; $\chi^2=30.8$, with 32 degrees of freedom
assuming Cash statistics). From this fit the unabsorbed count-rate-to-energy
conversion factor is
$5\times10^{-11}\,\mathrm{erg\,cm}^{-2}\,\mathrm{ct}^{-1}$.

From the count-rate conversion factor, we estimate an upper limit on the X-ray
flux of $2.9 \times 10^{-13}$\,erg\,cm$^{-2}$\,s$^{-1}$ at the first epoch of
\textit{Swift} observations. At the distance of \sn\ (see \S\ref{sec:host}),
this corresponds to an X-ray luminosity of $L_X < 6.2 \times
10^{40}$\,erg\,s$^{-1}$. This luminosity is significantly lower than the
$\sim$5$\times 10^{44}$\,erg\,s$^{-1}$ estimate from \citet{Kasen10a} for the
interaction between the SN ejecta and a nondegenerate companion. However, this
discrepancy is not surprising as the X-ray emission is only expected to last
for minutes to hours, and the \textit{Swift} observations occurred at least
1.1\,d after explosion (based on the initial detection from
\citealt{Itagaki19}).

\subsection{Optical Spectroscopy}

\begin{deluxetable}{lrcccc}
\tabletypesize{\scriptsize}
\tablewidth{0pt}
\tablecaption{Spectroscopic Observations of \sn\label{tab:spectra}}
\tablehead{
\colhead{$t_\mathrm{obs}$} &
\colhead{Phase} &
\colhead{Telescope/} &
\colhead{$R$} &
\colhead{Range} &
\colhead{Air} \\
\colhead{(MJD)} &
\colhead{(d)} &
\colhead{Instrument} &
\colhead{$(\Delta\lambda/\lambda)$} &
\colhead{(\AA)} & 
\colhead{Mass}
}
\startdata
58,848.27 & $-$14.9 & LT/SPRAT & 350 & 4020--7990 & 1.24 \\
58,850.28 & $-$12.9 & LT/SPRAT & 350 & 4020--7990 & 1.24 \\
58,851.21 & $-$12.0 & LT/SPRAT & 350 & 4020--7990 & 1.29 \\
58,852.07 & $-$11.2 & LT/SPRAT & 350 & 4020--7990 & 1.88 \\
58,853.07 & $-$10.2 & LT/SPRAT & 350 & 4020--7990 & 1.86 \\
58,854.22 &  $-$9.0 & LT/SPRAT & 350 & 4020--7990 & 1.27 \\
58,860.13 &  $-$3.2 & LT/SPRAT & 350 & 4020--7990 & 1.46 \\
58,860.34 &  $-$3.0 & P60/SEDM & 100 & 3850--9150 & 1.64 \\
58,863.38 &  $+$0.0 & P60/SEDM & 100 & 3850--9150 & 1.40 \\
58,866.50 &  $+$3.1 & MMT/Binospec & 4000 & 4645--6155 & 1.19 \\
58,872.61 &  $+$9.2 & Keck I/LRIS & 1100 & 3200--10250 & 1.41 \\
58,873.30 &  $+$9.9 & P60/SEDM & 100 & 3850--9150 & 1.64 \\
58,875.54 & $+$12.1 & P60/SEDM & 100 & 3850--9150 & 1.19 \\
58,878.09 & $+$14.6 & NOT/ALFOSC & 360 & 3760--9620 & 1.41 \\
58,880.39 & $+$16.9 & P60/SEDM & 100 & 3850--9150 & 1.26 \\
58,887.10 & $+$23.5 & LT/SPRAT & 350 & 4020--7990 & 1.32 \\
58,888.07 & $+$24.5 & LT/SPRAT & 350 & 4020--7990 & 1.39 \\
58,888.97 & $+$25.4 & LT/SPRAT & 350 & 4020--7990 & 1.87 \\
58,890.01 & $+$26.4 & LT/SPRAT & 350 & 4020--7990 & 1.60 \\
58,891.06 & $+$27.5 & LT/SPRAT & 350 & 4020--7990 & 1.40 \\
58,892.25 & $+$28.6 & P60/SEDM & 100 & 3850--9150 & 1.66 \\
58,900.22 & $+$36.5 & P60/SEDM & 100 & 3850--9150 & 1.69 \\
58,906.45 & $+$42.7 & P200/DBSP & 700 & 3410--9995 & 1.19 \\
58,908.32 & $+$44.6 & P60/SEDM & 100 & 3850--9150 & 1.25 \\
58,930.47 & $+$66.5 & Keck I/LRIS & 1100 & 3200--10250 & 1.42 \\
\enddata
\tablecomments{Phase is measured relative to \tbmax\ in the SN rest frame. The resolution $R$ is reported for the central region of the spectrum.
}
\end{deluxetable}

\begin{figure}
    \centering
    \includegraphics[width=\columnwidth]{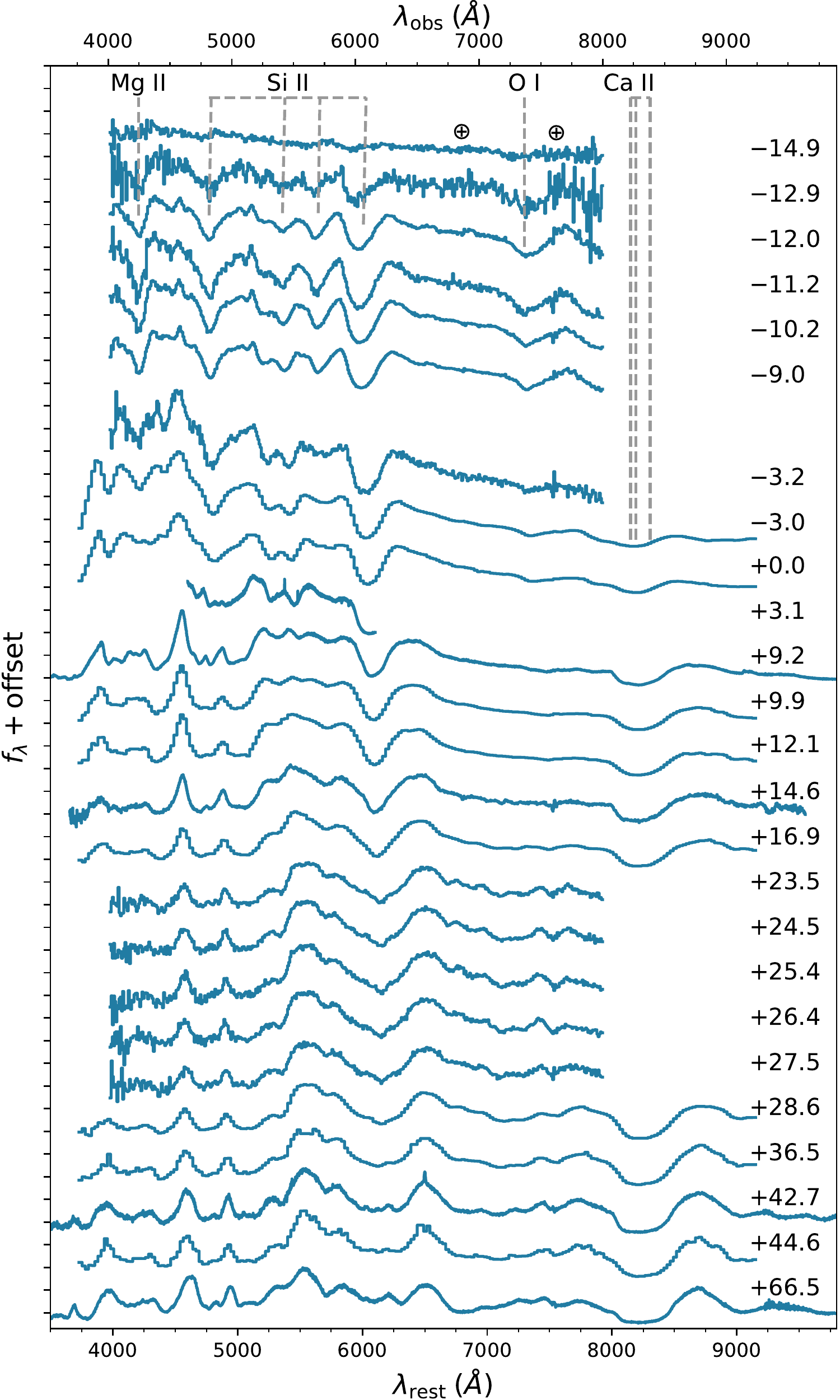}
    \caption{Observed spectral sequence of \sn. Spectra have been normalized
    by their median flux between 7200 and 7400\,\AA. The phase of each
    observation relative to \tbmax\ is shown to the right of the individual
    spectra. Prominent spectral features from intermediate mass elements
    (IMEs) are highlighted with vertical-dashed lines. Some of the spectra
    show imperfect Telluric subtractions, giving rise to the non-smooth
    features around $\lambda_\mathrm{obs} \approx 7600$\,\AA. The blue and red
    edges of the $-12.9$\,d spectrum are not shown for clarity.}
    \label{fig:spec_evo}
\end{figure}

Spectroscopic observations of \sn\ were initiated because the transient passed
the threshold criteria for both the ZTF Bright Transient Survey
\citep{Fremling20} and the ZTF Census of the Local Universe experiment
\citep{De20}. Our first spectrum, obtained $\sim$1.8\,d after the initial ZTF
detection with the SPectrograph for the Rapid Acquisition of Transients
\citep[SPRAT;][]{Piascik14} on the 2\,m Liverpool Telescope
\citep[LT;][]{Steele04}, had a blue and nearly featureless continuum. Further
spectroscopy was obtained with a variety of telescopes, including: the
Spectral Energy Density machine \citep[SEDM;][]{Blagorodnova18,Rigault19} on
the Palomar 60 inch telescope (P60), Binospec \citep{Fabricant19} on the
6.5\,m MMT telescope, the Low-Resolution Imaging Spectrometer
\citep[LRIS;][]{Oke95} on the 10\,m Keck I telescope, the Andalucia Faint
Object Spectrograph and Camera
(ALFOSC)\footnote{\href{http://www.not.iac.es/instruments/alfosc}
{\url{http://www.not.iac.es/instruments/alfosc}}} on the 2.5\,m Nordic Optical
Telescope (NOT), and the Double Spectrograph \citep[DBSP;][]{Oke82} on the
Palomar 200\,in Hale Telescope. The optical spectral evolution of \sn\ is
illustrated in Figure~\ref{fig:spec_evo}, with an accompanying observing log
listed in Table~\ref{tab:spectra}.

With the exception of SEDM, all observations were obtained with the
slit positioned along the parallactic angle, and the spectra were reduced
using standard procedures in \texttt{IDL}/\texttt{Python}/\texttt{Matlab}.
SEDM is a low-resolution ($R \approx 100$) integral field unit
\citep{Blagorodnova18}, and the observations are reduced using the custom
\texttt{pysedm} software package \citep{Rigault19}. While SEDM was designed
specifically for SN classification \citep[e.g.,][]{Fremling20}, the quality
for \sn\ is high enough to provide detailed absorption line measurements (see
\S\ref{sec:SiII}).

\section{NGC\,4441: The Host of \sn}\label{sec:host}

NGC\,4441 is the host galaxy of \sn. Sloan Digital Sky Survey (SDSS;
\citealt{York00}) spectroscopic measurements of the nucleus of NGC\,4441 yield
a heliocentric-recession velocity of 2663\,\kms\ ($z_\mathrm{helio} =
0.00888$; \citealt{Abolfathi18}) and a template-matched \texttt{STARBURST}
classification for NGC\,4441. Morphologically, NGC\,4441 is classified as a
peculiar, weakly barred, late-type lenticular galaxy (SABO$+$ pec;
\citealt{de-Vaucouleurs91}). SDSS images show a clear dust lane near the
center of the galaxy.

Using the 2M++ model of \citet{Carrick15}, we estimate a peculiar velocity
toward NGC\,4441 of $+328.6$\,\kms, which combined with the recession
velocity in the frame of the cosmic microwave background\footnote{See
\url{https://ned.ipac.caltech.edu/velocity_calculator}} (CMB; $v_\mathrm{CMB}
= 2770.6$\,\kms), yields a total recession velocity $= 3099.2 \pm 150$\,\kms.
The final uncertainty in the total recession velocity is dominated by
systematic uncertainties in the 2M++ model. The 2M++ estimate is consistent,
to within $\sim$5\%, with the Virgo and Great Attractor infall models of
\citet{Mould00}. Adopting $H_0 = 73$\,\kms\,Mpc$^{-1}$, we estimate the
distance to NGC\,4441 to be $42.5 \pm 2.1$\,Mpc, corresponding to a distance
modulus of $\mu = 33.14 \pm 0.11$\,mag, where the uncertainty on $\mu$ is
dominated by the uncertainty in the peculiar velocity correction. We hereafter
adopt $33.14 \pm 0.11$\,mag as the distance modulus to
NGC\,4441.\footnote{\citet{Tully13} estimated a significantly smaller distance
to NGC\,4441 ($\mu = 31.43 \pm 0.14$\,mag; $D = 19.0$\,Mpc) based on surface
brightness fluctuation measurements from \citet{Tonry01}. If NGC\,4441 is at
this distance, then \sn\ peaks at $M_g \approx -16.8$\,mag, which is
significantly underluminous for a SN\,Ia. Given that \sn\ has a normal rise
time $t_\mathrm{rise} \approx 18$\,d (\S\ref{sec:phot}), relatively normal
spectra at peak (\S\ref{sec:spec}), and lacks the spectral signatures of
intrinsically faint SNe\,Ia (\S\ref{sec:spec_comp}), we consider such a low
luminosity improbable. We therefore adopt the larger kinematic distance to
NGC\,4441.}

\begin{figure} \centering \includegraphics[width=\columnwidth]{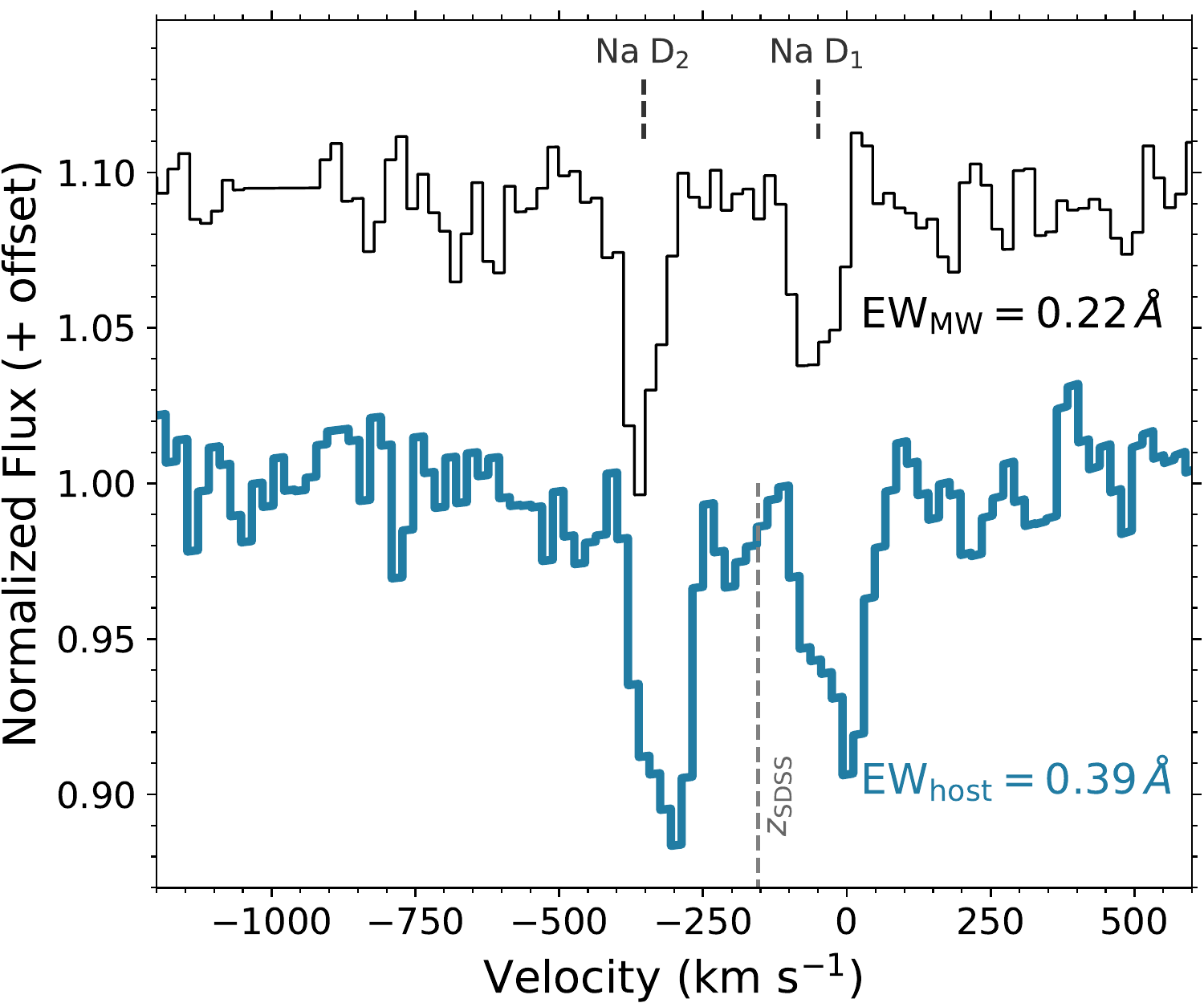}
    \caption{Zoom in on our moderate resolution ($R \approx 4000$)
    MMT+Binospec spectrum of \sn\ highlighting absorption due to \ion{Na}{I}~D
    in the host galaxy, NGC\,4441 (blue solid line), and the Milky Way (thin
    black line). The velocity scale is centered on the D$_1$ line in
    NGC\,4441, with the SDSS redshift shown via the vertical-dashed line. The
    velocity scale is centered on 5895.92\,\AA\ for the Milky Way absorption
    lines. The \ion{Na}{I} D lines, which serve as a proxy for interstellar
    dust-obscuration along the line of sight (e.g.,
    \citealt{Poznanski12,Phillips13}) are weak, indicating a relatively small
    amount of reddening.}
    \label{fig:NaD}
\end{figure}

We estimate the total reddening toward \sn\ to be small. There is relatively
little line-of-sight extinction due to the Milky Way, $E(B-V) \approx
0.018$\,mag \citep{Schlegel98,Schlafly11}. Furthermore, we do not find
significant evidence for strong interstellar extinction in NGC\,4441.
Figure~\ref{fig:NaD} highlights the \ion{Na}{I} D absorption in the spectrum
of \sn\ due to gas in NGC\,4441 and the Milky Way from our highest-resolution
spectrum, $R \approx 4000$, obtained with Binospec+MMT. The \ion{Na}{I} D
absorption is weak, and we estimate a total equivalent width (EW) $=
390$\,m\AA\ for NGC\,4441 and $220$\,m\AA\ for the Milky Way. There is a
systematic uncertainty of $\sim$10\% on each of these estimates due to
uncertainties in the continuum-fitting procedure.

Assuming similar properties for the dust in NGC\,4441 and the Milky Way, we
scale the color excess measurement for the Milky Way by the ratio of
\ion{Na}{I} D EWs to estimate $E(B-V) \approx 0.032$\,mag for \sn\ due to
interstellar absorption in NGC\,4441. This yields a total color excess toward
\sn\ of $E(B-V) \approx 0.05$\,mag, which we adopt for the subsequent analysis
in this study. We note that this estimate is consistent, to within the
uncertainties, with the EW(\ion{Na}{I} D)--$E(B-V)$ relations presented in
\citet{Poznanski12}. Further support for low interstellar extinction toward
\sn\ is the lack of a detection of the \ion{K}{I} $\lambda\lambda$7665, 7699
interstellar lines, or the diffuse interstellar band at 5780\,\AA, which also
serve as proxies for extinction \citep{Phillips13}.

The measured redshift of the \ion{Na}{I}~D doublet in the Binospec spectrum is
0.0094. We adopt this, rather than the SDSS measurement of 0.00888, as the
redshift of \sn, $z_\mathrm{SN}$. This choice does not ultimately play a
significant role in our final analysis, as our ejecta velocity measurements
and rest-frame time differences would change by $<1\%$ when using
$z_\mathrm{SDSS}$ versus our adopted $z_\mathrm{SN}$.

\section{Photometric Analysis}\label{sec:phot}

\subsection{The Time of First Light, \tfl}\label{sec:t_fl}

We estimate the time of first light, \tfl, for \sn\ following the procedure
described in \citet{Miller20}. Briefly, \citet{Miller20} model the early
emission from an SN\,Ia as a power law in time, $f \propto (t -
t_\mathrm{fl})^\alpha$, where $f$ is the flux, $t$ is time, and $\alpha$ is
the power-law index. \tfl\ is assumed to be the same everywhere in the
optical, allowing us to simultaneously fit observations in each of the ZTF
filters.

An important caveat for \sn\ is that the observed early decline in the light
curve clearly does not follow the simple power-law model, and thus these
observations must be masked when performing the fit. We conservatively exclude
observations from the first two nights of ZTF detection from the fit (this
choice is conservative as it is unclear when the mechanism that powers the
initial bump in \sn\ no longer significantly contributes to the flux in the
\gztf\ and \rztf\ filters). From the fit we find \tfl$ =
-17.5^{+1.0}_{-1.3}$\,d relative to \tbmax.\footnote{Here, and throughout this
study, times are reported in rest-frame days relative to \tfl\ or \tbmax.} We
know that the time of explosion must be $< -17.4$\,d based on the discovery
detection in \citealt{Itagaki19}, and, by definition $t_\mathrm{fl} \ge
t_\mathrm{exp}$, meaning a portion of the posterior distribution for our model
cannot be correct. We find $\alpha_g = 2.15^{+0.49}_{-0.36}$ and $\alpha_r =
1.91^{+0.42}_{-0.31}$, which are typical of the normal SNe Ia studied in
\citet{Miller20}. If we only exclude the first observation from the model fit
we find significantly different results with a rise time that increases by
$\sim$5\,d and power-law indices that increase by $\gtrsim 1$. We note that
such a long rise is unlikely, however, as our spectroscopic models (see
\S\ref{sec:tardis}) estimate the time of explosion, $t_\mathrm{exp}$, to be
$\sim$17.9\,d prior to \tbmax, fully consistent with our estimate of \tfl.

\subsection{Luminosity of the Initial UV/optical Flash}\label{sec:luminosity}

To estimate the luminosity and temperature of the initial UV/optical flash
from \sn, we model the broadband SED as a
blackbody. The assumed distance and reddening toward \sn\ are taken from
\S\ref{sec:host}. The ZTF optical and \textit{Swift} UV observations were not
simultaneous, so we interpolate the optical light curves to estimate the flux
during the same epochs as \textit{Swift} observations. While SNe\,Ia do not
emit as pure blackbodies, our initial spectrum of \sn\ shows a blue and nearly
featureless continuum largely consistent with blackbody emission. The
blackbody assumption is therefore reasonable for the early flash from \sn,
which is distinctly different from normal SNe.

Following interpolation to an epoch 1.24\,d after \tfl\ ($\mathrm{MJD} =
58$,846.93), and including the $uvw2$ filter, we estimate a blackbody
luminosity $L = (1.7^{+0.2}_{-0.1}) \times 10^{42}$\,erg\,s$^{-1}$ and
temperature $T_\mathrm{eff} = 14.8^{+0.9}_{-1.2}$\,kK. This estimate
represents a lower limit to the peak luminosity of the initial flash, as the
UV flux was already decreasing at this time (\textit{Swift} obtained two sets
of UV observations separated by $\sim$90\,min during the first epoch of
observations, and the $uvm2$ and $uvw1$ flux is clearly decreasing during this
time; see Figure~\ref{fig:p48}).

At an epoch 3.15\,d after \tfl, we estimate the luminosity and temperature to
have fallen to $L = (7.0^{+0.9}_{-0.6}) \times 10^{41}$\,erg\,s$^{-1}$ and
$T_\mathrm{eff} = 8.7^{+0.5}_{-0.4}$\,kK, respectively. For this epoch we have
excluded the $uvw2$ flux from the blackbody model due to the significantly
lower temperature, and known red leak for that filter (see \S\ref{sec:swift}).
This measurement of $T_\mathrm{eff}$ is consistent with our model spectrum
from 2.6\,d after \tfl\ (see \S\ref{sec:tardis}). At a similar epoch,
$\sim$4\,d after explosion, \citet{Cao15} estimated a UV flash luminosity of
$\sim$3$ \times 10^{41}$erg\,s$^{-1}$ in iPTF\,14atg, a factor of $\sim$2 less
than for \sn.

Finally, if we conservatively assume that the early flash peaked 1\,d after
\tfl\ (i.e., at the epoch of the first \textit{Swift} observation), and
abruptly ended 3\,d after \tfl\ (i.e., at the epoch of the second
\textit{Swift} observation), then the initial flash emitted a total integrated
energy of $\sim$4$\times 10^{42}$\,erg.
These assumed times are highly uncertain, however, it is likely that the SN
exploded before \tfl\ (see, e.g., \S\ref{sec:tardis}~and~\ref{sec:models}),
and the UV flux continues to decline $>$3\,d after \tfl\
(Figure~\ref{fig:p48}) suggesting the flash lasted longer than 3\,d.

\subsection{Bolometric Luminosity, \radni\ Mass, and Mass of the Ejecta}\label{sec:ni_mass}

While the early emission from \sn\ may be approximated as a blackbody, SNe\,Ia
do not emit as blackbodies around maximum light. To estimate the bolometric
luminosity of \sn, we model changes in the observed flux in the $uvm2$,
$uvw1$, \gztf, \rztf, and \iztf\ filters as a Gaussian process
\citep{Rasmussen06} using the \texttt{gaussian\_process} library in
\texttt{scikit-learn} \citep{Pedregosa11}. This allows us to interpolate flux
measurements in each of these filters to a grid of times between 1 and 70\,d
after \tfl, while also estimating an uncertainty on the interpolation. From
there, we can estimate the bolometric luminosity, $L_\mathrm{bol}$, via
trapezoidal integration of the SED.

There is emission blueward of the $uvm2$-band and redward of the \iztf-band
that is not constrained by our observations, and for fast-declining SNe the
near-infrared (NIR) provides a significant contribution to $L_\mathrm{bol}$
\citep[e.g.,][]{Taubenberger08}. To estimate the NIR flux, we extrapolate
redward from the \iztf-band to the $K_s$-band ($\lambda_\mathrm{eff} =
2.159\,\mu$m) by assuming the $K_s$-band flux is equal to the ratio of the
\iztf-band to the $K_s$-band flux for a 8500\,K blackbody. This choice of
temperature is reasonable based on our \texttt{TARDIS} spectral models (see
\S\ref{sec:tardis} and Table~\ref{tab:tardis}). While the true temperature is
not constant, we find that varying the temperature between 6000 and 12,500\,K
changes the peak $L_\mathrm{bol}$ by $\lesssim\,3$\%, which is significantly
less than the total systematic uncertainty. The assumed emission redward of
the \iztf-band results in a NIR contribution of $\sim$20\% to $L_\mathrm{bol}$
near maximum light, and $\sim$35\% at \tbmax\,$\approx +30$\,d. This is
similar to SN\,2004eo, an SN with an intermediate decline rate like \sn\ (see
\S\ref{sec:max_decline}), and other fast-declining SNe\,Ia
\citep{Taubenberger08}.

Similar to our procedure in the NIR, we estimate the flux in the far-UV
by extrapolating between the $uvm2$-band and 1000\,\AA\ assuming a 12,500\,K
blackbody. While such a high temperature is only appropriate for the early UV
flash from \sn\ (see \S\ref{sec:luminosity}), the far-UV contribution
to $L_\mathrm{bol}$ following this assumption is negligible ($\lesssim 1$\%)
around maximum light and later epochs.

In the near-UV, probed by the UVOT $uvm2$ and $uvw1$ filters, the SN is
only marginally detected at epochs $> 26.5$\,d after \tfl\ (see
Figure~\ref{fig:p48}). Given the low S/N in the
\textit{Swift} observations at these epochs, for $t > t_\mathrm{fl} + 26.5$\,d
we interpolate the $uvm2$ and $uvw1$ flux by assuming their ratio relative to
the \gztf\ flux, which is measured at very high S/N, is fixed and set by their
relative ratios at $t = t_\mathrm{fl} + 26.5$\,d. Fixing the UV flux in this
manner does not change our estimate of the \radni\ mass, and does not have a
significant effect on our estimate of the total ejecta mass.

\begin{figure}
    \centering
    \includegraphics[width=\columnwidth]{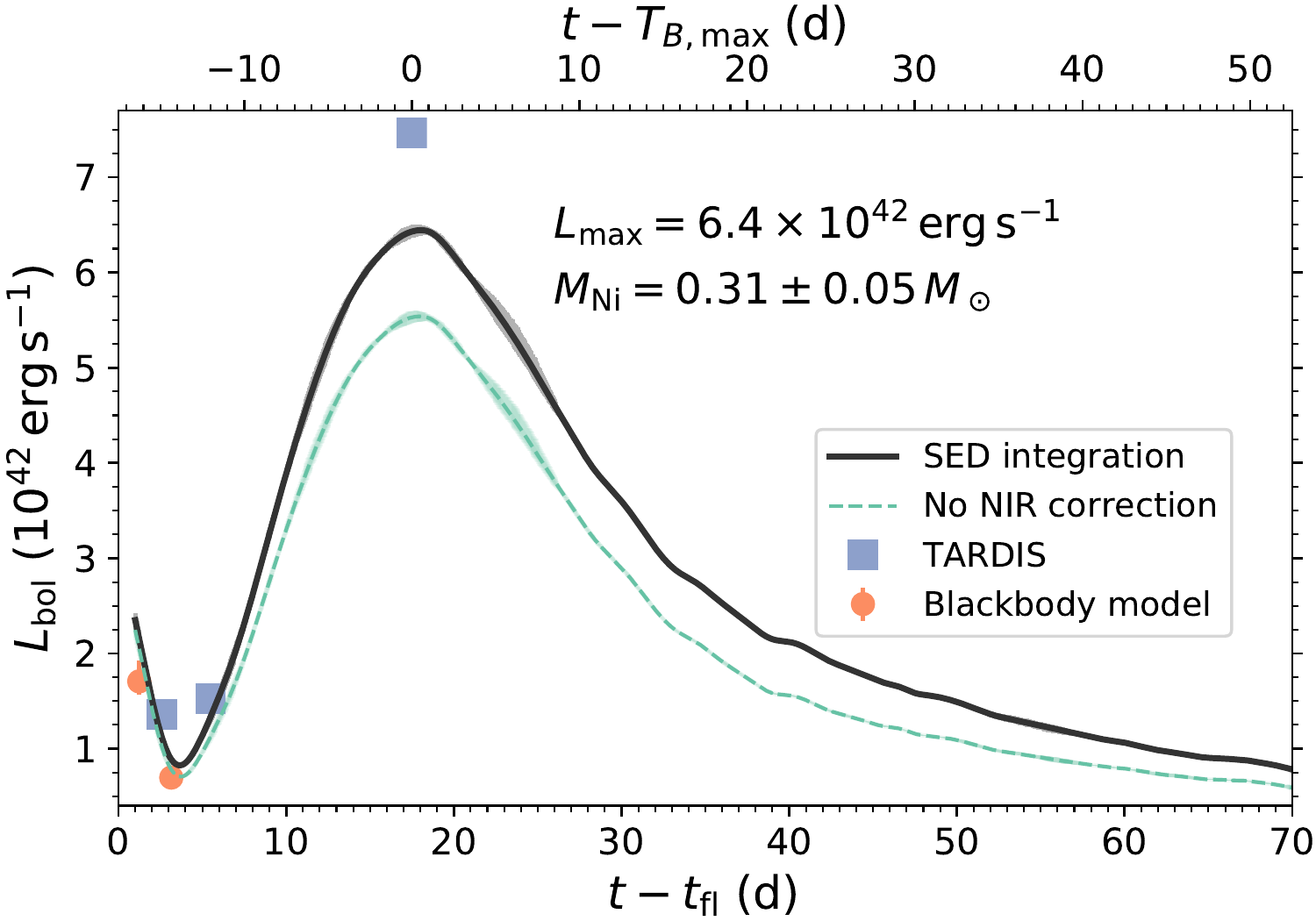}
    \caption{Bolometric luminosity, $L_\mathrm{bol}$, of \sn\ as a function of
    time. $L_\mathrm{bol}$ is estimated via SED integration (see text) and
    shown as a black line, with statistical uncertainties shown in light gray.
    Purple squares show luminosity estimates from spectral modeling (see
    \S\ref{sec:tardis}), orange circles show luminosity estimates from a
    blackbody fit to the SED (see \S\ref{sec:luminosity}). The methods
    generally agree, though the \texttt{TARDIS} spectral models likely
    overestimate the flux around maximum light (see text). The
    luminosity of \sn\ assuming no NIR correction is shown as a dashed green
    line. The total inferred mass of synthesized \radni\ is $0.31 \pm
    0.05\,M_\odot$.}
    \label{fig:Lbol}
\end{figure}

The bolometric luminosity of \sn\ is shown as a function of time in
Figure~\ref{fig:Lbol}. Statistical uncertainties in $L_\mathrm{bol}$ are
estimated via bootstrap resampling of the interpolated flux at each epoch, and
are typically on the order of a few percent. We estimate a systematic
uncertainty of $\sim$10\% based on the total procedure (including
interpolation, extrapolation, and integration). 

As shown in Figure~\ref{fig:Lbol} the method compares favorably with a
blackbody model (at early epochs) and spectroscopic modeling (during the SN
rise). The maximum-light luminosity estimate from the \texttt{TARDIS} spectral
model likely overestimates the flux in the NIR (see the third panel in
Figure~\ref{fig:tardis}), due to the model assumption that there is a single,
sharp photosphere that does not vary with wavelength. This explains the
discrepancy between SED integration and the \texttt{TARDIS} model at that
epoch.

From the SED integration we find that the bolometric luminosity of \sn\ peaked
18.1\,d after \tfl\ ($\sim$0.6\,d after \tbmax) at $L_\mathrm{bol,max} = 6.4
\pm 0.1\,(\mathrm{statistical}) \pm 0.6\,(\mathrm{systematic}) \times
10^{42}\,\mathrm{erg\,s}^{-1}$. This peak luminosity is $\sim$70\% larger than
the peak luminosity of iPTF\,14atg \citep[$3.8 \times
10^{42}\,\mathrm{erg\,s}^{-1}$;][]{Kromer16}.

From Arnett's rule \citep{Arnett82}, which states that the peak luminosity is
equal to the instantaneous rate of radioactive energy released by \radni, we
estimate the total mass of \radni, $M_\mathrm{^{56}Ni}$, synthesized in the
explosion. Using Equation~19 from \citet[][see also
\citealt{Stritzinger06,Howell09,Scalzo14}]{Nadyozhin94}, we find
$M_\mathrm{^{56}Ni} = 0.31 \pm 0.05\,M_\odot$, where the uncertainty is
dominated by the (assumed) systematic uncertainty on $L_\mathrm{bol}$. We note
that if our previous assumption about the NIR contribution to $L_\mathrm{bol}$
at maximum light is revised downward from $\sim$20\% to $\sim$5\%, as is
typical for normal SNe\,Ia \citep[e.g.,][]{Suntzeff96,Contardo00}, the total
\radni\ mass still agrees with the above estimate to within the uncertainties.
This yield is low for a normal SN\,Ia as typical explosions yield
$\sim$0.4--0.8\,$M_\odot$ of \radni~\citep[e.g.,][]{Scalzo14a}.

Following \citet{Jeffery99}, we can estimate the total mass ejected by
\sn, $M_\mathrm{ej}$, by calculating the transparency timescale, $t_0$, from
the decline of the bolometric light curve \citep[see
also][]{Stritzinger06,Scalzo14,Dhawan18}. Briefly, $t_0$ is a parameter that
governs the time-varying $\gamma$-ray optical depth of an SN, and it is related
to $M_\mathrm{ej}$ as follows \citep{Jeffery99,Dhawan18}:
\begin{equation}
    M_\mathrm{ej} = 1.38 \left(\frac{1/3}{q}\right)
    \left(\frac{v_e}{3000\,\mathrm{km\,s}^{-1}}\right)^2 
    \left(\frac{t_0}{36.80\,\mathrm{d}}\right)^2 M_\odot,
    \label{eq:ejecta_mass}
\end{equation}
where $v_e$ is the e-folding velocity of an exponential density profile, and
$q$ is a form factor that describes the distribution of \radni\ in the ejecta
($q = 1/3$ corresponds to an evenly distributed \radni\ profile). In
Equation~\ref{eq:ejecta_mass}, the $\gamma$-ray opacity has been assumed to be
0.025\,cm$^{2}$\,g$^{-1}$.

For \sn\ we estimate $t_0 = 42.0 \pm 1.0$\,d (the uncertainty is dominated by
the uncertainty on the rise time, for which we adopt 1\,d). Assuming $v_e =
3000 \pm 180$\,\kms\ and that $q = 0.45 \pm 0.08$, we find $M_\mathrm{ej} =
1.33 \pm 0.27\,M_\odot$. Unlike our estimate of $M_\mathrm{^{56}Ni}$, the
adopted NIR correction does affect the measurement of $t_0$. In addition to
the total luminosity, Figure~\ref{fig:Lbol} shows the SED-integrated
luminosity assuming no NIR flux (i.e., $\mathrm{flux} = 0$ for all $\lambda >
1\,\mu$m). From this light curve we estimate $t_0 = 38.1 \pm 1.0$\,d,
corresponding to $M_\mathrm{ej} = 1.10 \pm 0.22\,M_\odot$. Given the overall
uncertainty in the NIR correction, our observations broadly bracket the total
mass of ejecta to be somewhere between $\sim$0.9 and 1.5\,$M_\odot$.

\subsection{Maximum Light and Decline}\label{sec:max_decline}

\begin{figure}
    \centering
    \includegraphics[width=3.35in]{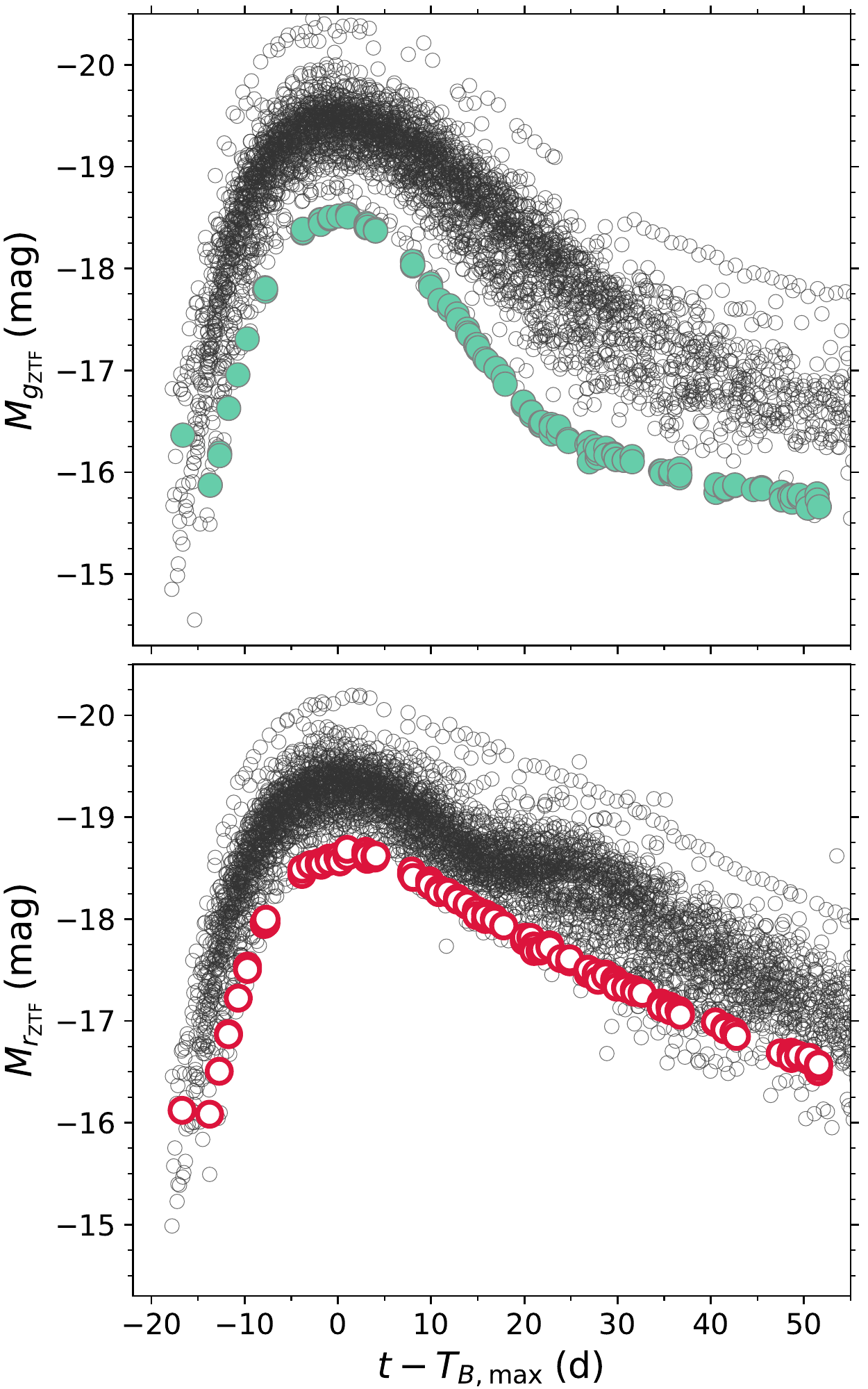}
    \caption{Photometric evolution of \sn\ compared to 121 normal SNe\,Ia
    observed by ZTF \citep{Yao19} in the \gztf\ (top) and \rztf\ (bottom)
    filters. The normal SNe are shown as open gray circles, while the symbols
    for \sn\ are the same as Figure~\ref{fig:p48}. Relative to normal SNe\,Ia,
    \sn\ is fainter, declines faster in \gztf, and lacks the ``shoulder''
    typically seen in the \rztf\ filter. Normal SNe light curves have been
    corrected for host-galaxy reddening and $K$-corrections have been applied,
    with both determined via \texttt{SNooPY} (see \citealt{Bulla20} for
    details of our implementation). $K$-corrections have not been applied to
    the light curve of \sn.}
    \label{fig:lc_comp}
\end{figure}

While the rise time and power-law indices of \sn\ are similar to other normal
SNe Ia (see \S\ref{sec:t_fl}), the full photometric evolution does not
resemble a normal SN\,Ia. The photometric evolution of \sn\ is highlighted in
Figure~\ref{fig:lc_comp}, where \sn\ is compared to 121 normal SNe\,Ia from
\citet{Yao19}.\footnote{For the purposes of this comparison we consider
SN\,1991T-like, SN\,1999a-like, and SN\,1986G-like events to all be normal
SNe\,Ia.} \sn\ is somewhat underluminous ($M_{g,\mathrm{max}} \approx
-18.5$\,mag), declines rapidly [$\Delta m_{15}(g) =
1.30^{+0.01}_{-0.02}$\,mag, uncertainties represent the 68\% credible region],
and does not exhibit a ``shoulder'' in the \rztf\ or a strong secondary
maximum in the \iztf\ light curves post maximum. The slightly underluminous
peak and moderately fast decline of \sn\ are very similar to SN\,1986G-like
SNe\,Ia, which represent a transitional group between normal SNe\,Ia and the
underluminous SN\,1991bg-like class (e.g., \citealt{Taubenberger17} and
references therein). While the photometric evolution of \sn\ is similar to
86G-like SNe, we show that \sn\ is spectroscopically distinct from these
transitional SNe (\S\ref{sec:spec_comp}).

We also find that standard SN\,Ia fitting techniques do not provide good
matches to the evolution of \sn. For example, a \texttt{SNooPY}
\citep{Burns11} fit to the optical light curve requires significant
host-galaxy extinction [$E(B-V)_\mathrm{host}\approx0.4$\,mag, see
\S\ref{sec:host}] to match the observed red colors, while predicting a
secondary maximum in the \iztf-band and a fast evolution after peak that is
not seen in \sn. A \texttt{SALT2} \citep{Guy07} fit leads to similar
inconsistencies to those in \texttt{SNooPy}. These inconsistencies support our
conclusion above that the photometric evolution of \sn\ does not match normal
SNe\,Ia.

\subsection{Color Evolution}

\begin{figure}
    \centering
    \includegraphics[width=3.35in]{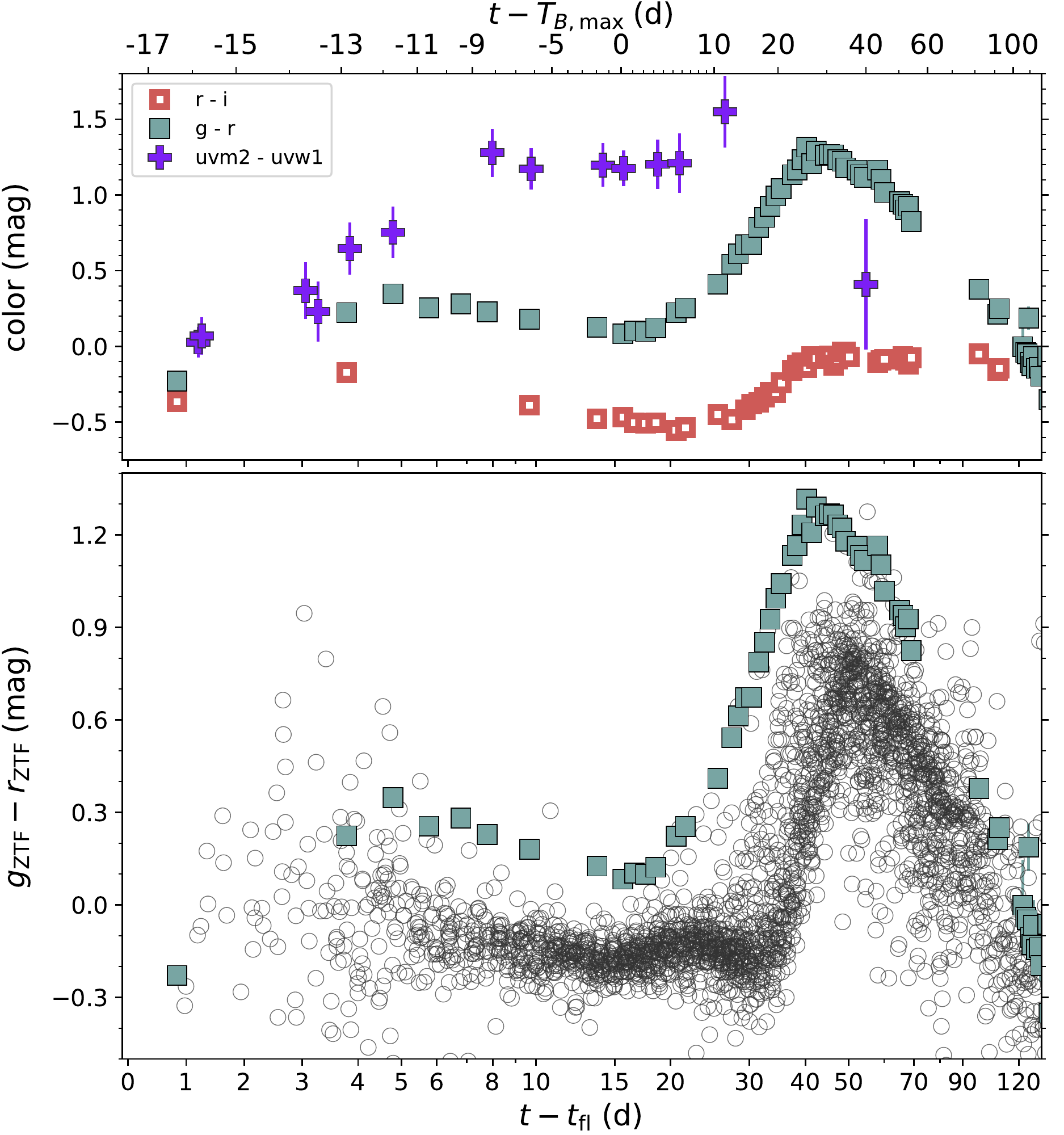}
    \caption{Photometric color evolution of \sn\ relative to \tfl\ (the
    timescale relative to \tbmax\ shown along the top axis only applies to
    \sn). Bottom: \gztf$ - $\rztf\ evolution of \sn\ (solid green squares),
    corrected for the total interstellar extinction (see \S\ref{sec:host}),
    and compared with the evolution of 62 normal SNe Ia (open circles)
    observed within 5\,d of \tfl\ by ZTF (from \citealt{Bulla20}). \sn\ is the
    reddest SN in the group, and it exhibits the fastest evolution to red
    colors post-\tbmax. Top: the $uvm2 - uvw1$ (purple crosses), \gztf$ -
    $\rztf\ (solid, green squares), and \rztf$ - $\iztf\ (open, red squares)
    color evolution of \sn.}
    \label{fig:colors}
\end{figure}

\sn\ is further distinguished from normal SNe Ia by its unusual color
evolution (Figure~\ref{fig:colors}). The lower panel of
Figure~\ref{fig:colors} shows the \gztf$ - $\rztf\ evolution of 62
spectroscopically normal SNe Ia with ZTF observations within 5\,d of \tfl\
(see \citealt{Bulla20}), with the color evolution of \sn\ over-plotted. The
initially blue colors in \sn\ rapidly evolve to the red over the first few
days of observation before gradually becoming bluer in the time leading up to
\tbmax\ (this behavior is deemed an early ``red bump'' in \citealt{Bulla20}).
Similar red bumps are only seen in 6 of the 62 normal SNe\,Ia ($\sim$10\%) in
the ZTF sample \citep{Bulla20}, and they are typically less pronounced than
what is observed in \sn.

Furthermore, while normal SNe Ia exhibit a large scatter in \gztf$ - $\rztf\
shortly after \tfl\ they evolve to form a tight locus between $\sim$10 and
30\,d after \tfl. \sn\ is redder at peak than each of the normal SNe Ia in the
\citet{Bulla20} sample. Post maximum, only one normal SN\,Ia, ZTF\,18abeegsl
(SN\,2018eag), exhibits a similarly rapid decline in \gztf$ - $\rztf\ color.
The \gztf$ - $\rztf\ color evolution of \sn\ is again intermediate between
normal SNe\,Ia and underluminous 91bg-like SNe. Figure~\ref{fig:colors} shows
that normal SNe\,Ia are reddest at $\sim$+30\,d, while 91bg-like SNe are
reddest between $\sim$+10 and 15\,d \citep{Burns14}. \sn\ reaches a
\gztf$-$\rztf\ maximum at an intermediate time of $\sim$+20\,d.

The offset in the \gztf$ - $\rztf\ color evolution of \sn\ relative to normal
SNe\,Ia would be reduced if the reddening toward \sn\ has been significantly
underestimated. A color excess of $E(B-V) \approx 0.25$\,mag, rather than the
0.05\,mag adopted in \S\ref{sec:host}, would roughly align the \gztf$ -
$\rztf\ color of \sn\ with the tight locus between $\sim$5 and 20\,d after
\tfl\ seen in Figure~\ref{fig:colors}. Such a correction would also bring the
peak optical brightness of \sn\ in line with normal SNe\,Ia [for $E(B-V)
\approx 0.25$\,mag, $M_g \approx -19.25$\,mag and $M_r \approx -19.1$\,mag for
\sn].

While the spectral appearance of \sn\ is similar to some normal SNe\,Ia (see
\S\ref{sec:spec_comp}), the observed rapid decline in the \gztf\ filter
provides strong evidence that \sn\ is not a normal luminosity SN\,Ia.
\citet{Phillips93} showed that in the optical SNe\,Ia follow a
brightness--width relation, whereby brighter explosions decline less rapidly.
Thus, with a typical peak in the optical, as would be implied with $E(B-V)
\approx 0.25$\,mag, the fast decline in \sn\ [$\Delta m_{15}(g) = 1.3$\,mag]
would be largely unprecedented.\footnote{Only two spectroscopically normal
SNe\,Ia in the \citet{Yao19} sample decline faster than \sn\ as measured by
$\Delta m_{15}(g)$. While the lack of \textit{Swift} $b$-band templates
prevents us from measuring $\Delta m_{15}(B)$, the relationship between that
and $\Delta m_{15}(g)$ for normal ZTF SNe\,Ia suggests $\Delta m_{15}(B)
\gtrsim 1.6$\,mag for \sn.} This conclusion is further corroborated by the
rapid evolution of the \gztf$-$\rztf\ color to the red after \tbmax\ and the
lack of a secondary maximum in the \iztf-band, each of which is typical of
lower luminosity SNe\,Ia \citep[see][and references therein]{Taubenberger17}.
We therefore conclude that the color excess toward \sn\ is not
underestimated, and that the SN is instead intrinsically red in the optical.

Even if one ignores the striking initial bump in the light curve of \sn, we
can still conclude that \sn\ is not a normal SN\,Ia based on its other
photometric properties (e.g., relatively faint peak optical brightness,
moderately fast decline, lack of a NIR secondary maximum, and red appearance
at peak).

\section{Spectral Evolution of \sn}\label{sec:spec}

Optical spectra of \sn\ were obtained at phases from $-$14.9\,d (2.6\,d after
\tfl) to 66.5\,d after \tbmax. Details of the spectra are presented in
Table~\ref{tab:spectra} and the spectral evolution is shown in
Figure~\ref{fig:spec_evo}. The absorption features in \sn\ are typical of
SNe\,Ia, including IMEs, primarily Si, Ca, and O, as well as iron-group
elements (IGEs).

\begin{deluxetable*}{lcrcccc}
\tabletypesize{\scriptsize}
\tablewidth{0pt}
\tablecaption{\texttt{TARDIS} input parameters\label{tab:tardis}}
\tablehead{
\colhead{Date} &
\colhead{MJD} &
\colhead{Phase} &
\colhead{$t - t_\mathrm{exp}$} &
\colhead{$L$} &
\colhead{$v_\mathrm{boundary}$\tablenotemark{{\scriptsize a}} }&
\colhead{$T_\mathrm{boundary}$\tablenotemark{{\scriptsize b}} } \\
\colhead{(UT) }&
\colhead{} &
\colhead{(d)} &
\colhead{(d)} &
\colhead{($\log L_{\odot}$)} &
\colhead{(\kms) } &
\colhead{(K)}
}
\startdata
2019 Dec 31.277 & 58,848.277 & $-14.9$ & 3.0 & 8.55 & 25,000 & 8173 \\
2020 Jan 03.217 & 58,851.217 & $-12.0$ & 6.0 & 8.60 & 16,500 & 7925 \\
2020 Jan 15.392 & 58,863.392 & $+0.0$ & 18.0 & 9.29 & 10,500 & 9696 \\
\enddata
\tablecomments{Phase is measured in rest-frame days relative to \tbmax. The time of explosion, $t_\mathrm{exp}$, is assumed to be 0.4\,d before \tfl\ for the \texttt{TARDIS} model. }
\tablenotetext{a}{Ejecta velocity at the inner boundary of the photosphere.}
\tablenotetext{b}{Temperature at the inner boundary of the photosphere. 
$T_\mathrm{boundary}$ is not explicitly an input parameter for
\texttt{TARDIS}, it is derived from the luminosity, time since explosion,
inner boundary velocity, and then iteratively updated throughout the simulation.}
\end{deluxetable*}

\subsection{TARDIS Models}\label{sec:tardis}

To determine the structure of the ejecta and relative contributions of
different ions at early and maximum-light phases, we have modeled the spectra
at $-$14.9, $-12.0$, and $+$0.0\,d using the one-dimensional (1D) Monte Carlo
radiative transfer code \texttt{TARDIS} \citep{Kerzendorf14}. We note that
\texttt{TARDIS} assumes a single, sharp photosphere between the optically
thick and thin regions. Therefore, if there is a contribution to the spectrum
from an underlying quasi-blackbody (at early times this could be due to
interaction, for example; see \S\ref{sec:companion_interaction}), this will
impact the ability of \texttt{TARDIS} to fully reproduce the observations.
Nevertheless, our models should provide a reasonable approximation of the
plasma state within the ejecta. The parameters of our \texttt{TARDIS} models
are given in Table~\ref{tab:tardis}.

\begin{figure*}[htb!]
    \centering
    \includegraphics[width=\textwidth]{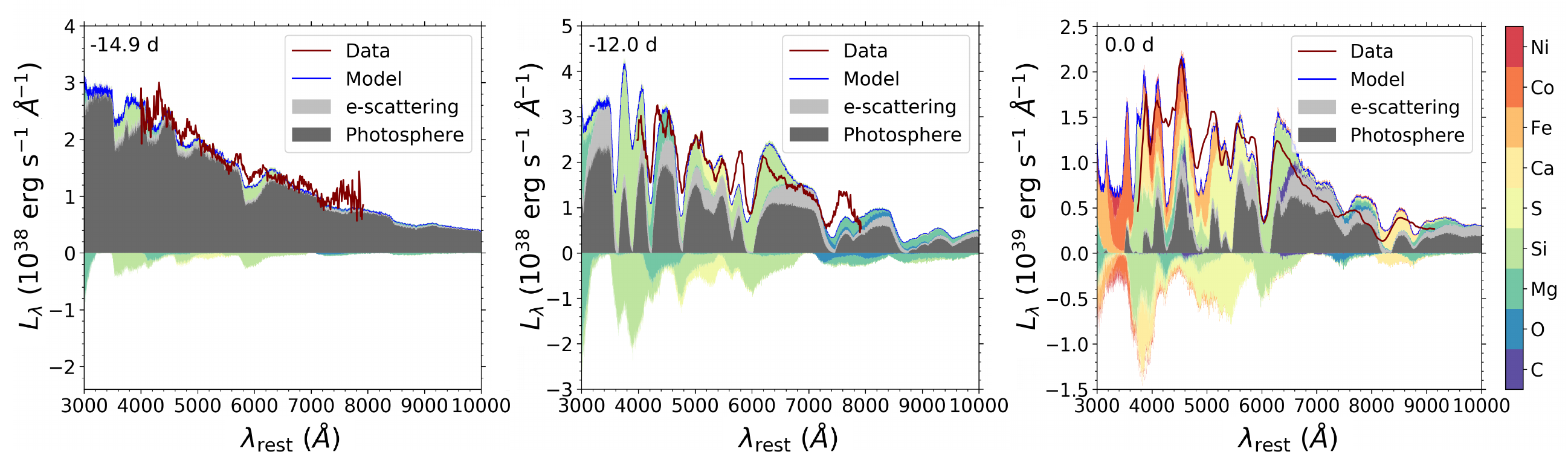}
    \caption{Comparison of \texttt{TARDIS} models to \sn\ at $-$14.9\,d
    (left), $-12.0$\,d (middle), and $+$0.0\,d (right), relative to \tbmax.
    For each model, we color code a histogram showing the contribution of each
    element to the spectroscopic features, based on the last element with
    which a Monte Carlo photon packet experienced an interaction. Photon
    packets may be absorbed and re-emitted at different wavelengths, with the
    exception of those packets that only experience electron scattering.
    During electron scattering, only the direction of propagation is changed.
    These packets are shown in light gray. Photon packets that did not
    interact during the simulation are shown in dark gray. Contributions below
    and above zero show the SED of packets before and after their last line
    interaction. Note that non-interacting (photosphere) and e-scattering
    photon packets are not shown below zero.}
    \label{fig:tardis}
\end{figure*}

The first spectrum of \sn\ at $-$14.9\,d (2.6\,d after \tfl, 3.0\,d after the
\texttt{TARDIS}-inferred $t_\mathrm{exp}$) shows shallow features consistent
with IMEs moving at extremely high velocities ($>$\,20,000\,\kms,
Figure~\ref{fig:spec_evo}). The best-fitting \texttt{TARDIS} model is shown in
Figure~\ref{fig:tardis}, along with the contribution of individual elements to
the spectral features. For this model, we have assumed a uniform composition
of O, Mg, Si, and S. Our model demonstrates that the shallow absorption
features observed at this phase can be reproduced solely by IMEs
(predominantly \ion{Si}{II}), and that the presence of IGE is not required to
match the data. Our model also confirms the high velocities of the ejecta --
we find the spectral features and temperature are best reproduced with a
photospheric velocity of $\sim$25,000\,\kms.

Similarly, for the $-12.0$\,d spectrum we find that a model that does not
contain IGE above $\sim$16,500\,\kms\ reproduces the majority of the
spectroscopic features. Again, our model contains a uniform composition of O,
Mg, Si, and S, and is shown in Figure~\ref{fig:tardis}. At this phase the
model suggests the photospheric temperature has not significantly changed,
however, the features have become much more prominent. Compared to modeling of
the spectroscopically similar SN\,2002bo (see \S\ref{sec:spec_comp}) at the
same epoch \citep{Stehle05}, we find \sn\ has a lower photospheric temperature
($\sim$8000\,K, compared to $\sim$9500\,K for SN\,2002bo).

While the early spectra of \sn\ are dominated by IMEs, there is no evidence in
the observed spectra for \ion{C}{II} absorption. However, in our
\texttt{TARDIS} models even if we increase the C abundance in the outer ejecta
to large amounts (50\%), the model spectra still lack any significant
\ion{C}{II} features at the time of our observations. Our ability to detect
\ion{C}{II} in the spectra of \sn\ is likely affected by the extremely high
ejecta velocities, which leads to blending with \ion{Si}{II}. Therefore,
despite the lack of a \ion{C}{II} detection in the observed spectra, we are
unable to place meaningful constraints on the C abundance in the very
outermost ejecta.

Given that our $+0$\,d maximum-light spectrum occurs 12\,d after our previous
model spectrum, we assume a composition for the inner ejecta
($\textless$\,16,500\,\kms) similar to that found for SN\,2002bo
\citep{Stehle05}. A more detailed ejecta structure could be achieved through
modeling additional pre-maximum spectra, but is beyond the scope of the work
presented here. As shown in Figure~\ref{fig:tardis}, our model is able to
broadly reproduce many of the features observed. Notable exceptions include
the features around $\sim$4200 and 4900\,\AA, which we attribute to Fe. Better
spectroscopic agreement could potentially be achieved if \sn\ had a lower
abundance of IGE within the inner ejecta relative to SN\,2002bo.

Overall, our \texttt{TARDIS} modeling demonstrates that \sn\ is consistent
with a low (or zero) fraction of IGE in the outer ejecta (i.e., there is
little mixing in the SN ejecta). Additionally, the earliest phases show little
change in temperature (see Table~\ref{tab:tardis}), as expected from the color
evolution.

\subsection{\ion{Si}{II} Evolution}\label{sec:SiII}

We have measured the velocity of the \ion{Si}{II} $\lambda$6355 absorption
feature following the procedure in \citet[][see their Section~2.5]{Maguire14}.
We have also estimated the pseudo-equivalent widths (pEWs) of the \ion{Si}{II}
$\lambda\lambda$5972, 6355 features, allowing us to measure their ratio, also
known as $\mathcal{R}($\ion{Si}{II}$)$; see \citet{Hachinger08} for the
updated definition relative to \citet{Nugent95}.

\begin{figure}
    \centering
    \includegraphics[width=3.35in]{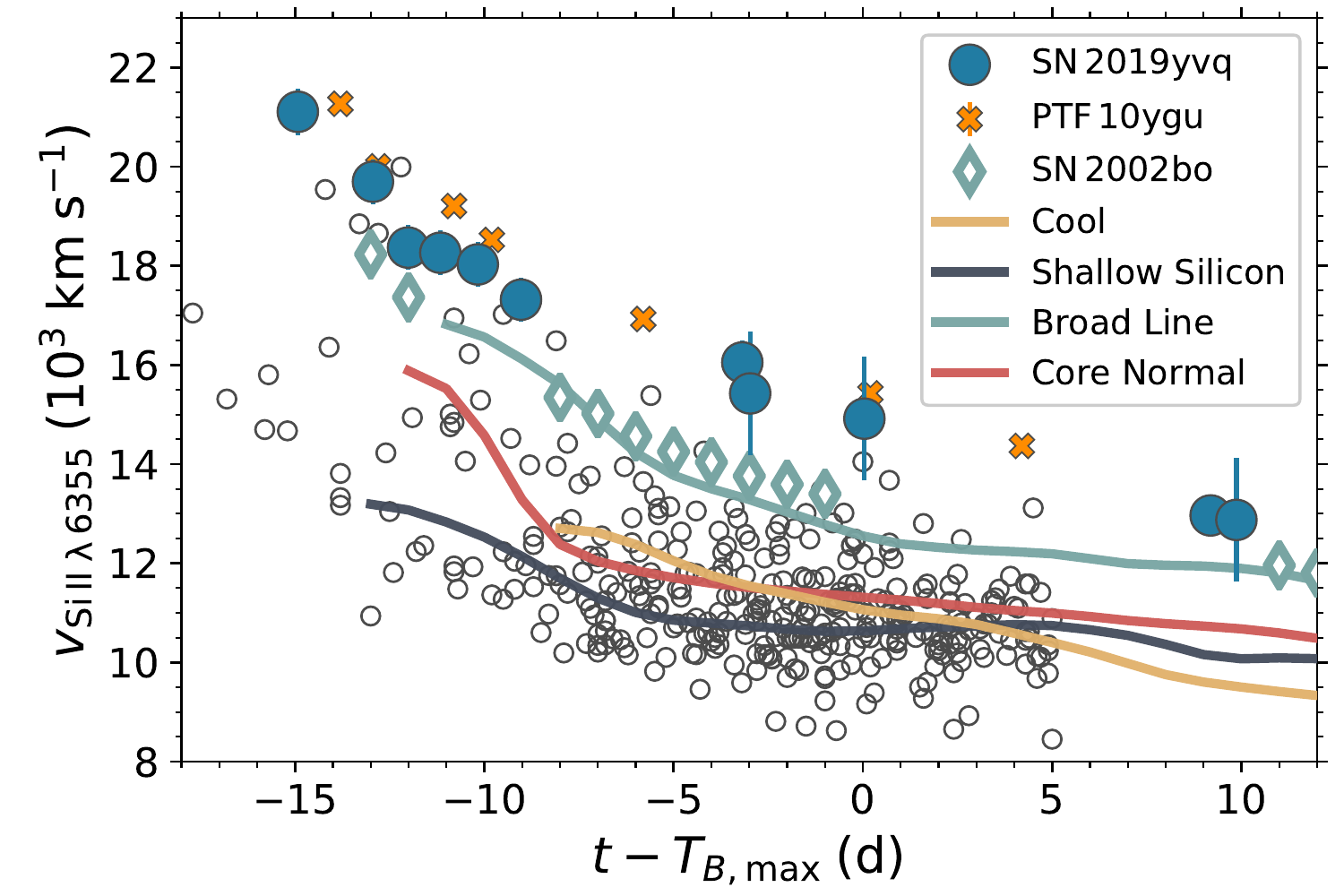}
    \caption{Velocity evolution of \ion{Si}{II} $\lambda$6355 absorption in
    \sn\ (large, filled circles). For comparison we also show the measurements
    for 264 SNe Ia observed by PTF \citep[data
    from][]{Maguire14} as open circles, with SN\,2010jn (PTF\,10ygu), the SN
    with the fastest moving ejecta in the PTF sample, highlighted via orange
    crosses. We additionally show the velocity evolution of SN\,2002bo
    \citep[data from][]{Benetti04}, an SN that is very similar to \sn, as open
    diamonds. The median velocity evolution of each of the spectroscopic
    classes defined by \citet[][Shallow Silicon, Core Normal, Broad Line, and
    Cool]{Branch06} are shown via solid lines. It is clear that \sn\ has
    exceptionally high-velocity ejecta.}
    \label{fig:vel_evo}
\end{figure}

The velocity evolution of \ion{Si}{II} $\lambda$6355 is shown in
Figure~\ref{fig:vel_evo}, compared to measurements for the Palomar Transient
Factory (PTF) SN\,Ia sample from \citet{Maguire14} and the median velocity
evolution of SNe\,Ia belonging to the four different classes (Shallow Silicon,
Core Normal, Broad Line, and Cool) identified in
\citet{Branch06};\footnote{The velocity measurements are from
\citet{Blondin12}, while the method to determine the median velocity is
described in \citet{Miller18}.} hereafter, the Branch~class. The \ion{Si}{II}
$\lambda$6355 velocity in \sn\ is $\sim$15,000\,\kms\ around \tbmax.

At \tbmax, the pEW measurements for the \ion{Si}{II} $\lambda$6355 and
$\lambda$5972 features are $183\pm1$\,\AA, and $13\pm2$\,\AA, respectively,
unambiguously classifying \sn\ as a Branch~Broad Line SN\,Ia. \sn\ stands out
in Figure~\ref{fig:vel_evo} with some of the highest \ion{Si}{II} velocities
that have ever been observed. Within the PTF sample, only SN\,2010jn
(PTF\,10ygu) exhibits a \ion{Si}{II} absorption velocity as high as \sn\ at
every phase in its evolution. Furthermore, we also find that the \ion{Ca}{II}
infrared triplet (IRT) velocities are high (we first detect this feature in
the $-3.0$\,d SEDM spectrum; see Table~\ref{tab:spectra}), with a photospheric
component velocity of $\sim$13,200\,\kms\ and a clear high-velocity component
at $\sim$23,500\,\kms. Within the PTF sample only one other SN (PTF\,09dnp)
has a \ion{Ca}{II} high-velocity component with a similarly large velocity at
the same phase.

\begin{figure}
    \centering
    \includegraphics[width=3.35in]{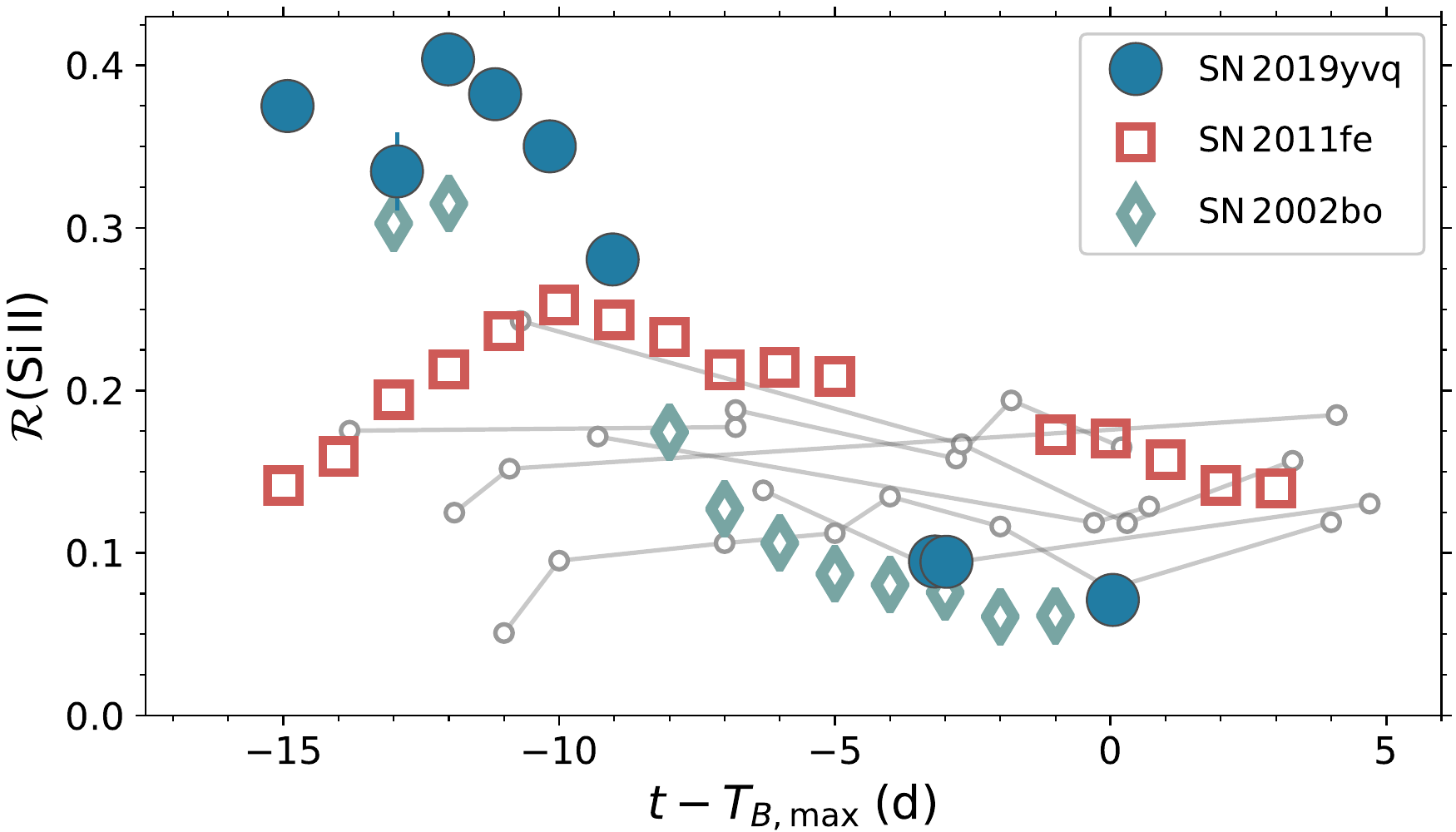}
    \caption{Evolution of the ratio of the pEW of \ion{Si}{II} $\lambda$5972
    to \ion{Si}{II} $\lambda$6355, $\mathcal{R}($\ion{Si}{II}$)$, in \sn\
    (large, filled circles). SN\,2002bo \citep[data from][]{Benetti04} and
    SN\,2011fe \citep[data from][]{Pereira13} are also highlighted as open
    diamonds and open squares, respectively. For comparison we also show the
    $\mathcal{R}($\ion{Si}{II}$)$ evolution for five PTF SNe\,Ia (10mwb, 10qjq,
    10tce, 10wof, 11hub) with $> 3$ measurements over a duration $> 8$\,d
    (data from \citealt{Maguire14}) and SN\,2017erp \citep[data
    from][]{Brown19} as connected, open circles. \sn\ and SN\,2002bo exhibit
    an unusual inversion in $\mathcal{R}($\ion{Si}{II}$)$ as they evolve
    toward maximum light.}
    \label{fig:r_evo}
\end{figure}

As first noted by \citet{Nugent95}, and later confirmed by
\citet{Hachinger08}, $\mathcal{R}($\ion{Si}{II}$)$ is a luminosity indicator,
with larger values of $\mathcal{R}($\ion{Si}{II}$)$ corresponding to lower
luminosities. This correlation is driven by the ionization balance of
\ion{Si}{II}/\ion{Si}{III}, with cooler objects having stronger \ion{Si}{II}
$\lambda$5972 features. In Figure~\ref{fig:r_evo} we show the evolution of
$\mathcal{R}($\ion{Si}{II}$)$ as a function of time for \sn, compared to
SN\,2011fe, SN\,2002bo, SN\,2017erp, and five SNe with multiple measurements
over a long baseline from the PTF SN\,Ia spectral sample.
Figure~\ref{fig:r_evo} shows that most SNe\,Ia have a relatively flat
evolution in $\mathcal{R}($\ion{Si}{II}$)$ in the time leading up to \tbmax\
\citep[see also][]{Riess98a}. \sn\ and SN\,2002bo, however, feature a very
different evolution with initially large values of
$\mathcal{R}($\ion{Si}{II}$)$ that rapidly decrease to very low values between
$\sim$10 and $\sim$5\,d before \tbmax.

At face value, the $\mathcal{R}($\ion{Si}{II}$)$ evolution in
Figure~\ref{fig:r_evo} suggests that the effective temperature of \sn\
increases significantly as it rises to maximum light. Both the optical colors,
which become bluer in the $\sim$14\,d leading up to \tbmax\ (see
Figure~\ref{fig:colors}), and the \texttt{TARDIS} modeling (see
Table~\ref{tab:tardis}) confirm an increase in temperature over the period in
which $\mathcal{R}($\ion{Si}{II}$)$ decreases. While the UV\,$-$\,optical
colors become redder over the same time period, this is likely due to the
increasing IGE fraction, and hence increased UV blanketing, as the photosphere
recedes (see \S\ref{sec:tardis}), and not a decrease in temperature.

This behavior is similar to, though less extreme than, SN\,2002bo, which
increases in temperature from $\sim$9,500\,K at $-12.9$\,d to $\sim$14,000\,K
at maximum light \citep{Stehle05}. The maximum-light temperature of SN\,2002bo
is similar to Branch~Core Normal SNe, such as SN\,2011fe, which
typically have temperatures of $\sim$14,500--15,000\,K at maximum light
\citep{Mazzali14}.

\citet{Benetti04} interpreted these competing effects as the result of
significant \ion{Si}{II} mixing in the ejecta of SN\,2002bo. Mixing or
synthesized Si in the outermost layers of the ejecta would (i) lead to larger
\ion{Si}{II} velocities, (ii) produce \ion{Si}{II} line ratios that indicate
cool temperatures (because there is less radioactive material to heat the
ejecta), before eventually (iii) producing low values of
$\mathcal{R}($\ion{Si}{II}$)$ as the photosphere recedes through the ejecta to
higher temperature regions. This picture is consistent with the
\citet{Stehle05} models of SN\,2002bo. In those models, Si completely
dominates the species at velocities above $\sim$23,000\,\kms, while there is
very little ($\sim$1\%) IGE above 1.35\,$M_\odot$ in radial mass
coordinates. A similar physical scenario likely explains the changes in
\ion{Si}{II} absorption seen in \sn. Although the temperature change in \sn\
is less dramatic than in SN\,2002bo, this may reflect slight differences in
the ejecta composition as we find \sn\ is consistent with no IGEs in the outer
layers of the SN ejecta.

\subsection{Spectral Comparison}\label{sec:spec_comp}

\begin{figure*}
    \centering
    \includegraphics[width=\textwidth]{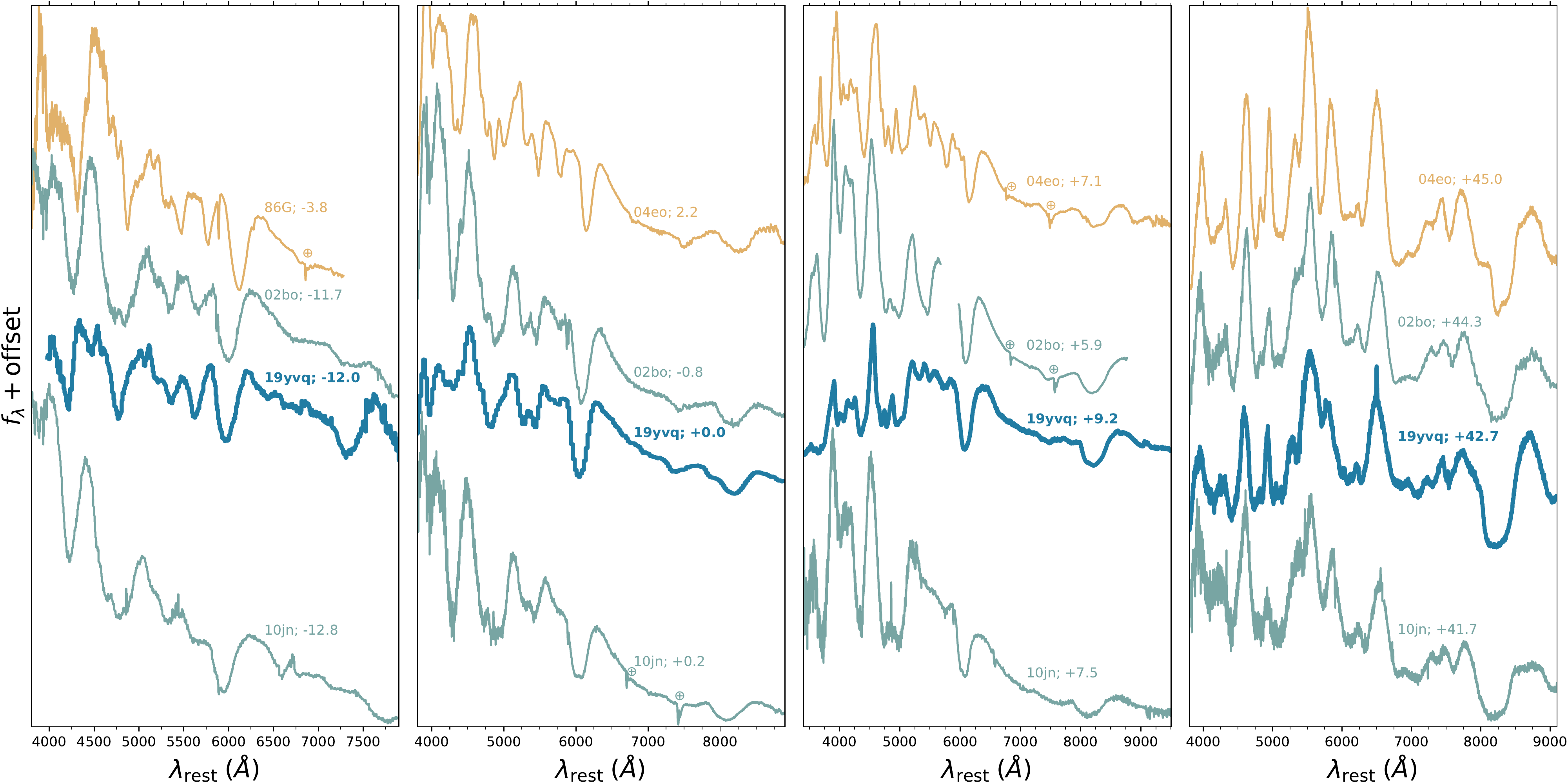}
    \caption{Spectral comparison of \sn\ to Branch~Broad Line and Cool
    SNe\,Ia. All spectra have been corrected for the total line-of-sight
    extinction with adopted $E(B-V)$ values of 0.9, 0.38, 0.39, 0.109, and
    0.05\,mag for SNe\,1986G \citep{Phillips87}, 2002bo \citep{Stehle05},
    2010jn \citep{Hachinger13}, 2004eo \citep{Pastorello07}, and \sn\ (this
    work), respectively. Left panel: pre-maximum spectra showing the
    similarity of \sn\ and SN\,2002bo. While the expansion velocities in the
    Cool SN\,1986G spectrum are considerably lower than those in the Broad
    Line SNe, the relative ratios of the \ion{Si}{II} features are similar to
    \sn. Second panel: Comparison of \sn\ to the Broad Line SNe\,2002bo and
    SN\,2010jn. These SNe all feature nearly identical maximum-light spectra.
    By this phase, the relative strength of the \ion{Si}{II} absorption
    features is no longer similar to Branch Cool SNe, as illustrated by
    SN\,2004eo. Third panel: $\sim$1 week post-maximum spectra. Fourth panel:
    Transitional phase spectra. Comparison spectra have been downloaded from
    WISeREP \citep{Yaron12}, with spectra for individual SNe from the
    following sources: SN\,1986G \citep{Cristiani92}, SN\,2002bo
    \citep{Benetti04,Silverman11}, SN\,2010jn (PTF\,10ygu)
    \citep{Hachinger13,Maguire14}, SN\,2004eo \citep{Pastorello07}.}
    \label{fig:spec_comp}
\end{figure*}

In Figure~\ref{fig:spec_comp}, we compare the spectral evolution of \sn\ to
two Branch~Broad~Line SNe\,Ia, SN\,2002bo and SN\,2010jn, and two Branch~Cool
SNe\,Ia, SN\,1986G and SN\,2004eo \citep{Cristiani92,
Benetti04,Pastorello07,Silverman11,Hachinger13,Maguire14} at four phases,
pre-maximum, maximum, $\sim$1 week post maximum, and $\sim$6 weeks post
maximum. The evolution of \sn\ and SN\,2002bo is remarkably similar at all
phases. The only significant difference between the two is the absorption
trough at $\sim$4800\,\AA\ in the pre-maximum and maximum-light spectra. This
feature, which is typically attributed to a combination of \ion{Fe}{II},
\ion{Fe}{III}, and \ion{Si}{II}, is extremely narrow in \sn. This is in
agreement with the \texttt{TARDIS} modeling results where no Fe is required in
the outer ejecta of \sn\ to match the observed spectra at early times.
SN\,2010jn, which exhibits large \ion{Si}{II} velocities like \sn, shows
weaker IME absorption and stronger IGE absorption than \sn. While the
Branch~Cool SNe\,1986G and 2004eo show lower velocities than \sn, there is
strong agreement in the relative \ion{Si}{II} line strengths of SN\,1986G and
the earliest spectra of \sn.

The maximum-light spectra shown in the second panel of
Figure~\ref{fig:spec_comp} reveal a higher temperature for \sn, as the
\ion{Si}{II} $\lambda$5972 absorption has nearly disappeared around \tbmax\
[see discussion of $\mathcal{R}($\ion{Si}{II}$)$ in \S\ref{sec:SiII}]. This
increase in temperature is consistent with the change in optical colors
(Figure~\ref{fig:colors}) and \texttt{TARDIS} spectral modeling
(\S\ref{sec:tardis}). The appearance of \sn, SN\,2002bo, and SN\,2010jn are
all similar at this epoch, with the exception of the 4800\,\AA\ feature
mentioned above. SN\,2004eo has a similar appearance to \sn, though it has
lower velocities and cooler temperatures (as traced by \ion{Si}{II}
$\lambda$5972).

The $+9.2$\,d spectrum of \sn, shown in the third panel of
Figure~\ref{fig:spec_comp}, shows absorption due to IGE. Additional
differences between \sn\ and SN\,2002bo can be seen at this phase. There is
stronger absorption in \sn\ blueward of \ion{Ca}{II} H\&K, and the \ion{S}{II}
``W'' absorption feature is still present in \sn\ and it cannot be identified
in SN\,2002bo or SN\,2010jn. SN\,2004eo maintains an appearance that is
somewhat similar to \sn, though as before, the temperature [as indicated by
\RSiII] and velocities are lower.

Spectra obtained $\sim$6 weeks after maximum light are shown in the fourth
panel of Figure~\ref{fig:spec_comp}. By this time, as the SNe are
transitioning into a nebular phase, the appearance of each spectrum is similar
modulo some minor differences in the relative line strengths of different
features.

\section{A Physical Explanation for \sn}\label{sec:models}

The most striking feature of \sn\ is the observed UV/optical peak that occurs
shortly after discovery (Figure~\ref{fig:p48}). Any model to explain \sn\ must
account for this highly unusual feature. A UV decline in the early phase of an
SN\,Ia has previously only been observed in a single event, iPTF\,14atg
\citep{Cao15}. Like \sn, iPTF\,14atg was a peculiar SN\,Ia, though it did not
resemble \sn\ in its peculiarity (iPTF\,14atg exhibited low expansion
velocities, and the spectra resembled SN\,2002es;
\citealt{Ganeshalingam12,Cao15}). Clearly resolved ``bumps'' in the early
optical emission of SNe\,Ia are also rare, having only been seen in a few
events: SN\,2014J \citep{Goobar15}, MUSSES1604D \citep{Jiang17}, SN\,2017cbv
\citep{Hosseinzadeh17} and SN\,2018oh \citep{Dimitriadis19,Shappee19}.

\sn\ features other properties, in addition to an initial peak $\sim$17\,d
prior to \tbmax, that separate it from normal SNe\,Ia. A good model should be
able to explain the following:
\begin{enumerate}
    \item The early UV/optical ``flash'' (Figure~\ref{fig:p48}).
    \item The moderately faint luminosity at peak  (\S\ref{sec:ni_mass}). 
    \item The relatively fast optical decline (\S\ref{sec:max_decline}). 
    \item The red optical colors at all epochs (Figure~\ref{fig:colors}). 
    \item The lack of IGE in the early spectra (\S\ref{sec:tardis}).
    \item The evolution in \RSiII\ (\S\ref{sec:SiII} and Figure~\ref{fig:r_evo}).
    \item The high \ion{Si}{II} velocities (Figure~\ref{fig:vel_evo}).
\end{enumerate}
The moderately faint peak combined with the high \ion{Si}{II} velocity is
particularly rare (see, e.g., Figure~11 in \citealt{Polin19}, where \sn\ would
be well isolated from all the other SNe\,Ia).

As noted in \S\ref{sec:max_decline}, the photometric evolution of \sn\ is
similar to intermediate 86G-like SNe, however, the spectra feature much weaker
\ion{Si}{II} $\lambda$5972 absorption and larger expansion velocities than
what is seen in 86G-like SNe (see Figure~\ref{fig:spec_comp}). Similarly,
while the spectral appearance and evolution of \sn\ is similar to SN\,2002bo,
and other Branch~Broad Line SNe, the photometric properties are entirely
different. SN\,2002bo features a relatively slow decline [$\Delta{m}_{15}(B) =
1.13$\,mag, which corresponds to $\Delta{m}_{15}(g) \approx 0.8$\,mag
\citep[see, e.g., Figure~2 in][]{Folatelli10}] with a clear secondary maximum
in the $I$-band \citep{Benetti04}, which stands in contrast to moderately fast
decline [$\Delta{m}_{15}(g) = 1.3$\,mag, roughly $\Delta{m}_{15}(B) \approx
1.55$\,mag \citep{Folatelli10}], and lack of \iztf\ secondary maximum in \sn.

If we otherwise ignore the early flash, several of the remaining features
(2--6) in the list above can be understood if the explosion that gave rise to
\sn\ produced a relatively small amount of \radni\ (\S\ref{sec:ni_mass}) that
is confined to the inner regions of the SN ejecta. A low \radni\ yield could
explain the underluminous light curve and red colors, while a centrally
concentrated IGE distribution could explain the IME-dominated early spectra,
as the IGE would not have been mixed to these outer layers. Furthermore, with
a centrally compact IGE ejecta composition, the photosphere would transition
somewhat rapidly from \radni\ poor to \radni\ rich, resulting in a significant
change in the temperature of the ejecta along the lines of what we see in the
evolution of \RSiII.

This interpretation is supported by photometric modeling of the rise of \sn.
\citet{Magee20} developed a suite of models featuring different \radni\
structures within the SN ejecta. These models were compared to early
observations of SNe\,Ia to see which ones replicate what is observed in
nature. Generally, it is found that centrally concentrated \radni\ models do
not match the early evolution of normal SNe\,Ia \citep{Magee20}. Using the
methods described in \citet{Magee20}, we have modeled the post-flash rise of
\sn\ using a new model with $M_{^{56}Ni} = 0.3\,M_\odot$ \citep[the models
in][all have $M_{^{56}Ni} > 0.4\,M_\odot$ and are therefore more luminous than
\sn]{Magee20}. After excluding the first two epochs of ZTF observations, as we
consider the mechanism that produces the early UV flash to be different from
the standard \radni\ decay that powers most SNe\,Ia, we find that \sn\ is best
matched with compact \radni\ distributions \citep[following the convention
of][an \texttt{EXP\_Ni0.3\_KE1.40\_P21} model provides the best match to \sn,
see also Figure~\ref{fig:Ni_bullet}]{Magee20}. We note, however, that
\citet{Magee20} demonstrate that the time of first detection can dramatically
alter the inferred model properties and it is unclear which epochs (if any)
should be excluded. Nevertheless, a scenario in which the \radni\ and other
IGEs are confined to the central regions of the ejecta is also consistent with
our spectroscopic analysis (see \S\ref{sec:tardis}).

On their own, a low-\radni\ yield that is centrally concentrated fails to
explain the blue UV/optical flash seen in \sn. A large number of scenarios
have been proposed to explain early ``bumps'' or ``flashes'' in SN\,Ia light
curves, including: shock cooling following the shock breakout from the surface
of the WD (e.g., \citealt{Piro10,Rabinak11}), interaction between the SN
ejecta and a nondegenerate binary companion \citep{Kasen10a}, extended clumps
of \radni\ in the SN ejecta (e.g., \citealt{Dimitriadis19,Shappee19}),
double-detonation explosions (e.g., \citealt{Noebauer17,Polin19}), and
interaction between the SN ejecta and circumstellar material (e.g.,
\citealt{Dessart14,Piro16,Levanon17}).

Below we discuss each of these models, aside from the shock breakout
model, and their ability to replicate observations of \sn. We do not
discuss shock breakout models as our initial detection of \sn\ occurred at
$M_g \approx -16.3$\,mag. A progenitor radius of $\sim$10$\,R_\odot$ is needed
to explain such a high luminosity \citep{Piro10,Rabinak11}, which we consider
implausible for a WD.

\subsection{Companion Interaction}\label{sec:companion_interaction}

For SD progenitors of SNe\,Ia, the SN ejecta will shock on the surface of the
nondegenerate companion giving rise to a short-lived transient in the days
after explosion. \citet{Kasen10a} provided models for the appearance of this
interaction, which is primarily dependent upon the binary separation of the
system (assuming Roche lobe overflow for the nondegenerate companion). The
observed emission following the ejecta-companion collision is dependent upon
the orientation of the binary system relative to the line of sight
\citep{Kasen10a}.

An analytic formulation for the luminosity and effective temperature of the
emission from the companion shock is given in Equations~(22) and (25) in
\citet{Kasen10a}. \citet{Brown12} provide an analytic function to approximate
the fractional decrease in the observed flux as a function of the orientation
of the system. We combine the equations from \citet{Kasen10a} and
\citet{Brown12} to model the early emission from \sn\ as an ejecta-companion
collision. We assume the interaction emits as a blackbody, and that the
electron scattering opacity $\kappa_e = 0.2$\,cm$^{2}$\,g$^{-1}$ (as in
\citealt{Kasen10a}). Assuming $z_\mathrm{SN} = 0.0094$, $E(B-V)_\mathrm{MW} =
0.018$\,mag, and $E(B-V)_\mathrm{host} = 0.032$\,mag, we compare observed flux
measurements with those predicted by the \citet{Kasen10a} model at epochs with
MJD$\,< 58$,849.2 (i.e., the first $\sim$2.5\,d after discovery when emission
from the companion interaction is significantly brighter than the luminosity
due to radioactive decay).\footnote{Given that \sn\ is an unusual SN, we make
no assumptions about the ``normal'' SN emission due to radioactive decay of
\radni. The companion-interaction model should therefore
\textit{underestimate} the observed flux after a few days as there will be a
growing contribution due to radioactive decay with time.} The model
parameters, summarized in Table~\ref{tab:companion}, are constrained via a
Gaussian likelihood and flat priors using an affine-invariant
\citep{Goodman10} Markov Chain Monte Carlo (MCMC) ensemble sampler
\citep{Foreman-Mackey13}.

\begin{deluxetable}{llcc}[htp]
\tablecaption{Model Parameters $\Theta$ and Their Priors and Posteriors \label{tab:companion}}
\tablehead{
\colhead{$\Theta$}
& \colhead{Description}
& \colhead{Prior}
& \colhead{Posterior}
} 
\startdata
$a$ & Companion separation & 
$\mathcal{U}(10^{10},10^{13})$  & $9.1 \pm 0.7 \times 10^{11}$\,cm \\
$M_\mathrm{ej}$ & Ejecta mass & 
$\mathcal{U}(0.6,1.5)$ & $1.1 \pm 0.3\,M_\odot$ \\
$v_\mathrm{ej}$ & Ejecta velocity & 
$\mathcal{U}(5 \times 10^8,3 \times 10^9)$ & $2.2 \pm^{0.5}_{0.3} \times 10^{9}$\,cm\,s$^{-1}$ \\
$\theta_\mathrm{obs}$ & Binary viewing angle & 
$\mathcal{U}(0,180)$ & $40 \pm 28^{\circ}$ \\
$t_\mathrm{exp}$ & Time of explosion &  
$\mathcal{U}(t_0 - 5,t_0)$\tablenotemark{{\scriptsize a}} & $58,845.82 \pm 0.04$ (MJD)
\enddata
\tablecomments{Marginalized 1D posterior values represent the 68\% credible region. $M_\mathrm{ej}$ is largely unconstrained by the observations. The posterior distribution on $\theta_\mathrm{obs}$ is effectively flat between $0^{\circ}$ and $\sim$70$^{\circ}$, and $\sim$0 for all angles above $\sim$85$^{\circ}$. There is a strong covariance between $a$ and $t_\mathrm{exp}$ and also between $v_\mathrm{ej}$ and $\theta_\mathrm{obs}$.
}
\tablenotetext{a}{$t_0$ is the time of the first ZTF observation ($\mathrm{MJD}=58,846.469942$).}
\end{deluxetable}

\begin{figure}
    \centering
    \includegraphics[width=3.35in]{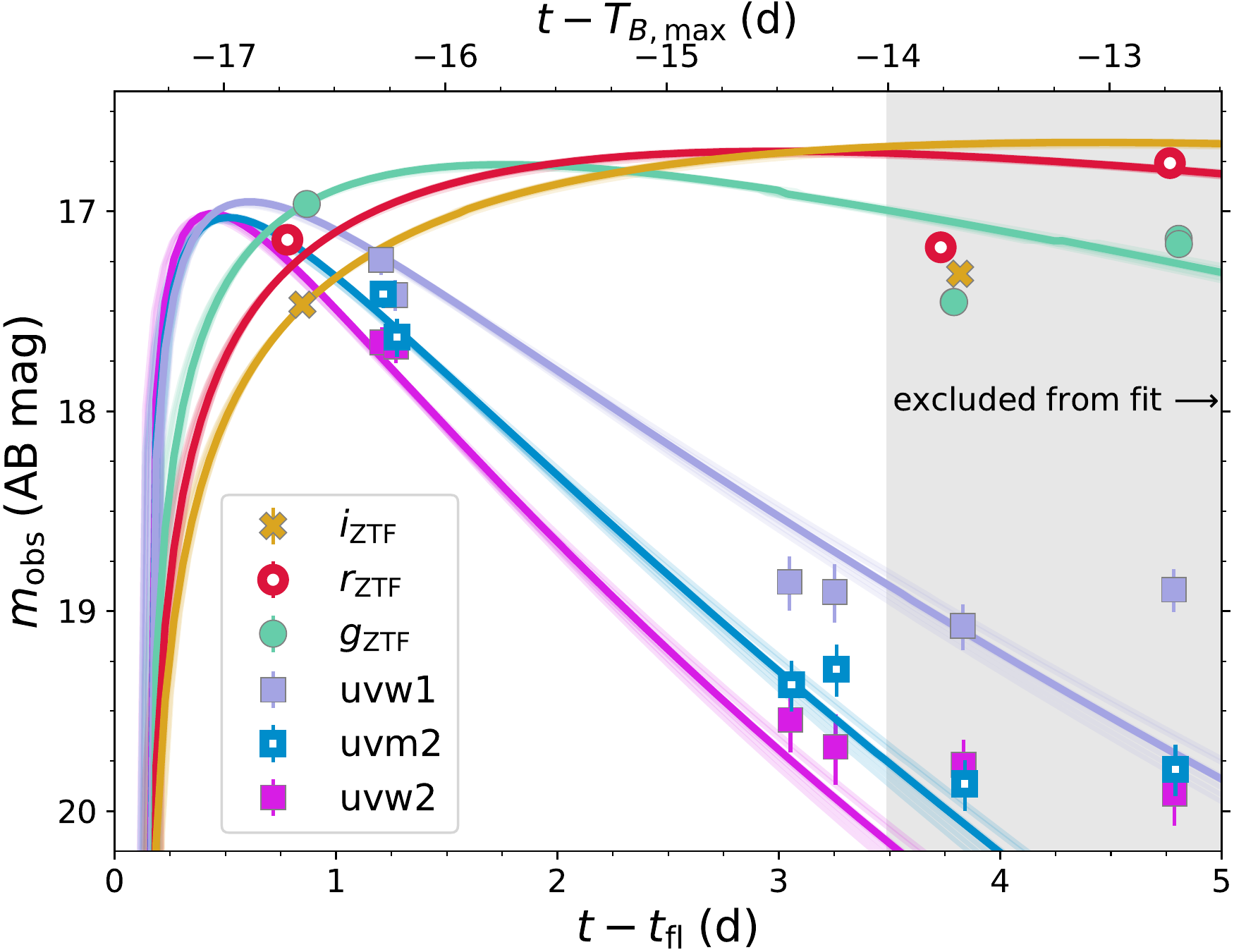}
    \caption{SN ejecta-companion-interaction models compared with the
    UV/optical observations of \sn. Observation symbols are the same as
    Figure~\ref{fig:p48} (solid magenta squares show \textit{Swift} $uvw2$
    observations that are not shown in Figure~\ref{fig:p48}). Solid lines show
    companion-interaction model predictions in each filter (the lines have the
    same colors as the corresponding symbols for each passband). The maximum a
    posteriori model is shown via the single bold lines, while other random
    draws from the posterior are shown as thin transparent lines. The shaded
    area shows observations that are excluded from the model fit. The
    overprediction of the optical flux $\sim$13.7\,d prior to \tbmax\ suggests
    that companion interaction does not explain the early flash in \sn\ (see
    text).}
    \label{fig:companion}
\end{figure}

The results of this procedure are shown in Figure~\ref{fig:companion}, where
it is clear that the model presented in \citet{Kasen10a} does an adequate job
of explaining the early UV/optical emission from \sn\ (constraints on the
model parameters are reported in Table~\ref{tab:companion}).

While the interaction models roughly approximate the SN emission in the
$\sim$3\,d after explosion, they significantly \textit{overestimate} the flux
immediately after the fitting window as shown in Figure~\ref{fig:companion}.
This problem is exacerbated by the fact that the models do not include
emission associated with radioactive decay, meaning the true discrepancy
between what is predicted and what is observed is even larger than what is
shown in Figure~\ref{fig:companion}. If we extend the fitting window to
include the optical observations obtained $\sim$13.7\,d before \tbmax, the
interaction models still overpredict the optical flux at this epoch. This
overprediction of the optical flux poses a challenge for the
companion-interaction scenario; an inability to simultaneously match both UV
and optical observations has been noted for other SNe\,Ia with early bumps or
linear rises \citep{Hosseinzadeh17,Miller18}.

\citet{Kasen10a} notes several assumptions and approximations in the
derivation of the equations used to estimate the emission from the companion
shock. It is possible that the inclusion of more detailed physics, or
additional complexity in the analytic formulation of the models,\footnote{For
example, \citet{Kasen10a} points out that the derived equation for the
luminosity of the shock interaction does not account for the advected
luminosity that would be seen in the observer frame.} could better reconcile
companion-interaction models with \sn. Such improvements are beyond the scope
of this paper, leading us to explore other explanations for the early flash.

Following arguments from \citet{Kromer16}, the evolution of \sn\ after the UV
flash also poses a challenge to the companion-interaction scenario. In the SD
scenario the WD explodes at, or very near, the Chandrasekhar mass. The leading
mechanism for such an explosion is the delayed detonation scenario, in which
the burning starts as a subsonic deflagration and later transitions to a
supersonic detonation \citep{Khokhlov91}. This scenario was explored in detail
via numerous 3D explosion models in \citet{Seitenzahl13}, with radiative
transfer calculations of the resulting emission presented in \citet{Sim13}.
While the faintest models presented in \citet{Sim13} have a similar luminosity
at peak as \sn, they feature \ion{Si}{II} velocities that are significantly
lower than \sn. The \citeauthor{Sim13} models with high \ion{Si}{II}
velocities are far too luminous to explain \sn. In addition to the delayed
detonation scenario, Chandrasekhar mass WDs can explode as pure deflagrations.
While the \radni\ yield and peak luminosity of pure deflagrations is more in
line with \sn\ than delayed detonation explosions, pure deflagrations result
in low expansion velocities and relatively weak \ion{Si}{II} absorption
\citep[e.g.,][]{Fink14} meaning they too provide a poor match to \sn.

\subsection{Ni Clumps in the SN Ejecta}

SN\,2018oh was observed with an exquisite 30\,minute cadence by the Kepler
spacecraft and showed a clearly delineated linear rise in flux followed by a
``standard'' $t^2$ power law $\sim$4\,d after \tfl. Models with extended
clumps of \radni\ just below the WD surface have been proposed as a possible
explanation for the initial linear rise in SN\,2018oh \citep{Dimitriadis19,
Shappee19}. The models considered in \citet{Shappee19} and
\citet{Dimitriadis19}, which build on the work of \citet{Piro16}, only cover
the first $\sim$10\,d after explosion and assume relatively simple gray
opacities. To further investigate this possibility, \citet{Magee20a} recently
performed more detailed radiative transfer calculations for SNe\,Ia with
extended clumps of \radni. They then compared these models to SN\,2018oh and
SN\,2017cbv, another event with a clearly resolved bump in the early light
curve \citep{Hosseinzadeh17}.

\begin{figure*}
    \centering
    \includegraphics[width=\textwidth]{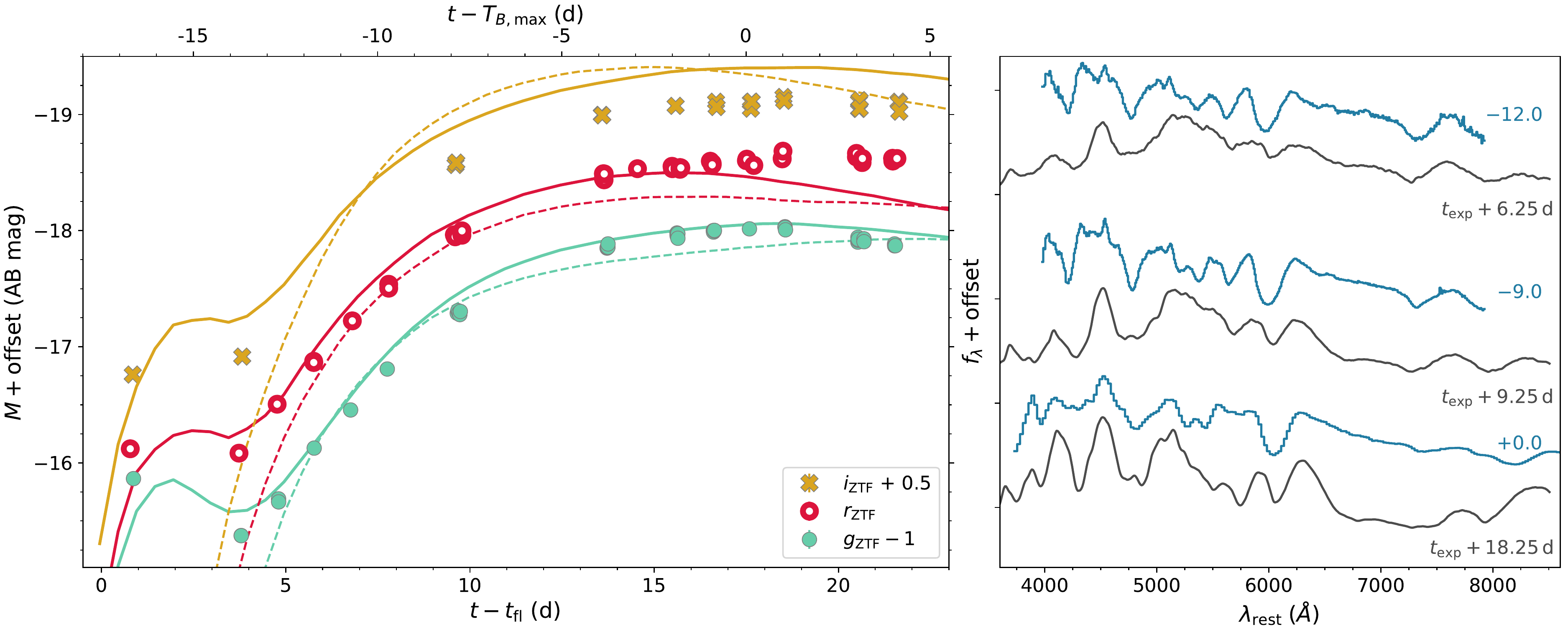}
    \caption{Comparison of \sn\ and our model with a 0.02\,$M_\odot$
    clump of \radni\ in the outer ejecta. For the comparison we have adopted a
    model explosion time $t_\mathrm{exp} = t_\mathrm{fl} - 0.8$\,d.
    Left: photometric comparison between \sn\ and the model. Symbols
    are the same as Figure~\ref{fig:p48}. The clump model is shown via solid
    lines, while the best-fit model for the ``normal'' rise is shown as dashed
    lines. The clump models have been offset by $-0.1$\,mag to account for the
    blanketing due to the clump (see text). The Ni-clump model provides an
    adequate match to the flash in the \gztf- and \rztf-bands.
    Right: spectroscopic comparison between \sn\ and the model.
    Observed spectra of \sn\ are shown in blue, with phases marked relative to
    \tbmax, whereas the model spectra are shown in dark gray, with phases
    marked relative to the modeled time of explosion. The modeled spectra have
    been smoothed with a Savitzky-Golay filter \citep{Savitzky64}. While an
    extended clump of \radni\ in the SN ejecta can produce an early optical
    flash, it leads to strong blanketing in the blue portion of the optical
    spectra ($\lambda \lesssim 4400$\,\AA) that is not observed around maximum
    light in \sn. }
    \label{fig:Ni_bullet}
\end{figure*}

For \sn\ we follow the procedure in \citet{Magee20a} to model the early flash
and rise of the SN. As described in the beginning of \S\ref{sec:models}, we
generate a ``baseline'' model that replicates the rise of \sn\ after the first
two epochs of ZTF optical detections. Following the generation of this
``baseline'' model, we add clumps of \radni\ to the outer layers of the SN
ejecta, and perform full radiative transfer calculations using \texttt{TURTLS}
\citep{Magee18}. 

We find that a model with a 0.02~$M_{\rm{\odot}}$ clump of \radni\ adequately
matches the early optical evolution of \sn\ in the \gztf\ and \rztf-bands, as
shown in Figure~\ref{fig:Ni_bullet}. The model flux in the \iztf-band
is overestimated, however, meaning the model is redder than what is observed.
In Figure~\ref{fig:Ni_bullet}, the Ni-clump models have been offset by
$-0.1$\,mag to better match the observations. This offset is necessary as the
Ni clump provides some blanketing around maximum light, and the parameter
space is too large to simultaneously optimize both the central Ni mass and the
clump Ni mass.

While a clump of \radni\ can produce an optical bump in the light curve, the
same challenges identified in \citet{Magee20a} apply to \sn. In particular, an
extended clump of \radni\ dramatically alters the appearance of the SN at
maximum light. Figure~\ref{fig:Ni_bullet} shows a comparison of the observed
spectra with our calculated models. The Ni-clump models feature strong
blanketing in the blue-optical region of the spectrum, which is simply not
present in the observed spectra of \sn. We therefore conclude that Ni clumps
cannot explain the early flash seen in \sn.

\subsection{Double-Detonation Models}

WDs that accrete a thin shell of He can explode via a ``double detonation''
whereby explosive burning in the He shell drives a shock into the C/O core of
the WD. This shock can ignite explosive C burning and trigger a detonation
that disrupts the entire star (e.g., \citealt{Nomoto82,Nomoto82a,Woosley94}).
Such explosions are even possible in C/O WDs that are well below the
Chandrasekhar mass (see \citealt{Fink07, Fink10} and references therein).

Recent models of double-detonation explosions presented in \citet{Polin19}
show that such explosions can replicate several of the peculiar properties of
\sn, including the early UV/optical flash, a blue to red to blue color
transition, the moderately faint optical peak, red colors at maximum, and a
lack of IGE in the early spectra.

The appearance of double-detonation SNe is effectively determined by two
properties: the mass of the C/O core and the mass of the He shell. The total
mass of the system determines the central density of the WD and thus the
amount of synthesized \radni. The \radni\ mass directly controls both the peak
luminosity and the kinetic energy of the explosion. High mass WDs ($\gtrsim
1.1\,M_\odot$) create enough \radni\ ($M_\mathrm{Ni} \gtrsim 0.5\,M_\odot$) to
produce large ($\gtrsim 1$4,000\,\kms) photospheric velocities and reach
normal brightness for an SN\,Ia, while low mass WDs ($\lesssim 0.9\,M_\odot$)
exhibit lower photospheric velocities ($\lesssim 1$0,000\,\kms) and produce
less \radni, therefore peaking at fainter luminosities \citep{Polin19}. That
we see both a high \ion{Si}{II} velocity and a low peak luminosity in \sn\
presents a challenge for the \citet{Polin19} double-detonation models (see
their Figure~11). Furthermore, thick He shells ($M_\mathrm{He} \gtrsim
0.05\,M_\odot$) produce more pronounced UV/optical flashes shortly after
explosion, particularly in conjunction with lower mass WDs, while thin He
shells ($M_\mathrm{He} \lesssim 0.02\,M_\odot$) produce a more extreme color
inversion in the days after explosion.

\begin{figure*}
    \centering
    \includegraphics[width=\textwidth]{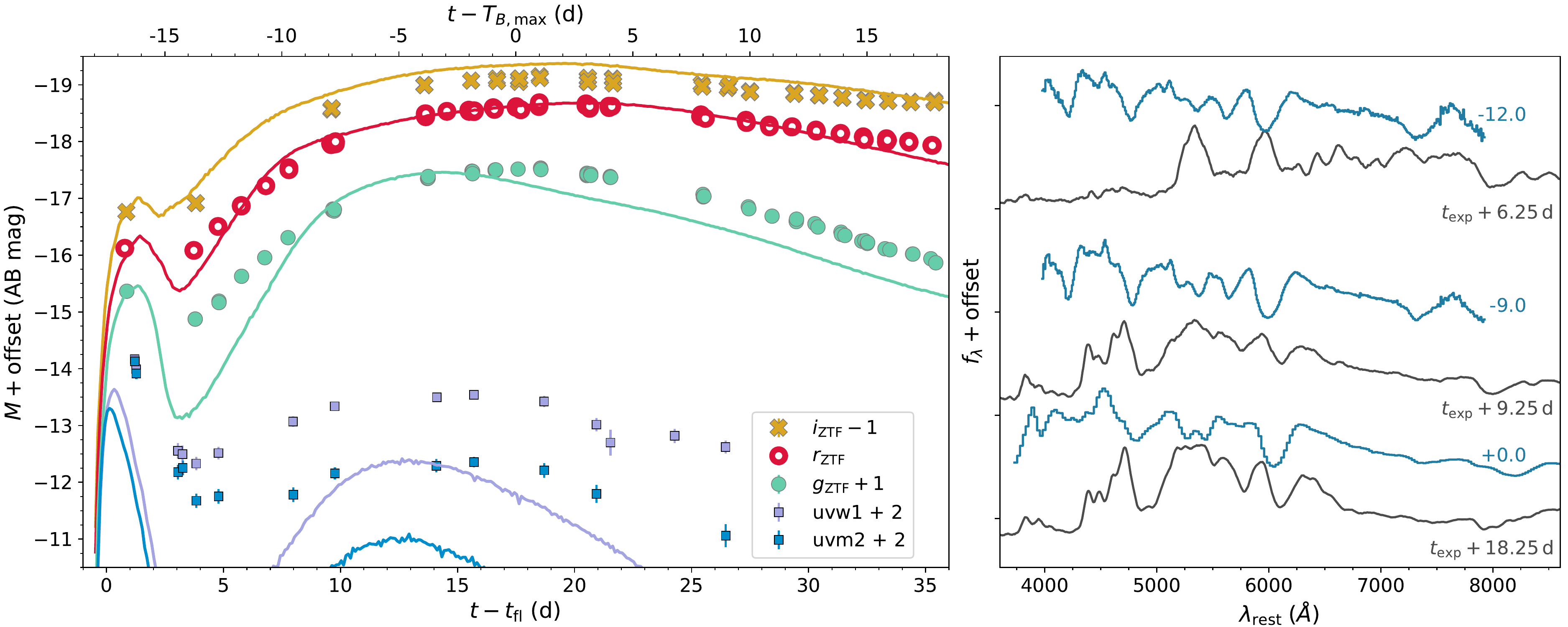}
    \caption{Comparison of \sn\ to a double-detonation model with a C/O core
    mass $M_\mathrm{C/O} = 0.92\,M_\odot$ and He shell mass $M_\mathrm{He} =
    0.04\,M_\odot$ (i.e., $M_\mathrm{WD} = 0.96\,M_\odot$). For the comparison
    we have adopted a model explosion time $t_\mathrm{exp} = t_\mathrm{fl} -
    0.72$\,d. Left: photometric comparison between \sn\ and the
    model. Symbols are the same as Figure~\ref{fig:p48}. The double-detonation
    model provides a good match to the \rztf-band evolution, though the flux
    in the \gztf- and \iztf-bands is under- and overpredicted, respectively.
    The UV emission is also underestimated by the double-detonation model.
    Right: spectroscopic comparison between \sn\ and the model.
    Observed spectra of \sn\ are shown in blue, with phases marked relative to
    \tbmax, whereas the model spectra are shown in dark gray, with phases
    marked relative to the modeled time of explosion. The modeled spectra
    have been smoothed with a Savitzky-Golay filter \citep{Savitzky64}. The
    photospheric velocity in the double-detonation model is lower than what is
    observed in \sn, and the models feature more absorption and blanketing in
    the blue portion of the optical than what is observed. }
    \label{fig:double_det}
\end{figure*}

We have attempted to model the evolution of \sn\ as a double-detonation
explosion, following the procedure in \citet{Polin19}. We have specifically
focused on matching the photometric evolution (as noted above no models create
high-velocity ejecta and underluminous optical peaks), with particular
attention to the colors during the early flash and at maximum light. We find
that a model with $M_\mathrm{C/O} = 0.92\,M_\odot$ C/O core and a
$M_\mathrm{He} = 0.04\,M_\odot$ He shell best match \sn, as shown in
Figure~\ref{fig:double_det}.

While this model adequately matches the evolution of \sn\ in the \rztf\
filter, the predictions in the \gztf- and \iztf-bands do not match what is
observed. We show for the first time that there is an expected UV flash
associated with these double-detonation models, however, our best-fit model
underestimates the flux that was observed in the UV.

Synthesized spectra from our double-detonation model exhibit features that are
not seen in \sn. The model spectra are dominated by \ion{Si}{II} absorption,
and show high-velocity absorption due to \ion{O}{I} and \ion{Ca}{II}, similar
to \sn. For our best-fit model, however, the \ion{Si}{II} velocities are too
slow, the \ion{Si}{II} $\lambda$5972 absorption is too strong, and the
\ion{S}{II} absorption is too weak. Nuclear burning in the He shell creates
heavy elements in the outermost ejecta of double-detonation explosions,
leading to deep \ion{Ti}{II} troughs and other blanketing in the blue-optical
portion of the spectrum. Our model exhibits a strong \ion{Ti}{II} absorption
trough blueward of $\sim$4400\,\AA\ (see the $t_\mathrm{exp} + 9.25$\,d
spectrum in Figure~\ref{fig:double_det}). As was the case for models with
extended clumps of \radni, the lack of such absorption in \sn\ poses a
challenge for the double-detonation model.

With observations that probe a previously unexplored phase in the evolution of
such explosions, \sn\ provides an opportunity to determine where the
double-detonation models must improve. It is possible that such improvements
could lead to better agreement with \sn. For instance, the nuclear reaction
networks and 1D models in \citet{Polin19} always burn the He shells to nuclear
statistical equilibrium. It is not unreasonable to think that 2D or 3D models,
with a more sophisticated nuclear reaction network, would create more IMEs and
less IGEs in the He shell, and that the ratio of the two created in the shell
could be highly dependent upon the line of sight. For example,
\citet{Townsley19} modeled the explosion of a $M_\mathrm{C/O} = 1.0\,M_\odot$
C/O core with a $M_\mathrm{He} = 0.02\,M_\odot$ He shell and found a higher
ratio of IME to IGE than the analogous 1D model presented in \citet[][though
see also \citealt{Gronow20}, which presents a 3D double-detonation explosion
with nuclear yields that are only mildly different from 1D models]{Polin19}.
This could explain the lack of IGEs and strong \ion{Si}{II} absorption seen in
the early spectra, while less IGEs in the outer layers would also reduce some
of the line blanketing seen around maximum light. This would lead to less
reprocessing of blue photons, perhaps creating better agreement between the
models and photometry, particularly in the \gztf-band. The velocity
discrepancy could also potentially be explained as a line-of-sight effect. If
the ignition of the WD occurred off center, then the ejecta aligned with the
site of the initial He ignition may receive a boost in velocity
\citep[e.g.,][]{Kromer10}. The discrepancies in the UV are less worrisome.
While we show a qualitative UV flash, the magnitude of this flash will be
highly sensitive to the precise temperature and composition in the very
outermost ejecta, and thus any of the changes discussed above could easily
boost the model flux in the UV.

\subsection{Violent Mergers and Circumstellar
Interaction}\label{sec:merger_csm}

\begin{figure*}[ht]
    \centering
    \includegraphics[width=\textwidth]{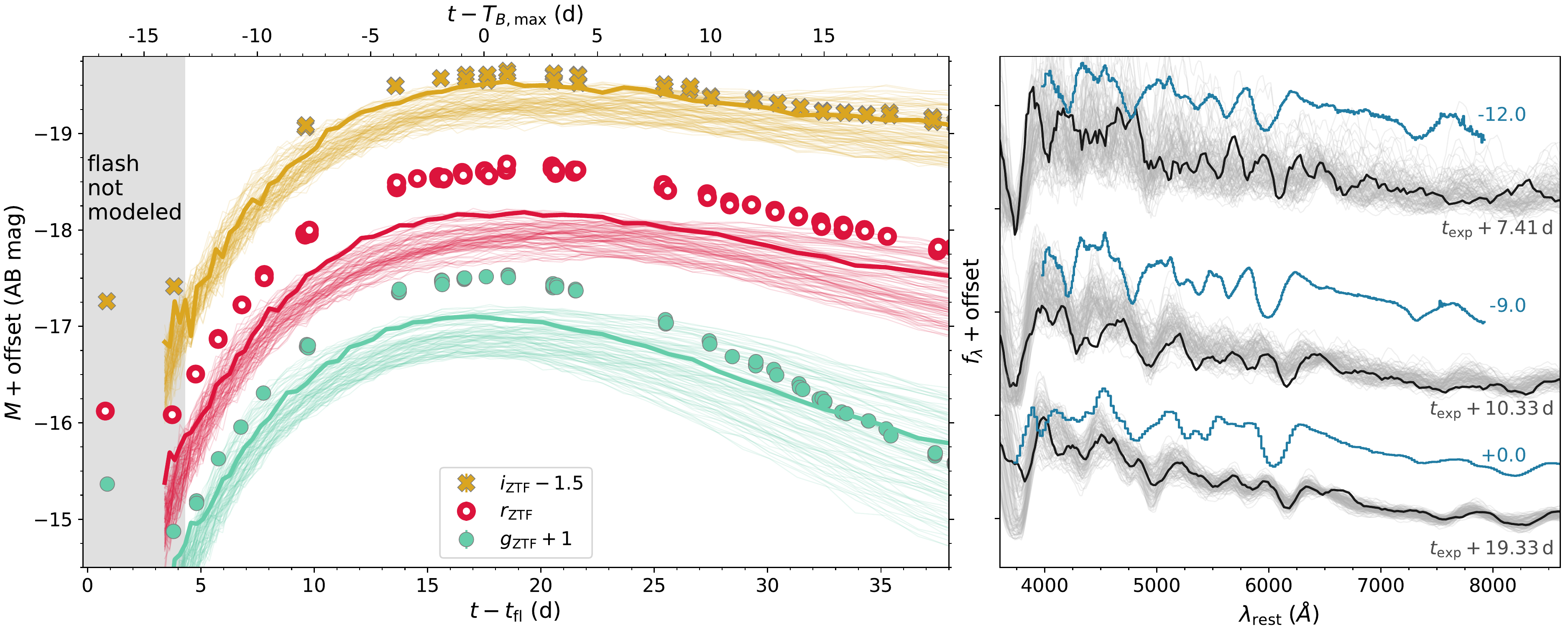}
    \caption{Comparison of \sn\ to the low-metallicity ($Z = 0.01 Z_\odot$)
    violent merger model of a 0.9 and 0.76\,$M_\odot$ WD from
    \citet{Kromer16}. This model provides a good match to iPTF\,14atg, and
    therefore significantly underestimates the brightness of \sn. For the
    comparison we have adopted a model explosion time of $t_\mathrm{exp} =
    t_\mathrm{fl} - 1.92$\,d. Thin lines represent one of 100 sightlines,
    while the bold lines represent a single sightline for illustrative
    purposes. Left: photometric comparison between \sn\ and the model. Symbols
    are the same as those in Figure~\ref{fig:p48}. The shaded area shows the
    UV/optical flash from \sn, which was not modeled by \citet{Kromer16}.
    Despite the underestimated optical flux, the qualitative behavior of the
    violent merger model, including red \gztf$ - $\rztf\ colors at peak and a
    lack of secondary maximum in the \iztf-band do match \sn. Right:
    Spectroscopic comparison between \sn\ and the violent merger model.
    Observed spectra of \sn\ are shown in blue, with phases marked relative to
    \tbmax, whereas the model spectra are shown as thin gray lines. The thick
    black line highlights a specific viewing angle. The modeled spectra have
    been smoothed with a Savitzky-Golay filter \citep{Savitzky64}. The
    photospheric velocity in the violent merger model features lower
    velocities than \sn, while the strength of the the IME absorption is
    weaker in the models than what is observed.}
    \label{fig:violent_merger}
\end{figure*}

\citet{Piro16} show that circumstellar material in the vicinity of a WD at the
time of explosion can give rise to an early flash or bump in the SN\,Ia light
curve. Using a 1D toy model, with an assumed circumstellar density profile
$\propto r^{-3}$ and gray opacities, \citet{Piro16} found that the peak of the
early emission is roughly proportional to the extent of the circumstellar
material, while the duration of the flash is proportional to the square root
of the circumstellar mass. While the brightest model from \citet{Piro16} has a
flash brightness that peaks at $M_V \approx -15$\,mag, circumstellar material
that extends beyond $\sim$10$^{12}$\,cm could give rise to a flash that peaks
at $M_g \lesssim -16.4$\,mag, as is observed in \sn.

There are few proposed WD explosion models that produce dense circumstellar
material in the vicinity of the WD at the time of explosion. A notable
exception is the violent merger, so called because the thermonuclear explosion
happens while the merger is still ongoing, of two C/O WDs
\citep{Pakmor10,Pakmor11,Pakmor12}. DD mergers should produce a wide variety
of circumstellar configurations, depending on the initial parameters of the
inspiralling binary, which would, in turn, produce different signals shortly
after explosion (e.g., \citealt{Raskin13,Levanon19}).\footnote{Indeed, the
large number of potential configurations makes it very difficult to rule out
or select any specific circumstellar interaction scenario.}

Given the vast parameter space populated by different circumstellar
configurations, we are going to proceed under the (potentially poor)
assumption that such interaction could reproduce the UV/optical flash seen in
\sn. Following this assumption, a relevant question is -- can violent mergers
reproduce the properties of \sn\ in the days before and weeks after \tbmax?

In \citet{Kromer16}, the violent merger of two C/O WDs with masses of 0.9 and
0.76\,$M_\odot$ produced a similar rise and maximum-light properties to
iPTF\,14atg, the other SN\,Ia with an observed early UV flash. A comparison of
\sn\ to the low-metallicity model from \citet{Kromer16}, which provides a good
match to iPTF\,14atg, is shown in
Figure~\ref{fig:violent_merger}.\footnote{The viewing angle dependent spectra
of this merger model are available on the Heidelberg Supernova Model Archive
\citep{Kromer17}.} We show that model here to illustrate the qualitative
behavior of such a merger; it is not meant to provide an optimal match to \sn.
The \citeauthor{Kromer16}~model was not designed to fit the early UV flash in
iPTF\,14atg.

The photometric evolution of this violent merger model qualitatively matches
\sn: (i) a moderately faint peak in the optical ($-17.6\,\mathrm{mag} \gtrsim
M_g \gtrsim -18.2$\,mag, depending on the viewing angle), (ii) red $g - r$
colors at peak, and (iii) a lack of a secondary maximum in the $i$-band.
Furthermore, the spectra lack significant IGE absorption in the days after
explosion (right panel of Figure~\ref{fig:violent_merger}), as is observed in
\sn. Interestingly, the violent merger model does show a decrease in the
relative strength of the \ion{Si}{II} $\lambda$5972 absorption with time,
similar to \sn\ and unlike the other models considered here. A critical
difference between \sn\ and violent merger models, is that the merger models
tend to produce relatively low expansion velocities (e.g.,
\citealt{Pakmor10,Kromer13a,Kromer16}). Indeed, this is one of the stark
differences between \sn\ and iPTF\,14atg, as iPTF\,14atg had a \ion{Si}{II}
$\lambda$6355 absorption velocity of $\sim$7500\,\kms\ at peak, or roughly
half that observed in \sn. It is also clear from
Figure~\ref{fig:violent_merger} that the violent merger model from
\citet{Kromer16} exhibits weaker IME absorption than what is seen in \sn.

It is clear that additional modeling, likely of a different WD binary
configuration, is needed to better match \sn. For example, it is known that a
higher mass primary WD can produce more \radni, and hence a brighter optical
peak \citep[e.g.,][]{Pakmor12}, which would be more in line with \sn. If, at
the same time, the mass of the secondary were slightly decreased, then the
kinetic energy of the ejecta would increase, perhaps bringing the model
velocity of \ion{Si}{II} and other IMEs in line with \sn. It would also be
beneficial to track the unbound material following the DD merger, to see if
the collision between this material and the SN ejecta can replicate the early
UV/optical flash seen in \sn. If this feature can readily be recreated, it is
possible that a violent merger is responsible for \sn.

\section{Discussion}\label{sec:conclusions}

\begin{deluxetable*}{lccccccc}[htp]
\tabletypesize{\small}
\tablecaption{Summary of Observational Properties of \sn \label{tab:models}}
\tablehead{
\colhead{} &
\multicolumn{7}{c}{Does the Model Replicate this Property?} \\
\cline{2-8}
\colhead{}
& \colhead{UV}
& \colhead{Low Peak}
& \colhead{Intermediate/Fast}
& \colhead{Red Colors}
& \colhead{Lack of IGE}
& \colhead{\RSiII}
& \colhead{High \ion{Si}{II}} \\
\colhead{Model}
& \colhead{Flash}
& \colhead{Luminosity}
& \colhead{Decline}
& \colhead{at All Epochs}
& \colhead{in Early Spectra}
& \colhead{Evolution}
& \colhead{Velocities}
} 
\startdata
Companion interaction & \cmark & \textbf{?} & \textbf{?} & \textbf{?} & \textbf{?} & \textbf{?} & \textbf{?} \\
\radni\ clumps & \cmark & \textbf{?} & \textbf{?} & \cmark & \xmark & \textbf{?} & \textbf{?} \\
He shell double detonation & \cmark & \cmark & \cmark & \cmark & \xmark & \xmark & \xmark \\
Violent merger & \textbf{?} & \cmark & \cmark & \cmark & \cmark & \cmark & \xmark
\enddata
\tablecomments{If a model replicates a specific property we show a \cmark,
whereas properties that are not matched are signified with an \xmark. Ambiguous
cases are shown as \textbf{?}. An important distinction for the
companion-interaction and \radni-clump models is that they are empirical,
whereas the double-detonation and violent merger models are based on a specific
realization of an exploding WD. Given that the companion-interaction and
\radni-clump models do not model the explosion itself, we label all properties
that are not generic to the class as \textbf{?}. While the double-detonation
and \radni-clump models produce a UV flash, it is unclear whether or not they
can match the magnitude of the observed flash in \sn. The violent merger model
does not track circumstellar material, and additional simulations are needed to
understand whether interaction between the ejecta and unbound material could
reproduce the UV flash (see text). In addition to showing evidence for IGE
absorption in the early spectra, the double-detonation and \radni-clump models
show strong IGE absorption and line blanketing around maximum light that is not
observed in \sn.}
\end{deluxetable*}

We have presented observations of the spectacular \sn, the second observed
SN\,Ia to exhibit a clear UV/optical flash in its early evolution. Despite
this dazzling, declarative display announcing \sn\ as a unique event among the
thousands of SNe\,Ia that have previously been cataloged, we find that \sn\
would be considered unusual even if the early flash had been missed.

The photometric evolution of \sn\ resembles that of the intermediate 86G-like
subclass of SNe\,Ia. With a moderately faint peak in the optical ($M_g \approx
-18.5$\,mag), relatively fast decline [$\Delta m_{15}(g) = 1.3$\,mag], and
lack of a secondary maximum in the \iztf\ filter, \sn\ is clearly
distinguished photometrically from normal SNe\,Ia. These photometric
properties typically correspond to Branch\ Cool SNe, yet the spectroscopic
evolution of \sn\ does not match such events. \sn\ is a Branch\ Broad Line SN,
with relatively weak \ion{Si}{II} $\lambda$5972 absorption and large
\ion{Si}{II} velocities. Furthermore, our \texttt{TARDIS} spectral models show
little to no IGE present in the outer layers of the SN ejecta, which further
distinguishes \sn, even relative to other Branch~Broad Line SNe. The fact that
\sn\ exhibits high-velocity \ion{Si}{II} $\lambda$6355 absorption and an
underluminous peak sets it apart from other SNe\,Ia.

\sn\ is one of a growing group of SNe\,Ia with photometric properties that may
or may not deviate from the standard width-luminosity relationship for normal
SNe\,Ia \citep[e.g.,][]{Phillips93,Phillips99}, but whose spectral evolution
is incongruous with their photometric properties. While these SNe all differ
in detail, many can be linked via the presence of 91bg-like spectroscopic
features, such as the \ion{Ti}{II} ``trough'' at $\sim$4200\,\AA\
\citep{Filippenko92,Leibundgut93}, despite relatively broad light curves that
are more consistent with normal or intermediate SNe\,Ia (examples include:
SN\,2006bt, \citealt{Foley10a}; PTF\,10ops, \citealt{Maguire11}; SN\,2006ot,
\citealt{Stritzinger11}; SN\,2010lp, \citealt{Kromer13a}; SN\,2002es,
\citealt{Ganeshalingam12}; and iPTF\,14atg, \citealt{Cao15}).

\citet{Benetti05} showed that photometric and spectroscopic properties of
SNe\,Ia are closely linked by connecting normal and subluminous 91bg-like
SNe\,Ia in a tight sequence in the \RSiII--$\Delta m_{15}(B)$ plane. As first
pointed out in \citet{Foley10a} and later confirmed by \citet{Maguire11} and
\citet{Ganeshalingam12}, the peculiar SNe mentioned above starkly standout
from the simple sequence found in \citet{Benetti05} as the peculiar SNe all
have \RSiII\ values that are much higher than expected given their decline
rate as parameterized by $\Delta m_{15}(B)$. \sn\ also stands out in this
plane, though in the opposite sense, the low \RSiII\ at maximum light
(\S\ref{sec:SiII}) suggests a slow decline, which is not observed
(\S\ref{sec:max_decline}). Whether these events all feature a common origin
remains to be seen, though it is interesting that the two events with observed
early UV flashes,\footnote{Evidence for excess optical emission in the early
light curve of PTF\,10ops is found in \citet{Jiang18}, though UV observations
are not available for PTF\,10ops making it impossible to know whether or not
there was an associated UV flash.} iPTF\,14atg and \sn, are both peculiar and
possibly connected as outliers in the \RSiII--$\Delta m_{15}(B)$ plane.

We have found that building a consistent physical model to explain all of the
observed properties of \sn\ is challenging. Most models either replicate the
early flash but fail to reproduce the observed behavior around maximum light,
or vice versa.

We have examined four models in detail to try to explain the dramatic early
UV/optical peak in \sn, including the collision of the SN ejecta with a
nondegenerate companion (e.g., \citealt{Kasen10a}), extended clumps of \radni\
in the outer layers of the SN ejecta \citep[e.g.,][]{Magee20a}, the
double-detonation explosion of a sub-Chandrasekhar mass WD
\citep[e.g.,][]{Polin19}, and the violent merger of two sub-Chandrasekhar mass
WDs \citep[e.g.,][]{Kromer16}. Table~\ref{tab:models} summarizes the
key observational properties of \sn\ listed in \S\ref{sec:models} and whether
or not these four models can explain the different aspects of \sn.

The SN ejecta-companion models, which can easily replicate the early UV flash
from \sn, simultaneously overpredict the optical flux at similar epochs.
Models with extended clumps of \radni\ produce significant blanketing in the
blue-optical region of the spectrum. While the double-detonation model
produces an early flash and \rztf\ evolution that provides a good match to
\sn, it too produces blanketing that is too strong relative to the
blue-optical spectra and features absorption velocities that are much lower
than what is observed. The specific WD merger model from \citet{Kromer16} that
we compare to \sn\ does a poor job of replicating the observations. Many of
the qualitative features match, however, so it is not unreasonable to think
that with some tuning (e.g., higher mass WDs) the merger model could better
reflect what is observed in \sn.

While we have focused on explaining the spectacular UV flash, we were also
unable to identify any models that match the maximum-light properties of \sn.
One possibility to explain the low \radni\ yield and large \ion{Si}{II}
velocities would be to terminate a lot of the nuclear burning at IMEs, which,
in turn, would result in a relatively low fraction of IGE. Such a scenario may
be possible at low central densities, which would keep the IGE fraction in the
ejecta low, if enough material burns (in order to release a sufficient amount
of energy to accelerate the ejecta to high velocities). Further work is
needed, however, to know whether such a scenario could be produced by
realistic binaries in nature.

Nebular spectra of \sn\ will play a crucial role in disambiguating between
these various scenarios. If the ejecta have collided with a nondegenerate
companion, then they will have stripped some surface material from the
companion, which will be revealed via narrow Balmer lines in the nebular phase
\citep[e.g.,][]{Wheeler75}. Alternatively, \citet{Polin19a} recently showed
that low mass ($M_\mathrm{WD} \lesssim 1.0\,M_\odot$) double-detonation
explosions do not create a significant amount of IGEs in their core. This
relative lack of IGEs means that [\ion{Ca}{II}] provides the best pathway for
the ejecta to cool, and as a result strong [\ion{Ca}{II}]
$\lambda\lambda$7291, 7324 emission is expected in the nebular phase. Finally,
violent mergers are expected to exhibit narrow [\ion{O}{I}]
$\lambda\lambda$6300, 6364 emission in their nebular spectra, as unburned O
from the disrupted WD is present at low velocities in the central ejecta
\citep{Taubenberger13,Kromer16}. Each of these predictions are unique to the
scenarios discussed here.

The critical challenge moving forward in understanding \sn-like events is the
rapid acquisition of UV observations shortly after explosion. ZTF, and other
similar surveys (ATLAS, ASAS-SN; \citealt{Tonry11,Holoien17}), have
demonstrated the ability to routinely find extremely young SNe\,Ia. Following
this the challenge is to (i) recognize these events as likely SNe\,Ia at the
epoch of discovery (i.e., without a significant delay to obtain a
spectroscopic classification) and (ii) promptly obtain \textit{Swift}
photometry. While the presence of an early UV flash may be intrinsically rare,
in the past $\sim$7\,yr it has only been observed twice, it seems more likely
that the above process (discovery, classification, \textit{Swift} ToO) is
highly incomplete. Furthermore, if a typical UV flash is either less luminous
or of a shorter duration than what was observed in iPTF\,14atg and \sn, then
the chain of events leading to \textit{Swift} observations may be insufficient
to regularly capture such a signal. It would be far more efficient to search
for such flashes directly using a wide-field UV telescope
\citep[e.g.,][]{Sagiv14}. Only after extremely early UV observations become as
routine as the discoveries themselves will we be able to statistically
constrain the models discussed herein, and, as a result, answer fundamental
questions about the nature of SN\,Ia progenitors.

\acknowledgements

The authors would like to thank the anonymous referee for helpful comments
that have improved this paper. We thank R.~Pakmor for useful conversations on
WD explosions, D.~M.~Scolnic for sharing the results of the 2M++ model, and
C.-C.~Ngeow for providing constructive comments on an early draft.

A.A.M. is funded by the Large Synoptic Survey Telescope Corporation, the
Brinson Foundation, and the Moore Foundation in support of the LSSTC Data
Science Fellowship Program; he also receives support as a CIERA Fellow by the
CIERA Postdoctoral Fellowship Program (Center for Interdisciplinary
Exploration and Research in Astrophysics, Northwestern University).

C.~F. gratefully acknowledges support of his research by the
Heising-Simons Foundation (\#2018-0907).

Ashish Mahabal acknowledges support from the NSF (1640818, AST-1815034).

E.S.P. was funded in part by the Gordon and Betty Moore Foundation
through grant GBMF5076.

M.R. and Y.-L.K. have received funding from the European Research
Council (ERC) under the European Union’s Horizon 2020 research and innovation
program (grant agreement No.\ 759194 — USNAC).

This work was supported by TCHPC (Research IT, Trinity College Dublin).
Calculations were performed on the Kelvin cluster maintained by the Trinity
Centre for High Performance Computing. This cluster was funded through grants
from the Higher Education Authority, through its PRTLI program.

This work was supported by the GROWTH project funded by the National Science
Foundation under grant No 1545949.

This work is based on observations obtained with the Samuel Oschin Telescope
48 inch and the 60 inch Telescope at the Palomar Observatory as part of the
Zwicky Transient Facility project. ZTF is supported by the National Science
Foundation under grant No. AST-1440341 and a collaboration, including Caltech,
IPAC, the Weizmann Institute for Science, the Oskar Klein Center at Stockholm
University, the University of Maryland, the University of Washington,
Deutsches Elektronen-Synchrotron and Humboldt University, Los Alamos National
Laboratories, the TANGO Consortium of Taiwan, the University of Wisconsin at
Milwaukee, and Lawrence Berkeley National Laboratories. Operations are
conducted by COO, IPAC, and UW.

SED Machine is based upon work supported by the National Science Foundation
under grant No 1106171.

This research made use of \texttt{TARDIS}, a community-developed software
package for spectral synthesis in SNe \citep{Kerzendorf14}. The development of
\texttt{TARDIS} received support from the Google Summer of Code initiative and
from ESA's Summer of Code in Space program. \texttt{TARDIS} makes extensive
use of \texttt{Astropy} and \texttt{PyNE}.

MMT Observatory access was supported by Northwestern University and the
Center for Interdisciplinary Exploration and Research in Astrophysics (CIERA).

The Liverpool Telescope is operated on the island of La Palma by Liverpool
John Moores University in the Spanish Observatorio del Roque de los Muchachos
of the Instituto de Astrofisica de Canarias with financial support from the UK
Science and Technology Facilities Council.

Partly based on observations made with the Nordic Optical Telescope, operated
at the Observatorio del Roque de los Muchachos, La Palma, Spain, of the
Instituto de Astrof\'isica de Canarias.

This work made use of data supplied by the UK \textit{Swift} Science Data
Centre at the University of Leicester.

\software{\texttt{astropy} \citep{Astropy-Collaboration13},
          \texttt{CASTRO} \citep{Almgren10},
          \texttt{corner} \citep{Foreman-Mackey16},
          \texttt{emcee} \citep{Foreman-Mackey13},
          \texttt{FRODOSpec L2 pipeline} \citep{Barnsley12},
          \texttt{LPipe} \citep{Perley19},
          \texttt{matplotlib} \citep{Hunter07}, 
          \texttt{pandas} \citep{McKinney10},
          \texttt{pyraf-dbsp} \citep{Bellm16b},
          \texttt{pysedm} \citep{Rigault19}
          \texttt{SALT2} \citep{Guy07},
          \texttt{scikit-learn} \citep{Pedregosa11},
          \texttt{scipy} \citep{2020SciPy-NMeth},
          \texttt{SEDONA} \citep{Kasen06a}, 
          \texttt{sncosmo} \citep{Barbary16},
          \texttt{SNooPY} \citep{Burns11},
          \texttt{TARDIS} \citep{Kerzendorf14},
          \texttt{TURTLS} \citep{Magee18}
          }


\bibliography{/Users/adamamiller/Documents/tex_stuff/papers}

\begin{thebibliography}{}
\expandafter\ifx\csname natexlab\endcsname\relax\def\natexlab#1{#1}\fi
\providecommand{\url}[1]{\href{#1}{#1}}
\providecommand{\dodoi}[1]{}
\providecommand{\doarXiv}[1]{\href{https://arxiv.org/abs/#1}{\nolinkurl{https://arxiv.org/abs/#1}}}

\bibitem[{{Abolfathi} {et~al.}(2018){Abolfathi}, {Aguado}, {Aguilar}, {Allende
  Prieto}, {Almeida}, {Ananna}, {Anders}, {Anderson}, {Andrews}, {Anguiano},
  {Arag{\'o}n-Salamanca}, {Argudo-Fern{\'a}ndez}, {Armengaud}, {Ata},
  {Aubourg}, {Avila-Reese}, {Badenes}, {Bailey}, {Balland}, {Barger},
  {Barrera-Ballesteros}, {Bartosz}, {Bastien}, {Bates}, {Baumgarten},
  {Bautista}, {Beaton}, {Beers}, {Belfiore}, {Bender}, {Bernardi}, {Bershady},
  {Beutler}, {Bird}, {Bizyaev}, {Blanc}, {Blanton}, {Blomqvist}, {Bolton},
  {Boquien}, {Borissova}, {Bovy}, {Bradna Diaz}, {Brandt}, {Brinkmann},
  {Brownstein}, {Bundy}, {Burgasser}, {Burtin}, {Busca}, {Ca{\~n}as},
  {Cano-D{\'\i}az}, {Cappellari}, {Carrera}, {Casey}, {Cervantes Sodi}, {Chen},
  {Cherinka}, {Chiappini}, {Choi}, {Chojnowski}, {Chuang}, {Chung}, {Clerc},
  {Cohen}, {Comerford}, {Comparat}, {Correa do Nascimento}, {da Costa},
  {Cousinou}, {Covey}, {Crane}, {Cruz-Gonzalez}, {Cunha}, {da Silva Ilha},
  {Damke}, {Darling}, {Davidson}, {Dawson}, {de Icaza Lizaola}, {de la
  Macorra}, {de la Torre}, {De Lee}, {de Sainte Agathe}, {Deconto Machado},
  {Dell'Agli}, {Delubac}, {Diamond-Stanic}, {Donor}, {Downes}, {Drory}, {du Mas
  des Bourboux}, {Duckworth}, {Dwelly}, {Dyer}, {Ebelke}, {Davis Eigenbrot},
  {Eisenstein}, {Elsworth}, {Emsellem}, {Eracleous}, {Erfanianfar},
  {Escoffier}, {Fan}, {Fern{\'a}ndez Alvar}, {Fernandez-Trincado}, {Fernand o
  Cirolini}, {Feuillet}, {Finoguenov}, {Fleming}, {Font-Ribera}, {Freischlad},
  {Frinchaboy}, {Fu}, {G{\'o}mez Maqueo Chew}, {Galbany}, {Garc{\'\i}a
  P{\'e}rez}, {Garcia-Dias}, {Garc{\'\i}a-Hern{\'a}ndez}, {Garma Oehmichen},
  {Gaulme}, {Gelfand }, {Gil-Mar{\'\i}n}, {Gillespie}, {Goddard}, {Gonz{\'a}lez
  Hern{\'a}ndez}, {Gonzalez-Perez}, {Grabowski}, {Green}, {Grier}, {Gueguen},
  {Guo}, {Guy}, {Hagen}, {Hall}, {Harding}, {Hasselquist}, {Hawley}, {Hayes},
  {Hearty}, {Hekker}, {Hernand ez}, {Hernandez Toledo}, {Hogg},
  {Holley-Bockelmann}, {Holtzman}, {Hou}, {Hsieh}, {Hunt}, {Hutchinson},
  {Hwang}, {Jimenez Angel}, {Johnson}, {Jones}, {J{\"o}nsson}, {Jullo}, {Khan},
  {Kinemuchi}, {Kirkby}, {Kirkpatrick}, {Kitaura}, {Knapp}, {Kneib},
  {Kollmeier}, {Lacerna}, {Lane}, {Lang}, {Law}, {Le Goff}, {Lee}, {Li}, {Li},
  {Lian}, {Liang}, {Lima}, {Lin}, {Long}, {Lucatello}, {Lundgren}, {Mackereth},
  {MacLeod}, {Mahadevan}, {Maia}, {Majewski}, {Manchado}, {Maraston},
  {Mariappan}, {Marques-Chaves}, {Masseron}, {Masters}, {McDermid}, {McGreer},
  {Melendez}, {Meneses-Goytia}, {Merloni}, {Merrifield}, {Meszaros}, {Meza},
  {Minchev}, {Minniti}, {Mueller}, {Muller-Sanchez}, {Muna}, {Mu{\~n}oz},
  {Myers}, {Nair}, {Nand ra}, {Ness}, {Newman}, {Nichol}, {Nidever},
  {Nitschelm}, {Noterdaeme}, {O'Connell}, {Oelkers}, {Oravetz}, {Oravetz},
  {Ort{\'\i}z}, {Osorio}, {Pace}, {Padilla}, {Palanque-Delabrouille},
  {Palicio}, {Pan}, {Pan}, {Parikh}, {P{\^a}ris}, {Park}, {Peirani},
  {Pellejero-Ibanez}, {Penny}, {Percival}, {Perez-Fournon}, {Petitjean},
  {Pieri}, {Pinsonneault}, {Pisani}, {Prada}, {Prakash}, {Queiroz}, {Raddick},
  {Raichoor}, {Barboza Rembold}, {Richstein}, {Riffel}, {Riffel}, {Rix},
  {Robin}, {Rodr{\'\i}guez Torres}, {Rom{\'a}n-Z{\'u}{\~n}iga}, {Ross},
  {Rossi}, {Ruan}, {Ruggeri}, {Ruiz}, {Salvato}, {S{\'a}nchez}, {S{\'a}nchez},
  {Sanchez Almeida}, {S{\'a}nchez-Gallego}, {Santana Rojas}, {Santiago},
  {Schiavon}, {Schimoia}, {Schlafly}, {Schlegel}, {Schneider}, {Schuster},
  {Schwope}, {Seo}, {Serenelli}, {Shen}, {Shen}, {Shetrone}, {Shull}, {Silva
  Aguirre}, {Simon}, {Skrutskie}, {Slosar}, {Smethurst}, {Smith}, {Sobeck},
  {Somers}, {Souter}, {Souto}, {Spindler}, {Stark}, {Stassun}, {Steinmetz},
  {Stello}, {Storchi-Bergmann}, {Streblyanska}, {Stringfellow}, {Su{\'a}rez},
  {Sun}, {Szigeti}, {Taghizadeh-Popp}, {Talbot}, {Tang}, {Tao}, {Tayar},
  {Tembe}, {Teske}, {Thakar}, {Thomas}, {Tissera}, {Tojeiro}, {Tremonti},
  {Troup}, {Urry}, {Valenzuela}, {van den Bosch}, {Vargas-Gonz{\'a}lez},
  {Vargas-Maga{\~n}a}, {Vazquez}, {Villanova}, {Vogt}, {Wake}, {Wang},
  {Weaver}, {Weijmans}, {Weinberg}, {Westfall}, {Whelan}, {Wilcots}, {Wild},
  {Williams}, {Wilson}, {Wood-Vasey}, {Wylezalek}, {Xiao}, {Yan}, {Yang},
  {Ybarra}, {Y{\`e}che}, {Zakamska}, {Zamora}, {Zarrouk}, {Zasowski}, {Zhang},
  {Zhao}, {Zhao}, {Zheng}, {Zheng}, {Zhou}, {Zhu}, {Zinn}, \&
  {Zou}}]{Abolfathi18}
{Abolfathi}, B., {Aguado}, D.~S., {Aguilar}, G., {et~al.} 2018,
  \href{http://dx.doi.org/10.3847/1538-4365/aa9e8a}{\color{magenta}\apjs},
  \href{https://ui.adsabs.harvard.edu/abs/2018ApJS..235...42A}{\color{blue}235},
  \href{https://ui.adsabs.harvard.edu/abs/2018ApJS..235...42A}{\color{blue}42}

\bibitem[{{Almgren} {et~al.}(2010){Almgren}, {Beckner}, {Bell}, {Day},
  {Howell}, {Joggerst}, {Lijewski}, {Nonaka}, {Singer}, \&
  {Zingale}}]{Almgren10}
{Almgren}, A.~S., {Beckner}, V.~E., {Bell}, J.~B., {et~al.} 2010,
  \href{http://dx.doi.org/10.1088/0004-637X/715/2/1221}{\color{magenta}\apj},
  \href{http://adsabs.harvard.edu/abs/2010ApJ...715.1221A}{\color{blue}715},
  \href{http://adsabs.harvard.edu/abs/2010ApJ...715.1221A}{\color{blue}1221}

\bibitem[{{Arnett}(1982)}]{Arnett82}
{Arnett}, W.~D. 1982,
  \href{http://dx.doi.org/10.1086/159681}{\color{magenta}\apj},
  \href{http://adsabs.harvard.edu/abs/1982ApJ...253..785A}{\color{blue}253},
  \href{http://adsabs.harvard.edu/abs/1982ApJ...253..785A}{\color{blue}785}

\bibitem[{{Astropy Collaboration} {et~al.}(2013){Astropy Collaboration},
  {Robitaille}, {Tollerud}, {Greenfield}, {Droettboom}, {Bray}, {Aldcroft},
  {Davis}, {Ginsburg}, {Price-Whelan}, {Kerzendorf}, {Conley}, {Crighton},
  {Barbary}, {Muna}, {Ferguson}, {Grollier}, {Parikh}, {Nair}, {Unther},
  {Deil}, {Woillez}, {Conseil}, {Kramer}, {Turner}, {Singer}, {Fox}, {Weaver},
  {Zabalza}, {Edwards}, {Azalee Bostroem}, {Burke}, {Casey}, {Crawford},
  {Dencheva}, {Ely}, {Jenness}, {Labrie}, {Lim}, {Pierfederici}, {Pontzen},
  {Ptak}, {Refsdal}, {Servillat}, \& {Streicher}}]{Astropy-Collaboration13}
{Astropy Collaboration}, {Robitaille}, T.~P., {Tollerud}, E.~J., {et~al.} 2013,
  \href{http://dx.doi.org/10.1051/0004-6361/201322068}{\color{magenta}\aap},
  \href{http://adsabs.harvard.edu/abs/2013A%26A...558A..33A}{\color{blue}558},
  \href{http://adsabs.harvard.edu/abs/2013A%26A...558A..33A}{\color{blue}A33}

\bibitem[{{Barbary} {et~al.}(2016){Barbary}, {Barclay}, {Biswas}, {Craig},
  {Feindt}, {Friesen}, {Goldstein}, {Jha}, {Rodney}, {Sofiatti}, {Thomas}, \&
  {Wood-Vasey}}]{Barbary16}
{Barbary}, K., {Barclay}, T., {Biswas}, R., {et~al.} 2016, {SNCosmo: Python
  library for supernova cosmology}

\bibitem[{{Barnsley} {et~al.}(2012){Barnsley}, {Smith}, \&
  {Steele}}]{Barnsley12}
{Barnsley}, R.~M., {Smith}, R.~J., \& {Steele}, I.~A. 2012,
  \href{http://dx.doi.org/10.1002/asna.201111634}{\color{magenta}AN},
  \href{https://ui.adsabs.harvard.edu/abs/2012AN....333..101B}{\color{blue}333},
  \href{https://ui.adsabs.harvard.edu/abs/2012AN....333..101B}{\color{blue}101}

\bibitem[{{Bellm} \& {Sesar}(2016)}]{Bellm16b}
{Bellm}, E.~C., \& {Sesar}, B. 2016, {pyraf-dbsp: Reduction pipeline for the
  Palomar Double Beam Spectrograph}

\bibitem[{{Bellm} {et~al.}(2019{\natexlab{a}}){Bellm}, {Kulkarni}, {Graham},
  {Dekany}, {Smith}, {Riddle}, {Masci}, {Helou}, {Prince}, {Adams},
  {Barbarino}, {Barlow}, {Bauer}, {Beck}, {Belicki}, {Biswas}, {Blagorodnova},
  {Bodewits}, {Bolin}, {Brinnel}, {Brooke}, {Bue}, {Bulla}, {Burruss}, {Cenko},
  {Chang}, {Connolly}, {Coughlin}, {Cromer}, {Cunningham}, {De}, {Delacroix},
  {Desai}, {Duev}, {Eadie}, {Farnham}, {Feeney}, {Feindt}, {Flynn},
  {Franckowiak}, {Frederick}, {Fremling}, {Gal-Yam}, {Gezari}, {Giomi},
  {Goldstein}, {Golkhou}, {Goobar}, {Groom}, {Hacopians}, {Hale}, {Henning},
  {Ho}, {Hover}, {Howell}, {Hung}, {Huppenkothen}, {Imel}, {Ip}, {Ivezi{\'c}},
  {Jackson}, {Jones}, {Juric}, {Kasliwal}, {Kaspi}, {Kaye}, {Kelley},
  {Kowalski}, {Kramer}, {Kupfer}, {Landry}, {Laher}, {Lee}, {Lin}, {Lin},
  {Lunnan}, {Giomi}, {Mahabal}, {Mao}, {Miller}, {Monkewitz}, {Murphy},
  {Ngeow}, {Nordin}, {Nugent}, {Ofek}, {Patterson}, {Penprase}, {Porter},
  {Rauch}, {Rebbapragada}, {Reiley}, {Rigault}, {Rodriguez}, {van Roestel},
  {Rusholme}, {van Santen}, {Schulze}, {Shupe}, {Singer}, {Soumagnac}, {Stein},
  {Surace}, {Sollerman}, {Szkody}, {Taddia}, {Terek}, {Van Sistine}, {van
  Velzen}, {Vestrand}, {Walters}, {Ward}, {Ye}, {Yu}, {Yan}, \&
  {Zolkower}}]{Bellm19}
{Bellm}, E.~C., {Kulkarni}, S.~R., {Graham}, M.~J., {et~al.}
  2019{\natexlab{a}},
  \href{http://dx.doi.org/10.1088/1538-3873/aaecbe}{\color{magenta}\pasp},
  \href{https://ui.adsabs.harvard.edu/\#abs/2019PASP..131a8002B}{\color{blue}131},
  \href{https://ui.adsabs.harvard.edu/\#abs/2019PASP..131a8002B}{\color{blue}018002}

\bibitem[{{Bellm} {et~al.}(2019{\natexlab{b}}){Bellm}, {Kulkarni}, {Barlow},
  {Feindt}, {Graham}, {Goobar}, {Kupfer}, {Ngeow}, {Nugent}, {Ofek}, {Prince},
  {Riddle}, {Walters}, \& {Ye}}]{Bellm19a}
{Bellm}, E.~C., {Kulkarni}, S.~R., {Barlow}, T., {et~al.} 2019{\natexlab{b}},
  \href{http://dx.doi.org/10.1088/1538-3873/ab0c2a}{\color{magenta}\pasp},
  \href{https://ui.adsabs.harvard.edu/abs/2019PASP..131f8003B}{\color{blue}131},
  \href{https://ui.adsabs.harvard.edu/abs/2019PASP..131f8003B}{\color{blue}068003}

\bibitem[{{Benetti} {et~al.}(2004){Benetti}, {Meikle}, {Stehle}, {Altavilla},
  {Desidera}, {Folatelli}, {Goobar}, {Mattila}, {Mendez}, {Navasardyan},
  {Pastorello}, {Patat}, {Riello}, {Ruiz-Lapuente}, {Tsvetkov}, {Turatto},
  {Mazzali}, \& {Hillebrand t}}]{Benetti04}
{Benetti}, S., {Meikle}, P., {Stehle}, M., {et~al.} 2004,
  \href{http://dx.doi.org/10.1111/j.1365-2966.2004.07357.x}{\color{magenta}\mnras},
  \href{https://ui.adsabs.harvard.edu/abs/2004MNRAS.348..261B}{\color{blue}348},
  \href{https://ui.adsabs.harvard.edu/abs/2004MNRAS.348..261B}{\color{blue}261}

\bibitem[{{Benetti} {et~al.}(2005){Benetti}, {Cappellaro}, {Mazzali},
  {Turatto}, {Altavilla}, {Bufano}, {Elias-Rosa}, {Kotak}, {Pignata}, {Salvo},
  \& {Stanishev}}]{Benetti05}
{Benetti}, S., {Cappellaro}, E., {Mazzali}, P.~A., {et~al.} 2005,
  \href{http://dx.doi.org/10.1086/428608}{\color{magenta}\apj},
  \href{https://ui.adsabs.harvard.edu/abs/2005ApJ...623.1011B}{\color{blue}623},
  \href{https://ui.adsabs.harvard.edu/abs/2005ApJ...623.1011B}{\color{blue}1011}

\bibitem[{{Bianco} {et~al.}(2011){Bianco}, {Howell}, {Sullivan}, {Conley},
  {Kasen}, {Gonz{\'a}lez-Gait{\'a}n}, {Guy}, {Astier}, {Balland}, {Carlberg},
  {Fouchez}, {Fourmanoit}, {Hardin}, {Hook}, {Lidman}, {Pain},
  {Palanque-Delabrouille}, {Perlmutter}, {Perrett}, {Pritchet}, {Regnault},
  {Rich}, \& {Ruhlmann-Kleider}}]{Bianco11}
{Bianco}, F.~B., {Howell}, D.~A., {Sullivan}, M., {et~al.} 2011,
  \href{http://dx.doi.org/10.1088/0004-637X/741/1/20}{\color{magenta}\apj},
  \href{https://ui.adsabs.harvard.edu/abs/2011ApJ...741...20B}{\color{blue}741},
  \href{https://ui.adsabs.harvard.edu/abs/2011ApJ...741...20B}{\color{blue}20}

\bibitem[{{Blagorodnova} {et~al.}(2018){Blagorodnova}, {Neill}, {Walters},
  {Kulkarni}, {Fremling}, {Ben-Ami}, {Dekany}, {Fucik}, {Konidaris}, {Nash},
  {Ngeow}, {Ofek}, {O'Sullivan}, {Quimby}, {Ritter}, \&
  {Vyhmeister}}]{Blagorodnova18}
{Blagorodnova}, N., {Neill}, J.~D., {Walters}, R., {et~al.} 2018,
  \href{http://dx.doi.org/10.1088/1538-3873/aaa53f}{\color{magenta}\pasp},
  \href{http://adsabs.harvard.edu/abs/2018PASP..130c5003B}{\color{blue}130},
  \href{http://adsabs.harvard.edu/abs/2018PASP..130c5003B}{\color{blue}035003}

\bibitem[{{Blondin} {et~al.}(2012){Blondin}, {Matheson}, {Kirshner}, {Mandel},
  {Berlind}, {Calkins}, {Challis}, {Garnavich}, {Jha}, {Modjaz}, {Riess}, \&
  {Schmidt}}]{Blondin12}
{Blondin}, S., {Matheson}, T., {Kirshner}, R.~P., {et~al.} 2012,
  \href{http://dx.doi.org/10.1088/0004-6256/143/5/126}{\color{magenta}\aj},
  \href{http://adsabs.harvard.edu/abs/2012AJ....143..126B}{\color{blue}143},
  \href{http://adsabs.harvard.edu/abs/2012AJ....143..126B}{\color{blue}126}

\bibitem[{{Bloom} {et~al.}(2012){Bloom}, {Kasen}, {Shen}, {Nugent}, {Butler},
  {Graham}, {Howell}, {Kolb}, {Holmes}, {Haswell}, {Burwitz}, {Rodriguez}, \&
  {Sullivan}}]{Bloom12a}
{Bloom}, J.~S., {Kasen}, D., {Shen}, K.~J., {et~al.} 2012,
  \href{http://dx.doi.org/10.1088/2041-8205/744/2/L17}{\color{magenta}\apjl},
  \href{http://adsabs.harvard.edu/abs/2012ApJ...744L..17B}{\color{blue}744},
  \href{http://adsabs.harvard.edu/abs/2012ApJ...744L..17B}{\color{blue}L17}

\bibitem[{{Branch} {et~al.}(2006){Branch}, {Dang}, {Hall}, {Ketchum},
  {Melakayil}, {Parrent}, {Troxel}, {Casebeer}, {Jeffery}, \&
  {Baron}}]{Branch06}
{Branch}, D., {Dang}, L.~C., {Hall}, N., {et~al.} 2006,
  \href{http://dx.doi.org/10.1086/502778}{\color{magenta}\pasp},
  \href{http://adsabs.harvard.edu/abs/2006PASP..118..560B}{\color{blue}118},
  \href{http://adsabs.harvard.edu/abs/2006PASP..118..560B}{\color{blue}560}

\bibitem[{{Breeveld} {et~al.}(2011){Breeveld}, {Landsman}, {Holland}, {Roming},
  {Kuin}, \& {Page}}]{Breeveld11}
{Breeveld}, A.~A., {Landsman}, W., {Holland}, S.~T., {et~al.} 2011,
  \href{http://dx.doi.org/10.1063/1.3621807}{\color{magenta}American Institute
  of Physics Conference Series},
  \href{http://adsabs.harvard.edu/abs/2011AIPC.1358..373B}{\color{blue}1358},
  \href{http://adsabs.harvard.edu/abs/2011AIPC.1358..373B}{\color{blue}373}

\bibitem[{{Brown} {et~al.}(2014){Brown}, {Breeveld}, {Holland}, {Kuin}, \&
  {Pritchard}}]{Brown14}
{Brown}, P.~J., {Breeveld}, A.~A., {Holland}, S., {et~al.} 2014,
  \href{http://dx.doi.org/10.1007/s10509-014-2059-8}{\color{magenta}\apss},
  \href{https://ui.adsabs.harvard.edu/abs/2014Ap&SS.354...89B}{\color{blue}354},
  \href{https://ui.adsabs.harvard.edu/abs/2014Ap&SS.354...89B}{\color{blue}89}

\bibitem[{{Brown} {et~al.}(2012){Brown}, {Dawson}, {Harris}, {Olmstead},
  {Milne}, \& {Roming}}]{Brown12}
{Brown}, P.~J., {Dawson}, K.~S., {Harris}, D.~W., {et~al.} 2012,
  \href{http://dx.doi.org/10.1088/0004-637X/749/1/18}{\color{magenta}\apj},
  \href{http://adsabs.harvard.edu/abs/2012ApJ...749...18B}{\color{blue}749},
  \href{http://adsabs.harvard.edu/abs/2012ApJ...749...18B}{\color{blue}18}

\bibitem[{{Brown} {et~al.}(2017){Brown}, {Landez}, {Milne}, \&
  {Stritzinger}}]{Brown17}
{Brown}, P.~J., {Landez}, N.~J., {Milne}, P.~A., \& {Stritzinger}, M.~D. 2017,
  \href{http://dx.doi.org/10.3847/1538-4357/aa5f5a}{\color{magenta}\apj},
  \href{https://ui.adsabs.harvard.edu/abs/2017ApJ...836..232B}{\color{blue}836},
  \href{https://ui.adsabs.harvard.edu/abs/2017ApJ...836..232B}{\color{blue}232}

\bibitem[{{Brown} {et~al.}(2019){Brown}, {Hosseinzadeh}, {Jha}, {Sand},
  {Vieira}, {Wang}, {Dai}, {Dettman}, {Mould}, {Uddin}, {Wang}, {Arcavi},
  {Bento}, {Burns}, {Diamond}, {Hiramatsu}, {Howell}, {Hsiao}, {Marion},
  {McCully}, {Milne}, {Mirzaqulov}, {Ruiter}, {Valenti}, \& {Xiang}}]{Brown19}
{Brown}, P.~J., {Hosseinzadeh}, G., {Jha}, S.~W., {et~al.} 2019,
  \href{http://dx.doi.org/10.3847/1538-4357/ab1a3f}{\color{magenta}\apj},
  \href{https://ui.adsabs.harvard.edu/abs/2019ApJ...877..152B}{\color{blue}877},
  \href{https://ui.adsabs.harvard.edu/abs/2019ApJ...877..152B}{\color{blue}152}

\bibitem[{{Bulla} {et~al.}(2020){Bulla}, {Miller}, {Yao}, {Dessart}, {Dhawan},
  {Papadogiannakis}, {Biswas}, {Goobar}, {Kulkarni}, {Nordin}, {Nugent},
  {Polin}, {Sollerman}, {Bellm}, {Coughlin}, {Dekany}, {Golkhou}, {Graham},
  {Kasliwal}, {Kupfer}, {Laher}, {Masci}, {Porter}, {Rusholme}, \&
  {Shupe}}]{Bulla20}
{Bulla}, M., {Miller}, A.~A., {Yao}, Y., {et~al.} 2020,
  \href{https://arxiv.org/abs/2001.00587}{\color{magenta}arXiv},
  \href{https://ui.adsabs.harvard.edu/abs/2020arXiv200100587B}{\color{blue}arXiv:2001.00587}

\bibitem[{{Burns} {et~al.}(2011){Burns}, {Stritzinger}, {Phillips}, {Kattner},
  {Persson}, {Madore}, {Freedman}, {Boldt}, {Campillay}, {Contreras},
  {Folatelli}, {Gonzalez}, {Krzeminski}, {Morrell}, {Salgado}, \&
  {Suntzeff}}]{Burns11}
{Burns}, C.~R., {Stritzinger}, M., {Phillips}, M.~M., {et~al.} 2011,
  \href{http://dx.doi.org/10.1088/0004-6256/141/1/19}{\color{magenta}\aj},
  \href{http://adsabs.harvard.edu/abs/2011AJ....141...19B}{\color{blue}141},
  \href{http://adsabs.harvard.edu/abs/2011AJ....141...19B}{\color{blue}19}

\bibitem[{{Burns} {et~al.}(2014){Burns}, {Stritzinger}, {Phillips}, {Hsiao},
  {Contreras}, {Persson}, {Folatelli}, {Boldt}, {Campillay}, {Castell{\'o}n},
  {Freedman}, {Madore}, {Morrell}, {Salgado}, \& {Suntzeff}}]{Burns14}
---. 2014,
  \href{http://dx.doi.org/10.1088/0004-637X/789/1/32}{\color{magenta}\apj},
  \href{https://ui.adsabs.harvard.edu/abs/2014ApJ...789...32B}{\color{blue}789},
  \href{https://ui.adsabs.harvard.edu/abs/2014ApJ...789...32B}{\color{blue}32}

\bibitem[{{Burrows} {et~al.}(2005){Burrows}, {Hill}, {Nousek}, {Kennea},
  {Wells}, {Osborne}, {Abbey}, {Beardmore}, {Mukerjee}, {Short}, {Chincarini},
  {Campana}, {Citterio}, {Moretti}, {Pagani}, {Tagliaferri}, {Giommi},
  {Capalbi}, {Tamburelli}, {Angelini}, {Cusumano}, {Br{\"a}uninger}, {Burkert},
  \& {Hartner}}]{Burrows05}
{Burrows}, D.~N., {Hill}, J.~E., {Nousek}, J.~A., {et~al.} 2005,
  \href{http://dx.doi.org/10.1007/s11214-005-5097-2}{\color{magenta}\ssr},
  \href{https://ui.adsabs.harvard.edu/abs/2005SSRv..120..165B}{\color{blue}120},
  \href{https://ui.adsabs.harvard.edu/abs/2005SSRv..120..165B}{\color{blue}165}

\bibitem[{{Cao} {et~al.}(2015){Cao}, {Kulkarni}, {Howell}, {Gal-Yam},
  {Kasliwal}, {Valenti}, {Johansson}, {Amanullah}, {Goobar}, {Sollerman},
  {Taddia}, {Horesh}, {Sagiv}, {Cenko}, {Nugent}, {Arcavi}, {Surace},
  {Wo{\'z}niak}, {Moody}, {Rebbapragada}, {Bue}, \& {Gehrels}}]{Cao15}
{Cao}, Y., {Kulkarni}, S.~R., {Howell}, D.~A., {et~al.} 2015,
  \href{http://dx.doi.org/10.1038/nature14440}{\color{magenta}\nat},
  \href{http://adsabs.harvard.edu/abs/2015Natur.521..328C}{\color{blue}521},
  \href{http://adsabs.harvard.edu/abs/2015Natur.521..328C}{\color{blue}328}

\bibitem[{{Carrick} {et~al.}(2015){Carrick}, {Turnbull}, {Lavaux}, \&
  {Hudson}}]{Carrick15}
{Carrick}, J., {Turnbull}, S.~J., {Lavaux}, G., \& {Hudson}, M.~J. 2015,
  \href{http://dx.doi.org/10.1093/mnras/stv547}{\color{magenta}\mnras},
  \href{https://ui.adsabs.harvard.edu/abs/2015MNRAS.450..317C}{\color{blue}450},
  \href{https://ui.adsabs.harvard.edu/abs/2015MNRAS.450..317C}{\color{blue}317}

\bibitem[{{Contardo} {et~al.}(2000){Contardo}, {Leibundgut}, \&
  {Vacca}}]{Contardo00}
{Contardo}, G., {Leibundgut}, B., \& {Vacca}, W.~D. 2000, \aap,
  \href{http://adsabs.harvard.edu/abs/2000A%26A...359..876C}{\color{blue}359},
  \href{http://adsabs.harvard.edu/abs/2000A%26A...359..876C}{\color{blue}876}

\bibitem[{{Cristiani} {et~al.}(1992){Cristiani}, {Cappellaro}, {Turatto},
  {Bergeron}, {Bues}, {Buson}, {Danziger}, {di Serego-Alighieri}, {Duerbeck},
  {Heydari-Malayeri}, {Krautter}, {Schmutz}, \&
  {Schulte-Ladbeck}}]{Cristiani92}
{Cristiani}, S., {Cappellaro}, E., {Turatto}, M., {et~al.} 1992, \aap,
  \href{https://ui.adsabs.harvard.edu/abs/1992A&A...259...63C}{\color{blue}259},
  \href{https://ui.adsabs.harvard.edu/abs/1992A&A...259...63C}{\color{blue}63}

\bibitem[{{De} {et~al.}(2019){De}, {Kasliwal}, {Polin}, {Nugent}, {Bildsten},
  {Adams}, {Bellm}, {Blagorodnova}, {Burdge}, {Cannella}, {Cenko}, {Dekany},
  {Feeney}, {Hale}, {Fremling}, {Graham}, {Ho}, {Jencson}, {Kulkarni}, {Laher},
  {Masci}, {Miller}, {Patterson}, {Rebbapragada}, {Riddle}, {Shupe}, \&
  {Smith}}]{De19}
{De}, K., {Kasliwal}, M.~M., {Polin}, A., {et~al.} 2019,
  \href{http://dx.doi.org/10.3847/2041-8213/ab0aec}{\color{magenta}\apjl},
  \href{http://adsabs.harvard.edu/abs/2019ApJ...873L..18D}{\color{blue}873},
  \href{http://adsabs.harvard.edu/abs/2019ApJ...873L..18D}{\color{blue}L18}

\bibitem[{{De} {et~al.}(2020){De}, {Kasliwal}, {Tzanidakis}, {Fremling},
  {Adams}, {Andreoni}, {Bagdasaryan}, {Bellm}, {Bildsten}, {Cannella}, {Cook},
  {Delacroix}, {Drake}, {Duev}, {Dugas}, {Frederick}, {Gal-Yam}, {Goldstein},
  {Golkhou}, {Graham}, {Hale}, {Hankins}, {Helou}, {Ho}, {Irani}, {Jencson},
  {Kaye}, {Kulkarni}, {Kupfer}, {Laher}, {Leadbeater}, {Lunnan}, {Masci},
  {Miller}, {Neill}, {Ofek}, {Perley}, {Polin}, {Prince}, {Quataert}, {Reiley},
  {Riddle}, {Rusholme}, {Sharma}, {Shupe}, {Sollerman}, {Tartaglia}, {Walters},
  {Yan}, \& {Yao}}]{De20}
{De}, K., {Kasliwal}, M.~M., {Tzanidakis}, A., {et~al.} 2020,
  \href{https://arxiv.org/abs/2004.09029}{\color{magenta}arXiv},
  \href{https://ui.adsabs.harvard.edu/abs/2020arXiv200409029D}{\color{blue}arXiv:2004.09029}

\bibitem[{{de Vaucouleurs} {et~al.}(1991){de Vaucouleurs}, {de Vaucouleurs},
  {Corwin}, {Buta}, {Paturel}, \& {Fouque}}]{de-Vaucouleurs91}
{de Vaucouleurs}, G., {de Vaucouleurs}, A., {Corwin}, Herold~G., J., {et~al.}
  1991, {Third Reference Catalogue of Bright Galaxies}

\bibitem[{{Dekany} {et~al.}(2020){Dekany}, {Smith}, {Riddle}, {Feeney},
  {Porter}, {Hale}, {Zolkower}, {Belicki}, {Kaye}, {Henning}, {Walters},
  {Cromer}, {Delacroix}, {Rodriguez}, {Reiley}, {Mao}, {Hover}, {Murphy},
  {Burruss}, {Baker}, {Kowalski}, {Reif}, {Mueller}, {Bellm}, {Graham}, \&
  {Kulkarni}}]{Dekany20}
{Dekany}, R., {Smith}, R.~M., {Riddle}, R., {et~al.} 2020,
  \href{http://dx.doi.org/10.1088/1538-3873/ab4ca2}{\color{magenta}\pasp},
  \href{https://ui.adsabs.harvard.edu/abs/2020PASP..132c8001D}{\color{blue}132},
  \href{https://ui.adsabs.harvard.edu/abs/2020PASP..132c8001D}{\color{blue}038001}

\bibitem[{{Dessart} {et~al.}(2014){Dessart}, {Blondin}, {Hillier}, \&
  {Khokhlov}}]{Dessart14}
{Dessart}, L., {Blondin}, S., {Hillier}, D.~J., \& {Khokhlov}, A. 2014,
  \href{http://dx.doi.org/10.1093/mnras/stu598}{\color{magenta}\mnras},
  \href{http://adsabs.harvard.edu/abs/2014MNRAS.441..532D}{\color{blue}441},
  \href{http://adsabs.harvard.edu/abs/2014MNRAS.441..532D}{\color{blue}532}

\bibitem[{{Dhawan} {et~al.}(2018){Dhawan}, {Bulla}, {Goobar}, {Lunnan},
  {Johansson}, {Fransson}, {Kulkarni}, {Papadogiannakis}, \&
  {Miller}}]{Dhawan18}
{Dhawan}, S., {Bulla}, M., {Goobar}, A., {et~al.} 2018,
  \href{http://dx.doi.org/10.1093/mnras/sty1908}{\color{magenta}\mnras},
  \href{https://ui.adsabs.harvard.edu/#abs/2018MNRAS.480.1445D}{\color{blue}480},
  \href{https://ui.adsabs.harvard.edu/#abs/2018MNRAS.480.1445D}{\color{blue}1445}

\bibitem[{{Dimitriadis} {et~al.}(2019){Dimitriadis}, {Foley}, {Rest}, {Kasen},
  {Piro}, {Polin}, {Jones}, {Villar}, {Narayan}, {Coulter}, {Kilpatrick},
  {Pan}, {Rojas-Bravo}, {Fox}, {Jha}, {Nugent}, {Riess}, {Scolnic}, {Drout},
  {K2 Mission Team}, {Barentsen}, {Dotson}, {Gully-Santiago}, {Hedges}, {Cody},
  {Barclay}, {Howell}, {KEGS}, {Garnavich}, {Tucker}, {Shaya}, {Mushotzky},
  {Olling}, {Margheim}, {Zenteno}, {Kepler spacecraft Team}, {Coughlin}, {Van
  Cleve}, {Cardoso}, {Larson}, {McCalmont-Everton}, {Peterson}, {Ross},
  {Reedy}, {Osborne}, {McGinn}, {Kohnert}, {Migliorini}, {Wheaton}, {Spencer},
  {Labonde}, {Castillo}, {Beerman}, {Steward}, {Hanley}, {Larsen},
  {Gangopadhyay}, {Kloetzel}, {Weschler}, {Nystrom}, {Moffatt}, {Redick},
  {Griest}, {Packard}, {Muszynski}, {Kampmeier}, {Bjella}, {Flynn},
  {Elsaesser}, {Pan-STARRS}, {Chambers}, {Flewelling}, {Huber}, {Magnier},
  {Waters}, {Schultz}, {Bulger}, {Lowe}, {Willman}, {Smartt}, {Smith}, {DECam},
  {Points}, {Strampelli}, {ASAS-SN}, {Brimacombe}, {Chen}, {Mu{\~n}oz},
  {Mutel}, {Shields}, {Vallely}, {Villanueva}, {PTSS/TNTS}, {Li}, {Wang},
  {Zhang}, {Lin}, {Mo}, {Zhao}, {Sai}, {Zhang}, {Zhang}, {Zhang}, {Wang},
  {Zhang}, {Baron}, {DerKacy}, {Li}, {Chen}, {Xiang}, {Rui}, {Wang}, {Huang},
  {Li}, {Cumbres Observatory}, {Hosseinzadeh}, {Howell}, {Arcavi}, {Hiramatsu},
  {Burke}, {Valenti}, {ATLAS}, {Tonry}, {Denneau}, {Heinze}, {Weiland},
  {Stalder}, {Konkoly}, {Vink{\'o}}, {S{\'a}rneczky}, {P{\'a}l}, {B{\'o}di},
  {Bogn{\'a}r}, {Cs{\'a}k}, {Cseh}, {Cs{\"o}rnyei}, {Hanyecz}, {Ign{\'a}cz},
  {Kalup}, {K{\"o}nyves-T{\'o}th}, {Kriskovics}, {Ordasi}, {Rajmon},
  {S{\'o}dor}, {Szab{\'o}}, {Szak{\'a}ts}, {Zsidi}, {ePESSTO}, {Williams},
  {Nordin}, {Cartier}, {Frohmaier}, {Galbany}, {Guti{\'e}rrez}, {Hook},
  {Inserra}, {Smith}, {Arizona}, {Sand}, {Andrews}, {Smith}, \&
  {Bilinski}}]{Dimitriadis19}
{Dimitriadis}, G., {Foley}, R.~J., {Rest}, A., {et~al.} 2019,
  \href{http://dx.doi.org/10.3847/2041-8213/aaedb0}{\color{magenta}\apjl},
  \href{https://ui.adsabs.harvard.edu/abs/2019ApJ...870L...1D}{\color{blue}870},
  \href{https://ui.adsabs.harvard.edu/abs/2019ApJ...870L...1D}{\color{blue}L1}

\bibitem[{{Evans} {et~al.}(2007){Evans}, {Beardmore}, {Page}, {Tyler},
  {Osborne}, {Goad}, {O'Brien}, {Vetere}, {Racusin}, {Morris}, {Burrows},
  {Capalbi}, {Perri}, {Gehrels}, \& {Romano}}]{Evans07}
{Evans}, P.~A., {Beardmore}, A.~P., {Page}, K.~L., {et~al.} 2007,
  \href{http://dx.doi.org/10.1051/0004-6361:20077530}{\color{magenta}\aap},
  \href{https://ui.adsabs.harvard.edu/abs/2007A&A...469..379E}{\color{blue}469},
  \href{https://ui.adsabs.harvard.edu/abs/2007A&A...469..379E}{\color{blue}379}

\bibitem[{{Evans} {et~al.}(2009){Evans}, {Beardmore}, {Page}, {Osborne},
  {O'Brien}, {Willingale}, {Starling}, {Burrows}, {Godet}, {Vetere}, {Racusin},
  {Goad}, {Wiersema}, {Angelini}, {Capalbi}, {Chincarini}, {Gehrels}, {Kennea},
  {Margutti}, {Morris}, {Mountford}, {Pagani}, {Perri}, {Romano}, \&
  {Tanvir}}]{Evans09}
---. 2009,
  \href{http://dx.doi.org/10.1111/j.1365-2966.2009.14913.x}{\color{magenta}\mnras},
  \href{https://ui.adsabs.harvard.edu/abs/2009MNRAS.397.1177E}{\color{blue}397},
  \href{https://ui.adsabs.harvard.edu/abs/2009MNRAS.397.1177E}{\color{blue}1177}

\bibitem[{{Fabricant} {et~al.}(2019){Fabricant}, {Fata}, {Epps}, {Gauron},
  {Mueller}, {Zajac}, {Amato}, {Barberis}, {Bergner}, {Brennan}, {Brown},
  {Chilingarian}, {Geary}, {Kradinov}, {McLeod}, {Smith}, \&
  {Woods}}]{Fabricant19}
{Fabricant}, D., {Fata}, R., {Epps}, H., {et~al.} 2019,
  \href{http://dx.doi.org/10.1088/1538-3873/ab1d78}{\color{magenta}\pasp},
  \href{https://ui.adsabs.harvard.edu/abs/2019PASP..131g5004F}{\color{blue}131},
  \href{https://ui.adsabs.harvard.edu/abs/2019PASP..131g5004F}{\color{blue}075004}

\bibitem[{{Filippenko} {et~al.}(1992){Filippenko}, {Richmond}, {Branch},
  {Gaskell}, {Herbst}, {Ford}, {Treffers}, {Matheson}, {Ho}, {Dey}, {Sargent},
  {Small}, \& {van Breugel}}]{Filippenko92}
{Filippenko}, A.~V., {Richmond}, M.~W., {Branch}, D., {et~al.} 1992,
  \href{http://dx.doi.org/10.1086/116339}{\color{magenta}\aj},
  \href{https://ui.adsabs.harvard.edu/abs/1992AJ....104.1543F}{\color{blue}104},
  \href{https://ui.adsabs.harvard.edu/abs/1992AJ....104.1543F}{\color{blue}1543}

\bibitem[{{Fink} {et~al.}(2007){Fink}, {Hillebrandt}, \& {R{\"o}pke}}]{Fink07}
{Fink}, M., {Hillebrandt}, W., \& {R{\"o}pke}, F.~K. 2007,
  \href{http://dx.doi.org/10.1051/0004-6361:20078438}{\color{magenta}\aap},
  \href{https://ui.adsabs.harvard.edu/abs/2007A&A...476.1133F}{\color{blue}476},
  \href{https://ui.adsabs.harvard.edu/abs/2007A&A...476.1133F}{\color{blue}1133}

\bibitem[{{Fink} {et~al.}(2010){Fink}, {R{\"o}pke}, {Hillebrandt},
  {Seitenzahl}, {Sim}, \& {Kromer}}]{Fink10}
{Fink}, M., {R{\"o}pke}, F.~K., {Hillebrandt}, W., {et~al.} 2010,
  \href{http://dx.doi.org/10.1051/0004-6361/200913892}{\color{magenta}\aap},
  \href{https://ui.adsabs.harvard.edu/abs/2010A&A...514A..53F}{\color{blue}514},
  \href{https://ui.adsabs.harvard.edu/abs/2010A&A...514A..53F}{\color{blue}A53}

\bibitem[{{Fink} {et~al.}(2014){Fink}, {Kromer}, {Seitenzahl},
  {Ciaraldi-Schoolmann}, {R{\"o}pke}, {Sim}, {Pakmor}, {Ruiter}, \&
  {Hillebrandt}}]{Fink14}
{Fink}, M., {Kromer}, M., {Seitenzahl}, I.~R., {et~al.} 2014,
  \href{http://dx.doi.org/10.1093/mnras/stt2315}{\color{magenta}\mnras},
  \href{http://adsabs.harvard.edu/abs/2014MNRAS.438.1762F}{\color{blue}438},
  \href{http://adsabs.harvard.edu/abs/2014MNRAS.438.1762F}{\color{blue}1762}

\bibitem[{{Folatelli} {et~al.}(2010){Folatelli}, {Phillips}, {Burns},
  {Contreras}, {Hamuy}, {Freedman}, {Persson}, {Stritzinger}, {Suntzeff},
  {Krisciunas}, {Boldt}, {Gonz{\'a}lez}, {Krzeminski}, {Morrell}, {Roth},
  {Salgado}, {Madore}, {Murphy}, {Wyatt}, {Li}, {Filippenko}, \&
  {Miller}}]{Folatelli10}
{Folatelli}, G., {Phillips}, M.~M., {Burns}, C.~R., {et~al.} 2010,
  \href{http://dx.doi.org/10.1088/0004-6256/139/1/120}{\color{magenta}\aj},
  \href{http://adsabs.harvard.edu/abs/2010AJ....139..120F}{\color{blue}139},
  \href{http://adsabs.harvard.edu/abs/2010AJ....139..120F}{\color{blue}120}

\bibitem[{{Foley} {et~al.}(2010){Foley}, {Narayan}, {Challis}, {Filippenko},
  {Kirshner}, {Silverman}, \& {Steele}}]{Foley10a}
{Foley}, R.~J., {Narayan}, G., {Challis}, P.~J., {et~al.} 2010,
  \href{http://dx.doi.org/10.1088/0004-637X/708/2/1748}{\color{magenta}\apj},
  \href{https://ui.adsabs.harvard.edu/abs/2010ApJ...708.1748F}{\color{blue}708},
  \href{https://ui.adsabs.harvard.edu/abs/2010ApJ...708.1748F}{\color{blue}1748}

\bibitem[{Foreman-Mackey(2016)}]{Foreman-Mackey16}
Foreman-Mackey, D. 2016,
  \href{http://dx.doi.org/10.21105/joss.00024}{\color{magenta}JOSS}, 24

\bibitem[{{Foreman-Mackey} {et~al.}(2013){Foreman-Mackey}, {Hogg}, {Lang}, \&
  {Goodman}}]{Foreman-Mackey13}
{Foreman-Mackey}, D., {Hogg}, D.~W., {Lang}, D., \& {Goodman}, J. 2013,
  \href{http://dx.doi.org/10.1086/670067}{\color{magenta}\pasp},
  \href{http://adsabs.harvard.edu/abs/2013PASP..125..306F}{\color{blue}125},
  \href{http://adsabs.harvard.edu/abs/2013PASP..125..306F}{\color{blue}306}

\bibitem[{{Fremling} {et~al.}(2020){Fremling}, {Miller}, {Sharma}, {Dugas},
  {Perley}, {Taggart}, {Sollerman}, {Goobar}, {Graham}, {Neill}, {Nordin},
  {Rigault}, {Walters}, {Andreoni}, {Bagdasaryan}, {Belicki}, {Cannella},
  {Bellm}, {Cenko}, {De}, {Dekany}, {Frederick}, {Golkhou}, {Graham}, {Helou},
  {Ho}, {Kasliwal}, {Kupfer}, {Laher}, {Mahabal}, {Masci}, {Riddle},
  {Rusholme}, {Schulze}, {Shupe}, {Smith}, {Velzen}, {Yan}, {Yao}, {Zhuang}, \&
  {Kulkarni}}]{Fremling20}
{Fremling}, C., {Miller}, A.~A., {Sharma}, Y., {et~al.} 2020,
  \href{http://dx.doi.org/10.3847/1538-4357/ab8943}{\color{magenta}\apj},
  \href{https://ui.adsabs.harvard.edu/abs/2020ApJ...895...32F}{\color{blue}895},
  \href{https://ui.adsabs.harvard.edu/abs/2020ApJ...895...32F}{\color{blue}32}

\bibitem[{{Ganeshalingam} {et~al.}(2011){Ganeshalingam}, {Li}, \&
  {Filippenko}}]{Ganeshalingam11}
{Ganeshalingam}, M., {Li}, W., \& {Filippenko}, A.~V. 2011,
  \href{http://dx.doi.org/10.1111/j.1365-2966.2011.19213.x}{\color{magenta}\mnras},
  \href{http://adsabs.harvard.edu/abs/2011MNRAS.416.2607G}{\color{blue}416},
  \href{http://adsabs.harvard.edu/abs/2011MNRAS.416.2607G}{\color{blue}2607}

\bibitem[{{Ganeshalingam} {et~al.}(2012){Ganeshalingam}, {Li}, {Filippenko},
  {Silverman}, {Chornock}, {Foley}, {Matheson}, {Kirshner}, {Milne}, {Calkins},
  \& {Shen}}]{Ganeshalingam12}
{Ganeshalingam}, M., {Li}, W., {Filippenko}, A.~V., {et~al.} 2012,
  \href{http://dx.doi.org/10.1088/0004-637X/751/2/142}{\color{magenta}\apj},
  \href{http://adsabs.harvard.edu/abs/2012ApJ...751..142G}{\color{blue}751},
  \href{http://adsabs.harvard.edu/abs/2012ApJ...751..142G}{\color{blue}142}

\bibitem[{{Gehrels} {et~al.}(2004){Gehrels}, {Chincarini}, {Giommi}, {Mason},
  {Nousek}, {Wells}, {White}, {Barthelmy}, {Burrows}, {Cominsky}, {Hurley},
  {Marshall}, {M{\'e}sz{\'a}ros}, {Roming}, {Angelini}, {Barbier}, {Belloni},
  {Campana}, {Caraveo}, {Chester}, {Citterio}, {Cline}, {Cropper}, {Cummings},
  {Dean}, {Feigelson}, {Fenimore}, {Frail}, {Fruchter}, {Garmire}, {Gendreau},
  {Ghisellini}, {Greiner}, {Hill}, {Hunsberger}, {Krimm}, {Kulkarni}, {Kumar},
  {Lebrun}, {Lloyd-Ronning}, {Markwardt}, {Mattson}, {Mushotzky}, {Norris},
  {Osborne}, {Paczynski}, {Palmer}, {Park}, {Parsons}, {Paul}, {Rees},
  {Reynolds}, {Rhoads}, {Sasseen}, {Schaefer}, {Short}, {Smale}, {Smith},
  {Stella}, {Tagliaferri}, {Takahashi}, {Tashiro}, {Townsley}, {Tueller},
  {Turner}, {Vietri}, {Voges}, {Ward}, {Willingale}, {Zerbi}, \&
  {Zhang}}]{Gehrels04}
{Gehrels}, N., {Chincarini}, G., {Giommi}, P., {et~al.} 2004,
  \href{http://dx.doi.org/10.1086/422091}{\color{magenta}\apj},
  \href{http://adsabs.harvard.edu/abs/2004ApJ...611.1005G}{\color{blue}611},
  \href{http://adsabs.harvard.edu/abs/2004ApJ...611.1005G}{\color{blue}1005}

\bibitem[{{Goobar} {et~al.}(2015){Goobar}, {Kromer}, {Siverd}, {Stassun},
  {Pepper}, {Amanullah}, {Kasliwal}, {Sollerman}, \& {Taddia}}]{Goobar15}
{Goobar}, A., {Kromer}, M., {Siverd}, R., {et~al.} 2015,
  \href{http://dx.doi.org/10.1088/0004-637X/799/1/106}{\color{magenta}\apj},
  \href{http://adsabs.harvard.edu/abs/2015ApJ...799..106G}{\color{blue}799},
  \href{http://adsabs.harvard.edu/abs/2015ApJ...799..106G}{\color{blue}106}

\bibitem[{{Goodman} \& {Weare}(2010)}]{Goodman10}
{Goodman}, J., \& {Weare}, J. 2010,
  \href{http://dx.doi.org/10.2140/camcos.2010.5.65}{\color{magenta}CAMCS},
  \href{https://ui.adsabs.harvard.edu/abs/2010CAMCS...5...65G}{\color{blue}5},
  \href{https://ui.adsabs.harvard.edu/abs/2010CAMCS...5...65G}{\color{blue}65}

\bibitem[{{Graham} {et~al.}(2019){Graham}, {Kulkarni}, {Bellm}, {Adams},
  {Barbarino}, {Blagorodnova}, {Bodewits}, {Bolin}, {Brady}, {Cenko}, {Chang},
  {Coughlin}, {De}, {Eadie}, {Farnham}, {Feindt}, {Franckowiak}, {Fremling},
  {Gezari}, {Ghosh}, {Goldstein}, {Golkhou}, {Goobar}, {Ho}, {Huppenkothen},
  {Ivezi{\'c}}, {Jones}, {Juric}, {Kaplan}, {Kasliwal}, {Kelley}, {Kupfer},
  {Lee}, {Lin}, {Lunnan}, {Mahabal}, {Miller}, {Ngeow}, {Nugent}, {Ofek},
  {Prince}, {Rauch}, {van Roestel}, {Schulze}, {Singer}, {Sollerman}, {Taddia},
  {Yan}, {Ye}, {Yu}, {Barlow}, {Bauer}, {Beck}, {Belicki}, {Biswas}, {Brinnel},
  {Brooke}, {Bue}, {Bulla}, {Burruss}, {Connolly}, {Cromer}, {Cunningham},
  {Dekany}, {Delacroix}, {Desai}, {Duev}, {Feeney}, {Flynn}, {Frederick},
  {Gal-Yam}, {Giomi}, {Groom}, {Hacopians}, {Hale}, {Helou}, {Henning},
  {Hover}, {Hillenbrand}, {Howell}, {Hung}, {Imel}, {Ip}, {Jackson}, {Kaspi},
  {Kaye}, {Kowalski}, {Kramer}, {Kuhn}, {Landry}, {Laher}, {Mao}, {Masci},
  {Monkewitz}, {Murphy}, {Nordin}, {Patterson}, {Penprase}, {Porter},
  {Rebbapragada}, {Reiley}, {Riddle}, {Rigault}, {Rodriguez}, {Rusholme}, {van
  Santen}, {Shupe}, {Smith}, {Soumagnac}, {Stein}, {Surace}, {Szkody}, {Terek},
  {Van Sistine}, {van Velzen}, {Vestrand}, {Walters}, {Ward}, {Zhang}, \&
  {Zolkower}}]{Graham19}
{Graham}, M.~J., {Kulkarni}, S.~R., {Bellm}, E.~C., {et~al.} 2019,
  \href{http://dx.doi.org/10.1088/1538-3873/ab006c}{\color{magenta}\pasp},
  \href{https://ui.adsabs.harvard.edu/abs/2019PASP..131g8001G}{\color{blue}131},
  \href{https://ui.adsabs.harvard.edu/abs/2019PASP..131g8001G}{\color{blue}078001}

\bibitem[{{Gronow} {et~al.}(2020){Gronow}, {Collins}, {Ohlmann}, {Pakmor},
  {Kromer}, {Seitenzahl}, {Sim}, \& {R{\"o}pke}}]{Gronow20}
{Gronow}, S., {Collins}, C., {Ohlmann}, S.~T., {et~al.} 2020,
  \href{http://dx.doi.org/10.1051/0004-6361/201936494}{\color{magenta}\aap},
  \href{https://ui.adsabs.harvard.edu/abs/2020A&A...635A.169G}{\color{blue}635},
  \href{https://ui.adsabs.harvard.edu/abs/2020A&A...635A.169G}{\color{blue}A169}

\bibitem[{{Guy} {et~al.}(2007){Guy}, {Astier}, {Baumont}, {Hardin}, {Pain},
  {Regnault}, {Basa}, {Carlberg}, {Conley}, {Fabbro}, {Fouchez}, {Hook},
  {Howell}, {Perrett}, {Pritchet}, {Rich}, {Sullivan}, {Antilogus}, {Aubourg},
  {Bazin}, {Bronder}, {Filiol}, {Palanque-Delabrouille}, {Ripoche}, \&
  {Ruhlmann-Kleider}}]{Guy07}
{Guy}, J., {Astier}, P., {Baumont}, S., {et~al.} 2007,
  \href{http://dx.doi.org/10.1051/0004-6361:20066930}{\color{magenta}\aap},
  \href{https://ui.adsabs.harvard.edu/abs/2007A&A...466...11G}{\color{blue}466},
  \href{https://ui.adsabs.harvard.edu/abs/2007A&A...466...11G}{\color{blue}11}

\bibitem[{{Hachinger} {et~al.}(2008){Hachinger}, {Mazzali}, {Tanaka},
  {Hillebrandt}, \& {Benetti}}]{Hachinger08}
{Hachinger}, S., {Mazzali}, P.~A., {Tanaka}, M., {et~al.} 2008,
  \href{http://dx.doi.org/10.1111/j.1365-2966.2008.13645.x}{\color{magenta}\mnras},
  \href{https://ui.adsabs.harvard.edu/abs/2008MNRAS.389.1087H}{\color{blue}389},
  \href{https://ui.adsabs.harvard.edu/abs/2008MNRAS.389.1087H}{\color{blue}1087}

\bibitem[{{Hachinger} {et~al.}(2013){Hachinger}, {Mazzali}, {Sullivan},
  {Ellis}, {Maguire}, {Gal-Yam}, {Howell}, {Nugent}, {Baron}, {Cooke},
  {Arcavi}, {Bersier}, {Dilday}, {James}, {Kasliwal}, {Kulkarni}, {Ofek},
  {Laher}, {Parrent}, {Surace}, {Yaron}, \& {Walker}}]{Hachinger13}
{Hachinger}, S., {Mazzali}, P.~A., {Sullivan}, M., {et~al.} 2013,
  \href{http://dx.doi.org/10.1093/mnras/sts492}{\color{magenta}\mnras},
  \href{https://ui.adsabs.harvard.edu/abs/2013MNRAS.429.2228H}{\color{blue}429},
  \href{https://ui.adsabs.harvard.edu/abs/2013MNRAS.429.2228H}{\color{blue}2228}

\bibitem[{{Hayden} {et~al.}(2010){Hayden}, {Garnavich}, {Kessler}, {Frieman},
  {Jha}, {Bassett}, {Cinabro}, {Dilday}, {Kasen}, {Marriner}, {Nichol},
  {Riess}, {Sako}, {Schneider}, {Smith}, \& {Sollerman}}]{Hayden10}
{Hayden}, B.~T., {Garnavich}, P.~M., {Kessler}, R., {et~al.} 2010,
  \href{http://dx.doi.org/10.1088/0004-637X/712/1/350}{\color{magenta}\apj},
  \href{http://adsabs.harvard.edu/abs/2010ApJ...712..350H}{\color{blue}712},
  \href{http://adsabs.harvard.edu/abs/2010ApJ...712..350H}{\color{blue}350}

\bibitem[{{HI4PI Collaboration} {et~al.}(2016){HI4PI Collaboration}, {Ben
  Bekhti}, {Fl{\"o}er}, {Keller}, {Kerp}, {Lenz}, {Winkel}, {Bailin},
  {Calabretta}, {Dedes}, {Ford}, {Gibson}, {Haud}, {Janowiecki}, {Kalberla},
  {Lockman}, {McClure-Griffiths}, {Murphy}, {Nakanishi}, {Pisano}, \&
  {Staveley-Smith}}]{HI4PI2016a}
{HI4PI Collaboration}, {Ben Bekhti}, N., {Fl{\"o}er}, L., {et~al.} 2016,
  \href{http://dx.doi.org/10.1051/0004-6361/201629178}{\color{magenta}\aap},
  \href{https://ui.adsabs.harvard.edu/abs/2016A&A...594A.116H}{\color{blue}594},
  \href{https://ui.adsabs.harvard.edu/abs/2016A&A...594A.116H}{\color{blue}A116}

\bibitem[{{Hillebrandt} {et~al.}(2013){Hillebrandt}, {Kromer}, {R{\"o}pke}, \&
  {Ruiter}}]{Hillebrandt13}
{Hillebrandt}, W., {Kromer}, M., {R{\"o}pke}, F.~K., \& {Ruiter}, A.~J. 2013,
  \href{http://dx.doi.org/10.1007/s11467-013-0303-2}{\color{magenta}Frontiers
  of Physics},
  \href{http://adsabs.harvard.edu/abs/2013FrPhy...8..116H}{\color{blue}8},
  \href{http://adsabs.harvard.edu/abs/2013FrPhy...8..116H}{\color{blue}116}

\bibitem[{{Holoien} {et~al.}(2017){Holoien}, {Stanek}, {Kochanek}, {Shappee},
  {Prieto}, {Brimacombe}, {Bersier}, {Bishop}, {Dong}, {Brown}, {Danilet},
  {Simonian}, {Basu}, {Beacom}, {Falco}, {Pojmanski}, {Skowron}, {Wo{\'z}niak},
  {{\'A}vila}, {Conseil}, {Contreras}, {Cruz}, {Fern{\'a}ndez}, {Koff}, {Guo},
  {Herczeg}, {Hissong}, {Hsiao}, {Jose}, {Kiyota}, {Long}, {Monard},
  {Nicholls}, {Nicolas}, \& {Wiethoff}}]{Holoien17}
{Holoien}, T.~W.~S., {Stanek}, K.~Z., {Kochanek}, C.~S., {et~al.} 2017,
  \href{http://dx.doi.org/10.1093/mnras/stw2273}{\color{magenta}\mnras},
  \href{https://ui.adsabs.harvard.edu/abs/2017MNRAS.464.2672H}{\color{blue}464},
  \href{https://ui.adsabs.harvard.edu/abs/2017MNRAS.464.2672H}{\color{blue}2672}

\bibitem[{{Hosseinzadeh} {et~al.}(2017){Hosseinzadeh}, {Sand}, {Valenti},
  {Brown}, {Howell}, {McCully}, {Kasen}, {Arcavi}, {Azalee Bostroem},
  {Tartaglia}, {Hsiao}, {Davis}, {Shahbandeh}, \&
  {Stritzinger}}]{Hosseinzadeh17}
{Hosseinzadeh}, G., {Sand}, D.~J., {Valenti}, S., {et~al.} 2017,
  \href{http://dx.doi.org/10.3847/2041-8213/aa8402}{\color{magenta}\apjl},
  \href{http://adsabs.harvard.edu/abs/2017ApJ...845L..11H}{\color{blue}845},
  \href{http://adsabs.harvard.edu/abs/2017ApJ...845L..11H}{\color{blue}L11}

\bibitem[{{Howell} {et~al.}(2009){Howell}, {Sullivan}, {Brown}, {Conley}, {Le
  Borgne}, {Hsiao}, {Astier}, {Balam}, {Balland}, {Basa}, {Carlberg},
  {Fouchez}, {Guy}, {Hardin}, {Hook}, {Pain}, {Perrett}, {Pritchet},
  {Regnault}, {Baumont}, {LeDu}, {Lidman}, {Perlmutter}, {Suzuki}, {Walker}, \&
  {Wheeler}}]{Howell09}
{Howell}, D.~A., {Sullivan}, M., {Brown}, E.~F., {et~al.} 2009,
  \href{http://dx.doi.org/10.1088/0004-637X/691/1/661}{\color{magenta}\apj},
  \href{https://ui.adsabs.harvard.edu/abs/2009ApJ...691..661H}{\color{blue}691},
  \href{https://ui.adsabs.harvard.edu/abs/2009ApJ...691..661H}{\color{blue}661}

\bibitem[{Hunter(2007)}]{Hunter07}
Hunter, J.~D. 2007,
  \href{http://dx.doi.org/10.1109/MCSE.2007.55}{\color{magenta}Computing In
  Science \& Engineering}, 9, 90

\bibitem[{{Itagaki}(2019)}]{Itagaki19}
{Itagaki}, K. 2019, Transient Name Server Discovery Report,
  \href{https://ui.adsabs.harvard.edu/abs/2019TNSTR2720....1I}{\color{blue}2019-2720},
  \href{https://ui.adsabs.harvard.edu/abs/2019TNSTR2720....1I}{\color{blue}1}

\bibitem[{{Jeffery}(1999)}]{Jeffery99}
{Jeffery}, D.~J. 1999,
  \href{https://arxiv.org/abs/astro-ph/9907015}{\color{magenta}arXiv},
  \href{https://ui.adsabs.harvard.edu/abs/1999astro.ph..7015J}{\color{blue}astro}

\bibitem[{{Jiang} {et~al.}(2018){Jiang}, {Doi}, {Maeda}, \&
  {Shigeyama}}]{Jiang18}
{Jiang}, J.-a., {Doi}, M., {Maeda}, K., \& {Shigeyama}, T. 2018,
  \href{http://dx.doi.org/10.3847/1538-4357/aadb9a}{\color{magenta}\apj},
  \href{https://ui.adsabs.harvard.edu/abs/2018ApJ...865..149J}{\color{blue}865},
  \href{https://ui.adsabs.harvard.edu/abs/2018ApJ...865..149J}{\color{blue}149}

\bibitem[{{Jiang} {et~al.}(2017){Jiang}, {Doi}, {Maeda}, {Shigeyama}, {Nomoto},
  {Yasuda}, {Jha}, {Tanaka}, {Morokuma}, {Tominaga}, {Ivezi{\'c}},
  {Ruiz-Lapuente}, {Stritzinger}, {Mazzali}, {Ashall}, {Mould}, {Baade},
  {Suzuki}, {Connolly}, {Patat}, {Wang}, {Yoachim}, {Jones}, {Furusawa}, \&
  {Miyazaki}}]{Jiang17}
{Jiang}, J.-A., {Doi}, M., {Maeda}, K., {et~al.} 2017,
  \href{http://dx.doi.org/10.1038/nature23908}{\color{magenta}\nat},
  \href{http://adsabs.harvard.edu/abs/2017Natur.550...80J}{\color{blue}550},
  \href{http://adsabs.harvard.edu/abs/2017Natur.550...80J}{\color{blue}80}

\bibitem[{{Kasen}(2010)}]{Kasen10a}
{Kasen}, D. 2010,
  \href{http://dx.doi.org/10.1088/0004-637X/708/2/1025}{\color{magenta}\apj},
  \href{http://adsabs.harvard.edu/abs/2010ApJ...708.1025K}{\color{blue}708},
  \href{http://adsabs.harvard.edu/abs/2010ApJ...708.1025K}{\color{blue}1025}

\bibitem[{{Kasen} {et~al.}(2006){Kasen}, {Thomas}, \& {Nugent}}]{Kasen06a}
{Kasen}, D., {Thomas}, R.~C., \& {Nugent}, P. 2006,
  \href{http://dx.doi.org/10.1086/506190}{\color{magenta}\apj},
  \href{https://ui.adsabs.harvard.edu/abs/2006ApJ...651..366K}{\color{blue}651},
  \href{https://ui.adsabs.harvard.edu/abs/2006ApJ...651..366K}{\color{blue}366}

\bibitem[{{Kawabata}(2020)}]{Kawabata20}
{Kawabata}, M. 2020, Transient Name Server Classification Report,
  \href{https://ui.adsabs.harvard.edu/abs/2020TNSCR..24....1K}{\color{blue}2020-24},
  \href{https://ui.adsabs.harvard.edu/abs/2020TNSCR..24....1K}{\color{blue}1}

\bibitem[{{Kerzendorf} \& {Sim}(2014)}]{Kerzendorf14}
{Kerzendorf}, W.~E., \& {Sim}, S.~A. 2014,
  \href{http://dx.doi.org/10.1093/mnras/stu055}{\color{magenta}\mnras},
  \href{https://ui.adsabs.harvard.edu/abs/2014MNRAS.440..387K}{\color{blue}440},
  \href{https://ui.adsabs.harvard.edu/abs/2014MNRAS.440..387K}{\color{blue}387}

\bibitem[{{Khokhlov}(1991)}]{Khokhlov91}
{Khokhlov}, A.~M. 1991, \aap,
  \href{https://ui.adsabs.harvard.edu/abs/1991A&A...245..114K}{\color{blue}245},
  \href{https://ui.adsabs.harvard.edu/abs/1991A&A...245..114K}{\color{blue}114}

\bibitem[{{Krisciunas} {et~al.}(2017){Krisciunas}, {Contreras}, {Burns},
  {Phillips}, {Stritzinger}, {Morrell}, {Hamuy}, {Anais}, {Boldt}, {Busta},
  {Campillay}, {Castell{\'o}n}, {Folatelli}, {Freedman}, {Gonz{\'a}lez},
  {Hsiao}, {Krzeminski}, {Persson}, {Roth}, {Salgado}, {Ser{\'o}n}, {Suntzeff},
  {Torres}, {Filippenko}, {Li}, {Madore}, {DePoy}, {Marshall}, {Rheault}, \&
  {Villanueva}}]{Krisciunas17}
{Krisciunas}, K., {Contreras}, C., {Burns}, C.~R., {et~al.} 2017,
  \href{http://dx.doi.org/10.3847/1538-3881/aa8df0}{\color{magenta}\aj},
  \href{https://ui.adsabs.harvard.edu/abs/2017AJ....154..211K}{\color{blue}154},
  \href{https://ui.adsabs.harvard.edu/abs/2017AJ....154..211K}{\color{blue}211}

\bibitem[{{Kromer} {et~al.}(2017){Kromer}, {Ohlmann}, \&
  {R{\"o}pke}}]{Kromer17}
{Kromer}, M., {Ohlmann}, S., \& {R{\"o}pke}, F.~K. 2017, \memsai,
  \href{https://ui.adsabs.harvard.edu/abs/2017MmSAI..88..312K}{\color{blue}88},
  \href{https://ui.adsabs.harvard.edu/abs/2017MmSAI..88..312K}{\color{blue}312}

\bibitem[{{Kromer} {et~al.}(2010){Kromer}, {Sim}, {Fink}, {R{\"o}pke},
  {Seitenzahl}, \& {Hillebrandt}}]{Kromer10}
{Kromer}, M., {Sim}, S.~A., {Fink}, M., {et~al.} 2010,
  \href{http://dx.doi.org/10.1088/0004-637X/719/2/1067}{\color{magenta}\apj},
  \href{https://ui.adsabs.harvard.edu/abs/2010ApJ...719.1067K}{\color{blue}719},
  \href{https://ui.adsabs.harvard.edu/abs/2010ApJ...719.1067K}{\color{blue}1067}

\bibitem[{{Kromer} {et~al.}(2013){Kromer}, {Pakmor}, {Taubenberger}, {Pignata},
  {Fink}, {R{\"o}pke}, {Seitenzahl}, {Sim}, \& {Hillebrandt}}]{Kromer13a}
{Kromer}, M., {Pakmor}, R., {Taubenberger}, S., {et~al.} 2013,
  \href{http://dx.doi.org/10.1088/2041-8205/778/1/L18}{\color{magenta}\apjl},
  \href{https://ui.adsabs.harvard.edu/abs/2013ApJ...778L..18K}{\color{blue}778},
  \href{https://ui.adsabs.harvard.edu/abs/2013ApJ...778L..18K}{\color{blue}L18}

\bibitem[{{Kromer} {et~al.}(2016){Kromer}, {Fremling}, {Pakmor},
  {Taubenberger}, {Amanullah}, {Cenko}, {Fransson}, {Goobar}, {Leloudas},
  {Taddia}, {R{\"o}pke}, {Seitenzahl}, {Sim}, \& {Sollerman}}]{Kromer16}
{Kromer}, M., {Fremling}, C., {Pakmor}, R., {et~al.} 2016,
  \href{http://dx.doi.org/10.1093/mnras/stw962}{\color{magenta}\mnras},
  \href{http://adsabs.harvard.edu/abs/2016MNRAS.459.4428K}{\color{blue}459},
  \href{http://adsabs.harvard.edu/abs/2016MNRAS.459.4428K}{\color{blue}4428}

\bibitem[{{Leibundgut} {et~al.}(1993){Leibundgut}, {Kirshner}, {Phillips},
  {Wells}, {Suntzeff}, {Hamuy}, {Schommer}, {Walker}, {Gonzalez}, {Ugarte},
  {Williams}, {Williger}, {Gomez}, {Marzke}, {Schmidt}, {Whitney}, {Caldwell},
  {Peters}, {Chaffee}, {Foltz}, {Rehner}, {Siciliano}, {Barnes}, {Cheng},
  {Hintzen}, {Kim}, {Maza}, {Parker}, {Porter}, {Schmidtke}, \&
  {Sonneborn}}]{Leibundgut93}
{Leibundgut}, B., {Kirshner}, R.~P., {Phillips}, M.~M., {et~al.} 1993,
  \href{http://dx.doi.org/10.1086/116427}{\color{magenta}\aj},
  \href{https://ui.adsabs.harvard.edu/abs/1993AJ....105..301L}{\color{blue}105},
  \href{https://ui.adsabs.harvard.edu/abs/1993AJ....105..301L}{\color{blue}301}

\bibitem[{{Levanon} \& {Soker}(2017)}]{Levanon17}
{Levanon}, N., \& {Soker}, N. 2017,
  \href{http://dx.doi.org/10.1093/mnras/stx1387}{\color{magenta}\mnras},
  \href{https://ui.adsabs.harvard.edu/abs/2017MNRAS.470.2510L}{\color{blue}470},
  \href{https://ui.adsabs.harvard.edu/abs/2017MNRAS.470.2510L}{\color{blue}2510}

\bibitem[{{Levanon} \& {Soker}(2019)}]{Levanon19}
---. 2019,
  \href{http://dx.doi.org/10.3847/2041-8213/ab0285}{\color{magenta}\apjl},
  \href{https://ui.adsabs.harvard.edu/abs/2019ApJ...872L...7L}{\color{blue}872},
  \href{https://ui.adsabs.harvard.edu/abs/2019ApJ...872L...7L}{\color{blue}L7}

\bibitem[{{Maeda} {et~al.}(2018){Maeda}, {Jiang}, {Shigeyama}, \&
  {Doi}}]{Maeda18}
{Maeda}, K., {Jiang}, J.-a., {Shigeyama}, T., \& {Doi}, M. 2018,
  \href{http://dx.doi.org/10.3847/1538-4357/aac8d8}{\color{magenta}\apj},
  \href{https://ui.adsabs.harvard.edu/abs/2018ApJ...861...78M}{\color{blue}861},
  \href{https://ui.adsabs.harvard.edu/abs/2018ApJ...861...78M}{\color{blue}78}

\bibitem[{{Magee} \& {Maguire}(2020)}]{Magee20a}
{Magee}, M.~R., \& {Maguire}, K. 2020,
  \href{https://arxiv.org/abs/2007.02101}{\color{magenta}arXiv},
  arXiv:2007.02101

\bibitem[{{Magee} {et~al.}(2020){Magee}, {Maguire}, {Kotak}, {Sim},
  {Gillanders}, {Prentice}, \& {Skillen}}]{Magee20}
{Magee}, M.~R., {Maguire}, K., {Kotak}, R., {et~al.} 2020,
  \href{http://dx.doi.org/10.1051/0004-6361/201936684}{\color{magenta}\aap},
  \href{https://ui.adsabs.harvard.edu/abs/2020A&A...634A..37M}{\color{blue}634},
  \href{https://ui.adsabs.harvard.edu/abs/2020A&A...634A..37M}{\color{blue}A37}

\bibitem[{{Magee} {et~al.}(2018){Magee}, {Sim}, {Kotak}, \&
  {Kerzendorf}}]{Magee18}
{Magee}, M.~R., {Sim}, S.~A., {Kotak}, R., \& {Kerzendorf}, W.~E. 2018,
  \href{http://dx.doi.org/10.1051/0004-6361/201832675}{\color{magenta}\aap},
  \href{https://ui.adsabs.harvard.edu/abs/2018A&A...614A.115M}{\color{blue}614},
  \href{https://ui.adsabs.harvard.edu/abs/2018A&A...614A.115M}{\color{blue}A115}

\bibitem[{{Maguire} {et~al.}(2011){Maguire}, {Sullivan}, {Thomas}, {Nugent},
  {Howell}, {Gal-Yam}, {Arcavi}, {Ben-Ami}, {Blake}, {Botyanszki}, {Buton},
  {Cooke}, {Ellis}, {Hook}, {Kasliwal}, {Pan}, {Pereira}, {Podsiadlowski},
  {Sternberg}, {Suzuki}, {Xu}, {Yaron}, {Bloom}, {Cenko}, {Kulkarni}, {Law},
  {Ofek}, {Poznanski}, \& {Quimby}}]{Maguire11}
{Maguire}, K., {Sullivan}, M., {Thomas}, R.~C., {et~al.} 2011,
  \href{http://dx.doi.org/10.1111/j.1365-2966.2011.19526.x}{\color{magenta}\mnras},
  \href{https://ui.adsabs.harvard.edu/abs/2011MNRAS.418..747M}{\color{blue}418},
  \href{https://ui.adsabs.harvard.edu/abs/2011MNRAS.418..747M}{\color{blue}747}

\bibitem[{{Maguire} {et~al.}(2014){Maguire}, {Sullivan}, {Pan}, {Gal-Yam},
  {Hook}, {Howell}, {Nugent}, {Mazzali}, {Chotard}, {Clubb}, {Filippenko},
  {Kasliwal}, {Kandrashoff}, {Poznanski}, {Saunders}, {Silverman}, {Walker}, \&
  {Xu}}]{Maguire14}
{Maguire}, K., {Sullivan}, M., {Pan}, Y.~C., {et~al.} 2014,
  \href{http://dx.doi.org/10.1093/mnras/stu1607}{\color{magenta}\mnras},
  \href{https://ui.adsabs.harvard.edu/abs/2014MNRAS.444.3258M}{\color{blue}444},
  \href{https://ui.adsabs.harvard.edu/abs/2014MNRAS.444.3258M}{\color{blue}3258}

\bibitem[{{Mahabal} {et~al.}(2019){Mahabal}, {Rebbapragada}, {Walters},
  {Masci}, {Blagorodnova}, {van Roestel}, {Ye}, {Biswas}, {Burdge}, {Chang},
  {Duev}, {Golkhou}, {Miller}, {Nordin}, {Ward}, {Adams}, {Bellm}, {Branton},
  {Bue}, {Cannella}, {Connolly}, {Dekany}, {Feindt}, {Hung}, {Fortson},
  {Frederick}, {Fremling}, {Gezari}, {Graham}, {Groom}, {Kasliwal}, {Kulkarni},
  {Kupfer}, {Lin}, {Lintott}, {Lunnan}, {Parejko}, {Prince}, {Riddle},
  {Rusholme}, {Saunders}, {Sedaghat}, {Shupe}, {Singer}, {Soumagnac}, {Szkody},
  {Tachibana}, {Tirumala}, {van Velzen}, \& {Wright}}]{Mahabal19}
{Mahabal}, A., {Rebbapragada}, U., {Walters}, R., {et~al.} 2019,
  \href{http://dx.doi.org/10.1088/1538-3873/aaf3fa}{\color{magenta}\pasp},
  \href{https://ui.adsabs.harvard.edu/abs/2019PASP..131c8002M}{\color{blue}131},
  \href{https://ui.adsabs.harvard.edu/abs/2019PASP..131c8002M}{\color{blue}038002}

\bibitem[{{Maoz} {et~al.}(2014){Maoz}, {Mannucci}, \& {Nelemans}}]{Maoz14}
{Maoz}, D., {Mannucci}, F., \& {Nelemans}, G. 2014,
  \href{http://dx.doi.org/10.1146/annurev-astro-082812-141031}{\color{magenta}\araa},
  \href{http://adsabs.harvard.edu/abs/2014ARA%26A..52..107M}{\color{blue}52},
  \href{http://adsabs.harvard.edu/abs/2014ARA%26A..52..107M}{\color{blue}107}

\bibitem[{{Marion} {et~al.}(2016){Marion}, {Brown}, {Vink{\'o}}, {Silverman},
  {Sand}, {Challis}, {Kirshner}, {Wheeler}, {Berlind}, {Brown}, {Calkins},
  {Camacho}, {Dhungana}, {Foley}, {Friedman}, {Graham}, {Howell}, {Hsiao},
  {Irwin}, {Jha}, {Kehoe}, {Macri}, {Maeda}, {Mandel}, {McCully}, {Pandya},
  {Rines}, {Wilhelmy}, \& {Zheng}}]{Marion16}
{Marion}, G.~H., {Brown}, P.~J., {Vink{\'o}}, J., {et~al.} 2016,
  \href{http://dx.doi.org/10.3847/0004-637X/820/2/92}{\color{magenta}\apj},
  \href{http://adsabs.harvard.edu/abs/2016ApJ...820...92M}{\color{blue}820},
  \href{http://adsabs.harvard.edu/abs/2016ApJ...820...92M}{\color{blue}92}

\bibitem[{{Masci} {et~al.}(2019){Masci}, {Laher}, {Rusholme}, {Shupe}, {Groom},
  {Surace}, {Jackson}, {Monkewitz}, {Beck}, {Flynn}, {Terek}, {Landry},
  {Hacopians}, {Desai}, {Howell}, {Brooke}, {Imel}, {Wachter}, {Ye}, {Lin},
  {Cenko}, {Cunningham}, {Rebbapragada}, {Bue}, {Miller}, {Mahabal}, {Bellm},
  {Patterson}, {Juri{\'c}}, {Golkhou}, {Ofek}, {Walters}, {Graham}, {Kasliwal},
  {Dekany}, {Kupfer}, {Burdge}, {Cannella}, {Barlow}, {Van Sistine}, {Giomi},
  {Fremling}, {Blagorodnova}, {Levitan}, {Riddle}, {Smith}, {Helou}, {Prince},
  \& {Kulkarni}}]{Masci19}
{Masci}, F.~J., {Laher}, R.~R., {Rusholme}, B., {et~al.} 2019,
  \href{http://dx.doi.org/10.1088/1538-3873/aae8ac}{\color{magenta}\pasp},
  \href{https://ui.adsabs.harvard.edu/abs/2019PASP..131a8003M}{\color{blue}131},
  \href{https://ui.adsabs.harvard.edu/abs/2019PASP..131a8003M}{\color{blue}018003}

\bibitem[{{Mazzali} {et~al.}(2014){Mazzali}, {Sullivan}, {Hachinger}, {Ellis},
  {Nugent}, {Howell}, {Gal-Yam}, {Maguire}, {Cooke}, {Thomas}, {Nomoto}, \&
  {Walker}}]{Mazzali14}
{Mazzali}, P.~A., {Sullivan}, M., {Hachinger}, S., {et~al.} 2014,
  \href{http://dx.doi.org/10.1093/mnras/stu077}{\color{magenta}\mnras},
  \href{https://ui.adsabs.harvard.edu/abs/2014MNRAS.439.1959M}{\color{blue}439},
  \href{https://ui.adsabs.harvard.edu/abs/2014MNRAS.439.1959M}{\color{blue}1959}

\bibitem[{McKinney(2010)}]{McKinney10}
McKinney, W. 2010, 51

\bibitem[{{Miller} {et~al.}(2018){Miller}, {Cao}, {Piro}, {Blagorodnova},
  {Bue}, {Cenko}, {Dhawan}, {Ferretti}, {Fox}, {Fremling}, {Goobar}, {Howell},
  {Hosseinzadeh}, {Kasliwal}, {Laher}, {Lunnan}, {Masci}, {McCully}, {Nugent},
  {Sollerman}, {Taddia}, \& {Kulkarni}}]{Miller18}
{Miller}, A.~A., {Cao}, Y., {Piro}, A.~L., {et~al.} 2018,
  \href{http://dx.doi.org/10.3847/1538-4357/aaa01f}{\color{magenta}\apj},
  \href{http://adsabs.harvard.edu/abs/2018ApJ...852..100M}{\color{blue}852},
  \href{http://adsabs.harvard.edu/abs/2018ApJ...852..100M}{\color{blue}100}

\bibitem[{{Miller} {et~al.}(2020){Miller}, {Yao}, {Bulla}, {Pankow}, {Bellm},
  {Cenko}, {Dekany}, {Fremling}, {Graham}, {Kupfer}, {Laher}, {Mahabal},
  {Masci}, {Nugent}, {Riddle}, {Rusholme}, {Smith}, {Shupe}, {van Roestel}, \&
  {Kulkarni}}]{Miller20}
{Miller}, A.~A., {Yao}, Y., {Bulla}, M., {et~al.} 2020,
  \href{https://arxiv.org/abs/2001.00598}{\color{magenta}arXiv},
  \href{https://ui.adsabs.harvard.edu/abs/2020arXiv200100598M}{\color{blue}arXiv:2001.00598}

\bibitem[{{Mould} {et~al.}(2000){Mould}, {Huchra}, {Freedman}, {Kennicutt},
  {Ferrarese}, {Ford}, {Gibson}, {Graham}, {Hughes}, {Illingworth}, {Kelson},
  {Macri}, {Madore}, {Sakai}, {Sebo}, {Silbermann}, \& {Stetson}}]{Mould00}
{Mould}, J.~R., {Huchra}, J.~P., {Freedman}, W.~L., {et~al.} 2000,
  \href{http://dx.doi.org/10.1086/308304}{\color{magenta}\apj},
  \href{http://adsabs.harvard.edu/abs/2000ApJ...529..786M}{\color{blue}529},
  \href{http://adsabs.harvard.edu/abs/2000ApJ...529..786M}{\color{blue}786}

\bibitem[{{Nadyozhin}(1994)}]{Nadyozhin94}
{Nadyozhin}, D.~K. 1994,
  \href{http://dx.doi.org/10.1086/192008}{\color{magenta}\apjs},
  \href{http://adsabs.harvard.edu/abs/1994ApJS...92..527N}{\color{blue}92},
  \href{http://adsabs.harvard.edu/abs/1994ApJS...92..527N}{\color{blue}527}

\bibitem[{{NASA High Energy Astrophysics Science Archive Research Center
  (HEASARC)}(2014)}]{Heasarc}
{NASA High Energy Astrophysics Science Archive Research Center (HEASARC)}.
  2014, {HEAsoft: Unified Release of FTOOLS and XANADU}

\bibitem[{{Noebauer} {et~al.}(2017){Noebauer}, {Kromer}, {Taubenberger},
  {Baklanov}, {Blinnikov}, {Sorokina}, \& {Hillebrandt}}]{Noebauer17}
{Noebauer}, U.~M., {Kromer}, M., {Taubenberger}, S., {et~al.} 2017,
  \href{http://dx.doi.org/10.1093/mnras/stx2093}{\color{magenta}\mnras},
  \href{http://adsabs.harvard.edu/abs/2017MNRAS.472.2787N}{\color{blue}472},
  \href{http://adsabs.harvard.edu/abs/2017MNRAS.472.2787N}{\color{blue}2787}

\bibitem[{{Nomoto}(1982{\natexlab{a}})}]{Nomoto82}
{Nomoto}, K. 1982{\natexlab{a}},
  \href{http://dx.doi.org/10.1086/159682}{\color{magenta}\apj},
  \href{https://ui.adsabs.harvard.edu/abs/1982ApJ...253..798N}{\color{blue}253},
  \href{https://ui.adsabs.harvard.edu/abs/1982ApJ...253..798N}{\color{blue}798}

\bibitem[{{Nomoto}(1982{\natexlab{b}})}]{Nomoto82a}
---. 1982{\natexlab{b}},
  \href{http://dx.doi.org/10.1086/160031}{\color{magenta}\apj},
  \href{https://ui.adsabs.harvard.edu/abs/1982ApJ...257..780N}{\color{blue}257},
  \href{https://ui.adsabs.harvard.edu/abs/1982ApJ...257..780N}{\color{blue}780}

\bibitem[{{Nugent} {et~al.}(1995){Nugent}, {Phillips}, {Baron}, {Branch}, \&
  {Hauschildt}}]{Nugent95}
{Nugent}, P., {Phillips}, M., {Baron}, E., {et~al.} 1995,
  \href{http://dx.doi.org/10.1086/309846}{\color{magenta}\apjl},
  \href{https://ui.adsabs.harvard.edu/abs/1995ApJ...455L.147N}{\color{blue}455},
  \href{https://ui.adsabs.harvard.edu/abs/1995ApJ...455L.147N}{\color{blue}L147}

\bibitem[{{Nugent} {et~al.}(2011){Nugent}, {Sullivan}, {Cenko}, {Thomas},
  {Kasen}, {Howell}, {Bersier}, {Bloom}, {Kulkarni}, {Kandrashoff},
  {Filippenko}, {Silverman}, {Marcy}, {Howard}, {Isaacson}, {Maguire},
  {Suzuki}, {Tarlton}, {Pan}, {Bildsten}, {Fulton}, {Parrent}, {Sand},
  {Podsiadlowski}, {Bianco}, {Dilday}, {Graham}, {Lyman}, {James}, {Kasliwal},
  {Law}, {Quimby}, {Hook}, {Walker}, {Mazzali}, {Pian}, {Ofek}, {Gal-Yam}, \&
  {Poznanski}}]{Nugent11}
{Nugent}, P.~E., {Sullivan}, M., {Cenko}, S.~B., {et~al.} 2011,
  \href{http://dx.doi.org/10.1038/nature10644}{\color{magenta}\nat},
  \href{http://adsabs.harvard.edu/abs/2011Natur.480..344N}{\color{blue}480},
  \href{http://adsabs.harvard.edu/abs/2011Natur.480..344N}{\color{blue}344}

\bibitem[{{Oke} \& {Gunn}(1982)}]{Oke82}
{Oke}, J.~B., \& {Gunn}, J.~E. 1982,
  \href{http://dx.doi.org/10.1086/131027}{\color{magenta}\pasp},
  \href{http://adsabs.harvard.edu/abs/1982PASP...94..586O}{\color{blue}94},
  \href{http://adsabs.harvard.edu/abs/1982PASP...94..586O}{\color{blue}586}

\bibitem[{{Oke} {et~al.}(1995){Oke}, {Cohen}, {Carr}, {Cromer}, {Dingizian},
  {Harris}, {Labrecque}, {Lucinio}, {Schaal}, {Epps}, \& {Miller}}]{Oke95}
{Oke}, J.~B., {Cohen}, J.~G., {Carr}, M., {et~al.} 1995, \pasp,
  \href{http://adsabs.harvard.edu/abs/1995PASP..107..375O}{\color{blue}107},
  \href{http://adsabs.harvard.edu/abs/1995PASP..107..375O}{\color{blue}375}

\bibitem[{{Olling} {et~al.}(2015){Olling}, {Mushotzky}, {Shaya}, {Rest},
  {Garnavich}, {Tucker}, {Kasen}, {Margheim}, \& {Filippenko}}]{Olling15}
{Olling}, R.~P., {Mushotzky}, R., {Shaya}, E.~J., {et~al.} 2015,
  \href{http://dx.doi.org/10.1038/nature14455}{\color{magenta}\nat},
  \href{https://ui.adsabs.harvard.edu/abs/2015Natur.521..332O}{\color{blue}521},
  \href{https://ui.adsabs.harvard.edu/abs/2015Natur.521..332O}{\color{blue}332}

\bibitem[{{Pakmor} {et~al.}(2011){Pakmor}, {Hachinger}, {R{\"o}pke}, \&
  {Hillebrand t}}]{Pakmor11}
{Pakmor}, R., {Hachinger}, S., {R{\"o}pke}, F.~K., \& {Hillebrand t}, W. 2011,
  \href{http://dx.doi.org/10.1051/0004-6361/201015653}{\color{magenta}\aap},
  \href{https://ui.adsabs.harvard.edu/abs/2011A&A...528A.117P}{\color{blue}528},
  \href{https://ui.adsabs.harvard.edu/abs/2011A&A...528A.117P}{\color{blue}A117}

\bibitem[{{Pakmor} {et~al.}(2010){Pakmor}, {Kromer}, {R{\"o}pke}, {Sim},
  {Ruiter}, \& {Hillebrandt}}]{Pakmor10}
{Pakmor}, R., {Kromer}, M., {R{\"o}pke}, F.~K., {et~al.} 2010,
  \href{http://dx.doi.org/10.1038/nature08642}{\color{magenta}\nat},
  \href{https://ui.adsabs.harvard.edu/abs/2010Natur.463...61P}{\color{blue}463},
  \href{https://ui.adsabs.harvard.edu/abs/2010Natur.463...61P}{\color{blue}61}

\bibitem[{{Pakmor} {et~al.}(2012){Pakmor}, {Kromer}, {Taubenberger}, {Sim},
  {R{\"o}pke}, \& {Hillebrandt}}]{Pakmor12}
{Pakmor}, R., {Kromer}, M., {Taubenberger}, S., {et~al.} 2012,
  \href{http://dx.doi.org/10.1088/2041-8205/747/1/L10}{\color{magenta}\apjl},
  \href{https://ui.adsabs.harvard.edu/abs/2012ApJ...747L..10P}{\color{blue}747},
  \href{https://ui.adsabs.harvard.edu/abs/2012ApJ...747L..10P}{\color{blue}L10}

\bibitem[{{Pastorello} {et~al.}(2007){Pastorello}, {Mazzali}, {Pignata},
  {Benetti}, {Cappellaro}, {Filippenko}, {Li}, {Meikle}, {Arkharov}, {Blanc},
  {Bufano}, {Derekas}, {Dolci}, {Elias-Rosa}, {Foley}, {Ganeshalingam},
  {Harutyunyan}, {Kiss}, {Kotak}, {Larionov}, {Lucey}, {Napoleone},
  {Navasardyan}, {Patat}, {Rich}, {Ryder}, {Salvo}, {Schmidt}, {Stanishev},
  {Sz{\'e}kely}, {Taubenberger}, {Temporin}, {Turatto}, \&
  {Hillebrandt}}]{Pastorello07}
{Pastorello}, A., {Mazzali}, P.~A., {Pignata}, G., {et~al.} 2007,
  \href{http://dx.doi.org/10.1111/j.1365-2966.2007.11700.x}{\color{magenta}\mnras},
  \href{https://ui.adsabs.harvard.edu/abs/2007MNRAS.377.1531P}{\color{blue}377},
  \href{https://ui.adsabs.harvard.edu/abs/2007MNRAS.377.1531P}{\color{blue}1531}

\bibitem[{{Patterson} {et~al.}(2019){Patterson}, {Bellm}, {Rusholme}, {Masci},
  {Juric}, {Krughoff}, {Golkhou}, {Graham}, {Kulkarni}, {Helou}, \& {Zwicky
  Transient Facility Collaboration}}]{Patterson19}
{Patterson}, M.~T., {Bellm}, E.~C., {Rusholme}, B., {et~al.} 2019,
  \href{http://dx.doi.org/10.1088/1538-3873/aae904}{\color{magenta}\pasp},
  \href{https://ui.adsabs.harvard.edu/abs/2019PASP..131a8001P}{\color{blue}131},
  \href{https://ui.adsabs.harvard.edu/abs/2019PASP..131a8001P}{\color{blue}018001}

\bibitem[{Pedregosa {et~al.}(2011)Pedregosa, Varoquaux, Gramfort, Michel,
  Thirion, Grisel, Blondel, Prettenhofer, Weiss, Dubourg, Vanderplas, Passos,
  Cournapeau, Brucher, Perrot, \& Duchesnay}]{Pedregosa11}
Pedregosa, F., Varoquaux, G., Gramfort, A., {et~al.} 2011, Journal of Machine
  Learning Research, 12, 2825

\bibitem[{{Pereira} {et~al.}(2013){Pereira}, {Thomas}, {Aldering}, {Antilogus},
  {Baltay}, {Benitez-Herrera}, {Bongard}, {Buton}, {Canto}, {Cellier-Holzem},
  {Chen}, {Childress}, {Chotard}, {Copin}, {Fakhouri}, {Fink}, {Fouchez},
  {Gangler}, {Guy}, {Hillebrandt}, {Hsiao}, {Kerschhaggl}, {Kowalski},
  {Kromer}, {Nordin}, {Nugent}, {Paech}, {Pain}, {P{\'e}contal}, {Perlmutter},
  {Rabinowitz}, {Rigault}, {Runge}, {Saunders}, {Smadja}, {Tao},
  {Taubenberger}, {Tilquin}, \& {Wu}}]{Pereira13}
{Pereira}, R., {Thomas}, R.~C., {Aldering}, G., {et~al.} 2013,
  \href{http://dx.doi.org/10.1051/0004-6361/201221008}{\color{magenta}\aap},
  \href{http://adsabs.harvard.edu/abs/2013A%26A...554A..27P}{\color{blue}554},
  \href{http://adsabs.harvard.edu/abs/2013A%26A...554A..27P}{\color{blue}A27}

\bibitem[{{Perley}(2019)}]{Perley19}
{Perley}, D.~A. 2019,
  \href{http://dx.doi.org/10.1088/1538-3873/ab215d}{\color{magenta}\pasp},
  \href{https://ui.adsabs.harvard.edu/abs/2019PASP..131h4503P}{\color{blue}131},
  \href{https://ui.adsabs.harvard.edu/abs/2019PASP..131h4503P}{\color{blue}084503}

\bibitem[{{Phillips}(1993)}]{Phillips93}
{Phillips}, M.~M. 1993,
  \href{http://dx.doi.org/10.1086/186970}{\color{magenta}\apjl},
  \href{http://adsabs.harvard.edu/abs/1993ApJ...413L.105P}{\color{blue}413},
  \href{http://adsabs.harvard.edu/abs/1993ApJ...413L.105P}{\color{blue}L105}

\bibitem[{{Phillips} {et~al.}(1999){Phillips}, {Lira}, {Suntzeff}, {Schommer},
  {Hamuy}, \& {Maza}}]{Phillips99}
{Phillips}, M.~M., {Lira}, P., {Suntzeff}, N.~B., {et~al.} 1999,
  \href{http://dx.doi.org/10.1086/301032}{\color{magenta}\aj},
  \href{http://adsabs.harvard.edu/abs/1999AJ....118.1766P}{\color{blue}118},
  \href{http://adsabs.harvard.edu/abs/1999AJ....118.1766P}{\color{blue}1766}

\bibitem[{{Phillips} {et~al.}(1987){Phillips}, {Phillips}, {Heathcote},
  {Blanco}, {Geisler}, {Hamilton}, {Suntzeff}, {Jablonski}, {Steiner},
  {Cowley}, {Schmidtke}, {Wyckoff}, {Hutchings}, {Tonry}, {Strauss},
  {Thorstensen}, {Honey}, {Maza}, {Ruiz}, {Landolt}, {Uomoto}, {Rich},
  {Grindlay}, {Cohn}, {Smith}, {Lutz}, {Lavery}, \& {Saha}}]{Phillips87}
{Phillips}, M.~M., {Phillips}, A.~C., {Heathcote}, S.~R., {et~al.} 1987,
  \href{http://dx.doi.org/10.1086/132020}{\color{magenta}\pasp},
  \href{https://ui.adsabs.harvard.edu/abs/1987PASP...99..592P}{\color{blue}99},
  \href{https://ui.adsabs.harvard.edu/abs/1987PASP...99..592P}{\color{blue}592}

\bibitem[{{Phillips} {et~al.}(2013){Phillips}, {Simon}, {Morrell}, {Burns},
  {Cox}, {Foley}, {Karakas}, {Patat}, {Sternberg}, {Williams}, {Gal-Yam},
  {Hsiao}, {Leonard}, {Persson}, {Stritzinger}, {Thompson}, {Campillay},
  {Contreras}, {Folatelli}, {Freedman}, {Hamuy}, {Roth}, {Shields}, {Suntzeff},
  {Chomiuk}, {Ivans}, {Madore}, {Penprase}, {Perley}, {Pignata}, {Preston}, \&
  {Soderberg}}]{Phillips13}
{Phillips}, M.~M., {Simon}, J.~D., {Morrell}, N., {et~al.} 2013,
  \href{http://dx.doi.org/10.1088/0004-637X/779/1/38}{\color{magenta}\apj},
  \href{https://ui.adsabs.harvard.edu/abs/2013ApJ...779...38P}{\color{blue}779},
  \href{https://ui.adsabs.harvard.edu/abs/2013ApJ...779...38P}{\color{blue}38}

\bibitem[{{Piascik} {et~al.}(2014){Piascik}, {Steele}, {Bates}, {Mottram},
  {Smith}, {Barnsley}, \& {Bolton}}]{Piascik14}
{Piascik}, A.~S., {Steele}, I.~A., {Bates}, S.~D., {et~al.} 2014,
  \href{http://dx.doi.org/10.1117/12.2055117}{\color{magenta}Society of
  Photo-Optical Instrumentation Engineers (SPIE) Conference Series}, Vol.
  \href{https://ui.adsabs.harvard.edu/abs/2014SPIE.9147E..8HP}{\color{blue}9147},
  {SPRAT: Spectrograph for the Rapid Acquisition of Transients},
  \href{https://ui.adsabs.harvard.edu/abs/2014SPIE.9147E..8HP}{\color{blue}91478H}

\bibitem[{{Piro} {et~al.}(2010){Piro}, {Chang}, \& {Weinberg}}]{Piro10}
{Piro}, A.~L., {Chang}, P., \& {Weinberg}, N.~N. 2010,
  \href{http://dx.doi.org/10.1088/0004-637X/708/1/598}{\color{magenta}\apj},
  \href{https://ui.adsabs.harvard.edu/abs/2010ApJ...708..598P}{\color{blue}708},
  \href{https://ui.adsabs.harvard.edu/abs/2010ApJ...708..598P}{\color{blue}598}

\bibitem[{{Piro} \& {Morozova}(2016)}]{Piro16}
{Piro}, A.~L., \& {Morozova}, V.~S. 2016,
  \href{http://dx.doi.org/10.3847/0004-637X/826/1/96}{\color{magenta}\apj},
  \href{http://adsabs.harvard.edu/abs/2016ApJ...826...96P}{\color{blue}826},
  \href{http://adsabs.harvard.edu/abs/2016ApJ...826...96P}{\color{blue}96}

\bibitem[{{Polin} {et~al.}(2019{\natexlab{a}}){Polin}, {Nugent}, \&
  {Kasen}}]{Polin19}
{Polin}, A., {Nugent}, P., \& {Kasen}, D. 2019{\natexlab{a}},
  \href{http://dx.doi.org/10.3847/1538-4357/aafb6a}{\color{magenta}\apj},
  \href{http://adsabs.harvard.edu/abs/2019ApJ...873...84P}{\color{blue}873},
  \href{http://adsabs.harvard.edu/abs/2019ApJ...873...84P}{\color{blue}84}

\bibitem[{{Polin} {et~al.}(2019{\natexlab{b}}){Polin}, {Nugent}, \&
  {Kasen}}]{Polin19a}
---. 2019{\natexlab{b}},
  \href{https://arxiv.org/abs/1910.12434}{\color{magenta}arXiv},
  \href{https://ui.adsabs.harvard.edu/abs/2019arXiv191012434P}{\color{blue}arXiv:1910.12434}

\bibitem[{{Poznanski} {et~al.}(2012){Poznanski}, {Prochaska}, \&
  {Bloom}}]{Poznanski12}
{Poznanski}, D., {Prochaska}, J.~X., \& {Bloom}, J.~S. 2012,
  \href{http://dx.doi.org/10.1111/j.1365-2966.2012.21796.x}{\color{magenta}\mnras},
  \href{http://adsabs.harvard.edu/abs/2012MNRAS.426.1465P}{\color{blue}426},
  \href{http://adsabs.harvard.edu/abs/2012MNRAS.426.1465P}{\color{blue}1465}

\bibitem[{{Rabinak} \& {Waxman}(2011)}]{Rabinak11}
{Rabinak}, I., \& {Waxman}, E. 2011,
  \href{http://dx.doi.org/10.1088/0004-637X/728/1/63}{\color{magenta}\apj},
  \href{https://ui.adsabs.harvard.edu/abs/2011ApJ...728...63R}{\color{blue}728},
  \href{https://ui.adsabs.harvard.edu/abs/2011ApJ...728...63R}{\color{blue}63}

\bibitem[{{Raskin} \& {Kasen}(2013)}]{Raskin13}
{Raskin}, C., \& {Kasen}, D. 2013,
  \href{http://dx.doi.org/10.1088/0004-637X/772/1/1}{\color{magenta}\apj},
  \href{https://ui.adsabs.harvard.edu/abs/2013ApJ...772....1R}{\color{blue}772},
  \href{https://ui.adsabs.harvard.edu/abs/2013ApJ...772....1R}{\color{blue}1}

\bibitem[{{Rasmussen} \& {Williams}(2006)}]{Rasmussen06}
{Rasmussen}, C.~E., \& {Williams}, C. K.~I. 2006, {Gaussian Processes for
  Machine Learning}

\bibitem[{{Riess} {et~al.}(1998){Riess}, {Nugent}, {Filippenko}, {Kirshner}, \&
  {Perlmutter}}]{Riess98a}
{Riess}, A.~G., {Nugent}, P., {Filippenko}, A.~V., {et~al.} 1998,
  \href{http://dx.doi.org/10.1086/306106}{\color{magenta}\apj},
  \href{https://ui.adsabs.harvard.edu/abs/1998ApJ...504..935R}{\color{blue}504},
  \href{https://ui.adsabs.harvard.edu/abs/1998ApJ...504..935R}{\color{blue}935}

\bibitem[{{Rigault} {et~al.}(2019){Rigault}, {Neill}, {Blagorodnova}, {Dugas},
  {Feeney}, {Walters}, {Brinnel}, {Copin}, {Fremling}, {Nordin}, \&
  {Sollerman}}]{Rigault19}
{Rigault}, M., {Neill}, J.~D., {Blagorodnova}, N., {et~al.} 2019,
  \href{http://dx.doi.org/10.1051/0004-6361/201935344}{\color{magenta}\aap},
  \href{https://ui.adsabs.harvard.edu/abs/2019A&A...627A.115R}{\color{blue}627},
  \href{https://ui.adsabs.harvard.edu/abs/2019A&A...627A.115R}{\color{blue}A115}

\bibitem[{{Roming} {et~al.}(2005){Roming}, {Kennedy}, {Mason}, {Nousek}, {Ahr},
  {Bingham}, {Broos}, {Carter}, {Hancock}, {Huckle}, {Hunsberger}, {Kawakami},
  {Killough}, {Koch}, {McLelland}, {Smith}, {Smith}, {Soto}, {Boyd},
  {Breeveld}, {Holland}, {Ivanushkina}, {Pryzby}, {Still}, \&
  {Stock}}]{Roming05}
{Roming}, P.~W.~A., {Kennedy}, T.~E., {Mason}, K.~O., {et~al.} 2005,
  \href{http://dx.doi.org/10.1007/s11214-005-5095-4}{\color{magenta}Space
  Science Reviews},
  \href{http://adsabs.harvard.edu/abs/2005SSRv..120...95R}{\color{blue}120},
  \href{http://adsabs.harvard.edu/abs/2005SSRv..120...95R}{\color{blue}95}

\bibitem[{{R{\"o}pke} \& {Sim}(2018)}]{Ropke18}
{R{\"o}pke}, F.~K., \& {Sim}, S.~A. 2018,
  \href{http://dx.doi.org/10.1007/s11214-018-0503-8}{\color{magenta}\ssr},
  \href{https://ui.adsabs.harvard.edu/abs/2018SSRv..214...72R}{\color{blue}214},
  \href{https://ui.adsabs.harvard.edu/abs/2018SSRv..214...72R}{\color{blue}72}

\bibitem[{{R{\"o}pke} {et~al.}(2012){R{\"o}pke}, {Kromer}, {Seitenzahl},
  {Pakmor}, {Sim}, {Taubenberger}, {Ciaraldi-Schoolmann}, {Hillebrandt},
  {Aldering}, {Antilogus}, {Baltay}, {Benitez-Herrera}, {Bongard}, {Buton},
  {Canto}, {Cellier-Holzem}, {Childress}, {Chotard}, {Copin}, {Fakhouri},
  {Fink}, {Fouchez}, {Gangler}, {Guy}, {Hachinger}, {Hsiao}, {Chen},
  {Kerschhaggl}, {Kowalski}, {Nugent}, {Paech}, {Pain}, {Pecontal}, {Pereira},
  {Perlmutter}, {Rabinowitz}, {Rigault}, {Runge}, {Saunders}, {Smadja},
  {Suzuki}, {Tao}, {Thomas}, {Tilquin}, \& {Wu}}]{Ropke12}
{R{\"o}pke}, F.~K., {Kromer}, M., {Seitenzahl}, I.~R., {et~al.} 2012,
  \href{http://dx.doi.org/10.1088/2041-8205/750/1/L19}{\color{magenta}\apjl},
  \href{https://ui.adsabs.harvard.edu/abs/2012ApJ...750L..19R}{\color{blue}750},
  \href{https://ui.adsabs.harvard.edu/abs/2012ApJ...750L..19R}{\color{blue}L19}

\bibitem[{{Sagiv} {et~al.}(2014){Sagiv}, {Gal-Yam}, {Ofek}, {Waxman},
  {Aharonson}, {Kulkarni}, {Nakar}, {Maoz}, {Trakhtenbrot}, {Phinney}, {Topaz},
  {Beichman}, {Murthy}, \& {Worden}}]{Sagiv14}
{Sagiv}, I., {Gal-Yam}, A., {Ofek}, E.~O., {et~al.} 2014,
  \href{http://dx.doi.org/10.1088/0004-6256/147/4/79}{\color{magenta}\aj},
  \href{https://ui.adsabs.harvard.edu/abs/2014AJ....147...79S}{\color{blue}147},
  \href{https://ui.adsabs.harvard.edu/abs/2014AJ....147...79S}{\color{blue}79}

\bibitem[{{Savitzky} \& {Golay}(1964)}]{Savitzky64}
{Savitzky}, A., \& {Golay}, M.~J.~E. 1964, Analytical Chemistry,
  \href{http://adsabs.harvard.edu/abs/1964AnaCh..36.1627S}{\color{blue}36},
  \href{http://adsabs.harvard.edu/abs/1964AnaCh..36.1627S}{\color{blue}1627}

\bibitem[{{Scalzo} {et~al.}(2014{\natexlab{a}}){Scalzo}, {Aldering},
  {Antilogus}, {Aragon}, {Bailey}, {Baltay}, {Bongard}, {Buton},
  {Cellier-Holzem}, {Childress}, {Chotard}, {Copin}, {Fakhouri}, {Gangler},
  {Guy}, {Kim}, {Kowalski}, {Kromer}, {Nordin}, {Nugent}, {Paech}, {Pain},
  {Pecontal}, {Pereira}, {Perlmutter}, {Rabinowitz}, {Rigault}, {Runge},
  {Saunders}, {Sim}, {Smadja}, {Tao}, {Taubenberger}, {Thomas}, {Weaver}, \&
  {Nearby Supernova Factory}}]{Scalzo14}
{Scalzo}, R., {Aldering}, G., {Antilogus}, P., {et~al.} 2014{\natexlab{a}},
  \href{http://dx.doi.org/10.1093/mnras/stu350}{\color{magenta}\mnras},
  \href{http://adsabs.harvard.edu/abs/2014MNRAS.440.1498S}{\color{blue}440},
  \href{http://adsabs.harvard.edu/abs/2014MNRAS.440.1498S}{\color{blue}1498}

\bibitem[{{Scalzo} {et~al.}(2014{\natexlab{b}}){Scalzo}, {Ruiter}, \&
  {Sim}}]{Scalzo14a}
{Scalzo}, R.~A., {Ruiter}, A.~J., \& {Sim}, S.~A. 2014{\natexlab{b}},
  \href{http://dx.doi.org/10.1093/mnras/stu1808}{\color{magenta}\mnras},
  \href{http://adsabs.harvard.edu/abs/2014MNRAS.445.2535S}{\color{blue}445},
  \href{http://adsabs.harvard.edu/abs/2014MNRAS.445.2535S}{\color{blue}2535}

\bibitem[{{Schlafly} \& {Finkbeiner}(2011)}]{Schlafly11}
{Schlafly}, E.~F., \& {Finkbeiner}, D.~P. 2011,
  \href{http://dx.doi.org/10.1088/0004-637X/737/2/103}{\color{magenta}\apj},
  \href{http://adsabs.harvard.edu/abs/2011ApJ...737..103S}{\color{blue}737},
  \href{http://adsabs.harvard.edu/abs/2011ApJ...737..103S}{\color{blue}103}

\bibitem[{{Schlegel} {et~al.}(1998){Schlegel}, {Finkbeiner}, \&
  {Davis}}]{Schlegel98}
{Schlegel}, D.~J., {Finkbeiner}, D.~P., \& {Davis}, M. 1998,
  \href{http://dx.doi.org/10.1086/305772}{\color{magenta}\apj},
  \href{http://adsabs.harvard.edu/abs/1998ApJ...500..525S}{\color{blue}500},
  \href{http://adsabs.harvard.edu/abs/1998ApJ...500..525S}{\color{blue}525}

\bibitem[{{Seitenzahl} {et~al.}(2013){Seitenzahl}, {Ciaraldi-Schoolmann},
  {R{\"o}pke}, {Fink}, {Hillebrandt}, {Kromer}, {Pakmor}, {Ruiter}, {Sim}, \&
  {Taubenberger}}]{Seitenzahl13}
{Seitenzahl}, I.~R., {Ciaraldi-Schoolmann}, F., {R{\"o}pke}, F.~K., {et~al.}
  2013, \href{http://dx.doi.org/10.1093/mnras/sts402}{\color{magenta}\mnras},
  \href{http://adsabs.harvard.edu/abs/2013MNRAS.429.1156S}{\color{blue}429},
  \href{http://adsabs.harvard.edu/abs/2013MNRAS.429.1156S}{\color{blue}1156}

\bibitem[{{Shappee} {et~al.}(2018){Shappee}, {Piro}, {Stanek}, {Patel},
  {Margutti}, {Lipunov}, \& {Pogge}}]{Shappee18}
{Shappee}, B.~J., {Piro}, A.~L., {Stanek}, K.~Z., {et~al.} 2018,
  \href{http://dx.doi.org/10.3847/1538-4357/aaa1e9}{\color{magenta}\apj},
  \href{https://ui.adsabs.harvard.edu/abs/2018ApJ...855....6S}{\color{blue}855},
  \href{https://ui.adsabs.harvard.edu/abs/2018ApJ...855....6S}{\color{blue}6}

\bibitem[{{Shappee} {et~al.}(2019){Shappee}, {Holoien}, {Drout}, {Auchettl},
  {Stritzinger}, {Kochanek}, {Stanek}, {Shaya}, {Narayan}, \&
  {ASAS-SN}}]{Shappee19}
{Shappee}, B.~J., {Holoien}, T.~W.~S., {Drout}, M.~R., {et~al.} 2019,
  \href{http://dx.doi.org/10.3847/1538-4357/aaec79}{\color{magenta}\apj},
  \href{https://ui.adsabs.harvard.edu/abs/2019ApJ...870...13S}{\color{blue}870},
  \href{https://ui.adsabs.harvard.edu/abs/2019ApJ...870...13S}{\color{blue}13}

\bibitem[{{Shen} \& {Bildsten}(2014)}]{Shen14}
{Shen}, K.~J., \& {Bildsten}, L. 2014,
  \href{http://dx.doi.org/10.1088/0004-637X/785/1/61}{\color{magenta}\apj},
  \href{https://ui.adsabs.harvard.edu/abs/2014ApJ...785...61S}{\color{blue}785},
  \href{https://ui.adsabs.harvard.edu/abs/2014ApJ...785...61S}{\color{blue}61}

\bibitem[{{Silverman} {et~al.}(2011){Silverman}, {Ganeshalingam}, {Li},
  {Filippenko}, {Miller}, \& {Poznanski}}]{Silverman11}
{Silverman}, J.~M., {Ganeshalingam}, M., {Li}, W., {et~al.} 2011,
  \href{http://dx.doi.org/10.1111/j.1365-2966.2010.17474.x}{\color{magenta}\mnras},
  \href{http://adsabs.harvard.edu/abs/2011MNRAS.410..585S}{\color{blue}410},
  \href{http://adsabs.harvard.edu/abs/2011MNRAS.410..585S}{\color{blue}585}

\bibitem[{{Sim} {et~al.}(2013){Sim}, {Seitenzahl}, {Kromer},
  {Ciaraldi-Schoolmann}, {R{\"o}pke}, {Fink}, {Hillebrandt}, {Pakmor},
  {Ruiter}, \& {Taubenberger}}]{Sim13}
{Sim}, S.~A., {Seitenzahl}, I.~R., {Kromer}, M., {et~al.} 2013,
  \href{http://dx.doi.org/10.1093/mnras/stt1574}{\color{magenta}\mnras},
  \href{https://ui.adsabs.harvard.edu/abs/2013MNRAS.436..333S}{\color{blue}436},
  \href{https://ui.adsabs.harvard.edu/abs/2013MNRAS.436..333S}{\color{blue}333}

\bibitem[{{Steele} {et~al.}(2004){Steele}, {Smith}, {Rees}, {Baker}, {Bates},
  {Bode}, {Bowman}, {Carter}, {Etherton}, {Ford}, {Fraser}, {Gomboc}, {Lett},
  {Mansfield}, {Marchant}, {Medrano-Cerda}, {Mottram}, {Raback}, {Scott},
  {Tomlinson}, \& {Zamanov}}]{Steele04}
{Steele}, I.~A., {Smith}, R.~J., {Rees}, P.~C., {et~al.} 2004,
  \href{http://dx.doi.org/10.1117/12.551456}{\color{magenta}Society of
  Photo-Optical Instrumentation Engineers (SPIE) Conference Series}, Vol.
  \href{https://ui.adsabs.harvard.edu/abs/2004SPIE.5489..679S}{\color{blue}5489},
  {The Liverpool Telescope: performance and first results}, ed. J.~{Oschmann},
  Jacobus~M.,
  \href{https://ui.adsabs.harvard.edu/abs/2004SPIE.5489..679S}{\color{blue}679}

\bibitem[{{Stehle} {et~al.}(2005){Stehle}, {Mazzali}, {Benetti}, \& {Hillebrand
  t}}]{Stehle05}
{Stehle}, M., {Mazzali}, P.~A., {Benetti}, S., \& {Hillebrand t}, W. 2005,
  \href{http://dx.doi.org/10.1111/j.1365-2966.2005.09116.x}{\color{magenta}\mnras},
  \href{https://ui.adsabs.harvard.edu/abs/2005MNRAS.360.1231S}{\color{blue}360},
  \href{https://ui.adsabs.harvard.edu/abs/2005MNRAS.360.1231S}{\color{blue}1231}

\bibitem[{{Stritzinger} {et~al.}(2006){Stritzinger}, {Leibundgut}, {Walch}, \&
  {Contardo}}]{Stritzinger06}
{Stritzinger}, M., {Leibundgut}, B., {Walch}, S., \& {Contardo}, G. 2006,
  \href{http://dx.doi.org/10.1051/0004-6361:20053652}{\color{magenta}\aap},
  \href{http://adsabs.harvard.edu/abs/2006A%26A...450..241S}{\color{blue}450},
  \href{http://adsabs.harvard.edu/abs/2006A%26A...450..241S}{\color{blue}241}

\bibitem[{{Stritzinger} {et~al.}(2011){Stritzinger}, {Phillips}, {Boldt},
  {Burns}, {Campillay}, {Contreras}, {Gonzalez}, {Folatelli}, {Morrell},
  {Krzeminski}, {Roth}, {Salgado}, {DePoy}, {Hamuy}, {Freedman}, {Madore},
  {Marshall}, {Persson}, {Rheault}, {Suntzeff}, {Villanueva}, {Li}, \&
  {Filippenko}}]{Stritzinger11}
{Stritzinger}, M.~D., {Phillips}, M.~M., {Boldt}, L.~N., {et~al.} 2011,
  \href{http://dx.doi.org/10.1088/0004-6256/142/5/156}{\color{magenta}\aj},
  \href{http://adsabs.harvard.edu/abs/2011AJ....142..156S}{\color{blue}142},
  \href{http://adsabs.harvard.edu/abs/2011AJ....142..156S}{\color{blue}156}

\bibitem[{{Suntzeff}(1996)}]{Suntzeff96}
{Suntzeff}, N.~B. 1996,
  \href{https://ui.adsabs.harvard.edu/abs/1996ssr..conf...41S}{\color{blue}41}

\bibitem[{{Taubenberger}(2017)}]{Taubenberger17}
{Taubenberger}, S. 2017, {The Extremes of Thermonuclear Supernovae},
  \href{https://ui.adsabs.harvard.edu/abs/2017hsn..book..317T}{\color{blue}317}

\bibitem[{{Taubenberger} {et~al.}(2013){Taubenberger}, {Kromer}, {Pakmor},
  {Pignata}, {Maeda}, {Hachinger}, {Leibundgut}, \& {Hillebrand
  t}}]{Taubenberger13}
{Taubenberger}, S., {Kromer}, M., {Pakmor}, R., {et~al.} 2013,
  \href{http://dx.doi.org/10.1088/2041-8205/775/2/L43}{\color{magenta}\apjl},
  \href{https://ui.adsabs.harvard.edu/abs/2013ApJ...775L..43T}{\color{blue}775},
  \href{https://ui.adsabs.harvard.edu/abs/2013ApJ...775L..43T}{\color{blue}L43}

\bibitem[{{Taubenberger} {et~al.}(2008){Taubenberger}, {Hachinger}, {Pignata},
  {Mazzali}, {Contreras}, {Valenti}, {Pastorello}, {Elias-Rosa},
  {B{\"a}rnbantner}, {Barwig}, {Benetti}, {Dolci}, {Fliri}, {Folatelli},
  {Freedman}, {Gonzalez}, {Hamuy}, {Krzeminski}, {Morrell}, {Navasardyan},
  {Persson}, {Phillips}, {Ries}, {Roth}, {Suntzeff}, {Turatto}, \&
  {Hillebrandt}}]{Taubenberger08}
{Taubenberger}, S., {Hachinger}, S., {Pignata}, G., {et~al.} 2008,
  \href{http://dx.doi.org/10.1111/j.1365-2966.2008.12843.x}{\color{magenta}\mnras},
  \href{https://ui.adsabs.harvard.edu/abs/2008MNRAS.385...75T}{\color{blue}385},
  \href{https://ui.adsabs.harvard.edu/abs/2008MNRAS.385...75T}{\color{blue}75}

\bibitem[{{Tonry}(2011)}]{Tonry11}
{Tonry}, J.~L. 2011,
  \href{http://dx.doi.org/10.1086/657997}{\color{magenta}\pasp},
  \href{http://adsabs.harvard.edu/abs/2011PASP..123...58T}{\color{blue}123},
  \href{http://adsabs.harvard.edu/abs/2011PASP..123...58T}{\color{blue}58}

\bibitem[{{Tonry} {et~al.}(2001){Tonry}, {Dressler}, {Blakeslee}, {Ajhar},
  {Fletcher}, {Luppino}, {Metzger}, \& {Moore}}]{Tonry01}
{Tonry}, J.~L., {Dressler}, A., {Blakeslee}, J.~P., {et~al.} 2001,
  \href{http://dx.doi.org/10.1086/318301}{\color{magenta}\apj},
  \href{https://ui.adsabs.harvard.edu/abs/2001ApJ...546..681T}{\color{blue}546},
  \href{https://ui.adsabs.harvard.edu/abs/2001ApJ...546..681T}{\color{blue}681}

\bibitem[{{Townsley} {et~al.}(2019){Townsley}, {Miles}, {Shen}, \&
  {Kasen}}]{Townsley19}
{Townsley}, D.~M., {Miles}, B.~J., {Shen}, K.~J., \& {Kasen}, D. 2019,
  \href{http://dx.doi.org/10.3847/2041-8213/ab27cd}{\color{magenta}\apjl},
  \href{https://ui.adsabs.harvard.edu/abs/2019ApJ...878L..38T}{\color{blue}878},
  \href{https://ui.adsabs.harvard.edu/abs/2019ApJ...878L..38T}{\color{blue}L38}

\bibitem[{{Tully} {et~al.}(2013){Tully}, {Courtois}, {Dolphin}, {Fisher},
  {H{\'e}raudeau}, {Jacobs}, {Karachentsev}, {Makarov}, {Makarova},
  {Mitronova}, {Rizzi}, {Shaya}, {Sorce}, \& {Wu}}]{Tully13}
{Tully}, R.~B., {Courtois}, H.~M., {Dolphin}, A.~E., {et~al.} 2013,
  \href{http://dx.doi.org/10.1088/0004-6256/146/4/86}{\color{magenta}\aj},
  \href{https://ui.adsabs.harvard.edu/abs/2013AJ....146...86T}{\color{blue}146},
  \href{https://ui.adsabs.harvard.edu/abs/2013AJ....146...86T}{\color{blue}86}

\bibitem[{{Virtanen} {et~al.}(2020){Virtanen}, {Gommers}, {Oliphant},
  {Haberland}, {Reddy}, {Cournapeau}, {Burovski}, {Peterson}, {Weckesser},
  {Bright}, {van der Walt}, {Brett}, {Wilson}, {Jarrod Millman}, {Mayorov},
  {Nelson}, {Jones}, {Kern}, {Larson}, {Carey}, {Polat}, {Feng}, {Moore}, {Vand
  erPlas}, {Laxalde}, {Perktold}, {Cimrman}, {Henriksen}, {Quintero}, {Harris},
  {Archibald}, {Ribeiro}, {Pedregosa}, {van Mulbregt}, \&
  {Contributors}}]{2020SciPy-NMeth}
{Virtanen}, P., {Gommers}, R., {Oliphant}, T.~E., {et~al.} 2020,
  \href{http://dx.doi.org/https://doi.org/10.1038/s41592-019-0686-2}{\color{magenta}Nature
  Methods}, \href{https://rdcu.be/b08Wh}{\color{blue}17},
  \href{https://rdcu.be/b08Wh}{\color{blue}261}

\bibitem[{{Webbink}(1984)}]{Webbink84}
{Webbink}, R.~F. 1984,
  \href{http://dx.doi.org/10.1086/161701}{\color{magenta}\apj},
  \href{https://ui.adsabs.harvard.edu/abs/1984ApJ...277..355W}{\color{blue}277},
  \href{https://ui.adsabs.harvard.edu/abs/1984ApJ...277..355W}{\color{blue}355}

\bibitem[{{Wheeler} {et~al.}(1975){Wheeler}, {Lecar}, \& {McKee}}]{Wheeler75}
{Wheeler}, J.~C., {Lecar}, M., \& {McKee}, C.~F. 1975,
  \href{http://dx.doi.org/10.1086/153771}{\color{magenta}\apj},
  \href{https://ui.adsabs.harvard.edu/abs/1975ApJ...200..145W}{\color{blue}200},
  \href{https://ui.adsabs.harvard.edu/abs/1975ApJ...200..145W}{\color{blue}145}

\bibitem[{{Whelan} \& {Iben}(1973)}]{Whelan73}
{Whelan}, J., \& {Iben}, Icko, J. 1973,
  \href{http://dx.doi.org/10.1086/152565}{\color{magenta}\apj},
  \href{https://ui.adsabs.harvard.edu/abs/1973ApJ...186.1007W}{\color{blue}186},
  \href{https://ui.adsabs.harvard.edu/abs/1973ApJ...186.1007W}{\color{blue}1007}

\bibitem[{{Woosley} \& {Weaver}(1994)}]{Woosley94}
{Woosley}, S.~E., \& {Weaver}, T.~A. 1994,
  \href{http://dx.doi.org/10.1086/173813}{\color{magenta}\apj},
  \href{https://ui.adsabs.harvard.edu/abs/1994ApJ...423..371W}{\color{blue}423},
  \href{https://ui.adsabs.harvard.edu/abs/1994ApJ...423..371W}{\color{blue}371}

\bibitem[{{Yao} {et~al.}(2019){Yao}, {Miller}, {Kulkarni}, {Bulla}, {Masci},
  {Goldstein}, {Goobar}, {Nugent}, {Dugas}, {Blagorodnova}, {Neill}, {Rigault},
  {Sollerman}, {Nordin}, {Bellm}, {Cenko}, {De}, {Dhawan}, {Feindt},
  {Fremling}, {Gatkine}, {Graham}, {Graham}, {Ho}, {Hung}, {Kasliwal},
  {Kupfer}, {Laher}, {Perley}, {Rusholme}, {Shupe}, {Soumagnac}, {Taggart},
  {Walters}, \& {Yan}}]{Yao19}
{Yao}, Y., {Miller}, A.~A., {Kulkarni}, S.~R., {et~al.} 2019,
  \href{http://dx.doi.org/10.3847/1538-4357/ab4cf5}{\color{magenta}\apj},
  \href{https://ui.adsabs.harvard.edu/abs/2019ApJ...886..152Y}{\color{blue}886},
  \href{https://ui.adsabs.harvard.edu/abs/2019ApJ...886..152Y}{\color{blue}152}

\bibitem[{{Yaron} \& {Gal-Yam}(2012)}]{Yaron12}
{Yaron}, O., \& {Gal-Yam}, A. 2012,
  \href{http://dx.doi.org/10.1086/666656}{\color{magenta}\pasp},
  \href{http://adsabs.harvard.edu/abs/2012PASP..124..668Y}{\color{blue}124},
  \href{http://adsabs.harvard.edu/abs/2012PASP..124..668Y}{\color{blue}668}

\bibitem[{{York} {et~al.}(2000){York}, {Adelman}, {Anderson}, {Anderson},
  {Annis}, {Bahcall}, {Bakken}, {Barkhouser}, {Bastian}, {Berman}, {Boroski},
  {Bracker}, {Briegel}, {Briggs}, {Brinkmann}, {Brunner}, {Burles}, {Carey},
  {Carr}, {Castander}, {Chen}, {Colestock}, {Connolly}, {Crocker}, {Csabai},
  {Czarapata}, {Davis}, {Doi}, {Dombeck}, {Eisenstein}, {Ellman}, {Elms},
  {Evans}, {Fan}, {Federwitz}, {Fiscelli}, {Friedman}, {Frieman}, {Fukugita},
  {Gillespie}, {Gunn}, {Gurbani}, {de Haas}, {Haldeman}, {Harris}, {Hayes},
  {Heckman}, {Hennessy}, {Hindsley}, {Holm}, {Holmgren}, {Huang}, {Hull},
  {Husby}, {Ichikawa}, {Ichikawa}, {Ivezi{\'c}}, {Kent}, {Kim}, {Kinney},
  {Klaene}, {Kleinman}, {Kleinman}, {Knapp}, {Korienek}, {Kron}, {Kunszt},
  {Lamb}, {Lee}, {Leger}, {Limmongkol}, {Lindenmeyer}, {Long}, {Loomis},
  {Loveday}, {Lucinio}, {Lupton}, {MacKinnon}, {Mannery}, {Mantsch}, {Margon},
  {McGehee}, {McKay}, {Meiksin}, {Merelli}, {Monet}, {Munn}, {Narayanan},
  {Nash}, {Neilsen}, {Neswold}, {Newberg}, {Nichol}, {Nicinski}, {Nonino},
  {Okada}, {Okamura}, {Ostriker}, {Owen}, {Pauls}, {Peoples}, {Peterson},
  {Petravick}, {Pier}, {Pope}, {Pordes}, {Prosapio}, {Rechenmacher}, {Quinn},
  {Richards}, {Richmond}, {Rivetta}, {Rockosi}, {Ruthmansdorfer}, {Sandford},
  {Schlegel}, {Schneider}, {Sekiguchi}, {Sergey}, {Shimasaku}, {Siegmund},
  {Smee}, {Smith}, {Snedden}, {Stone}, {Stoughton}, {Strauss}, {Stubbs},
  {SubbaRao}, {Szalay}, {Szapudi}, {Szokoly}, {Thakar}, {Tremonti}, {Tucker},
  {Uomoto}, {Vanden Berk}, {Vogeley}, {Waddell}, {Wang}, {Watanabe},
  {Weinberg}, {Yanny}, {Yasuda}, \& {SDSS Collaboration}}]{York00}
{York}, D.~G., {Adelman}, J., {Anderson}, Jr., J.~E., {et~al.} 2000,
  \href{http://dx.doi.org/10.1086/301513}{\color{magenta}\aj},
  \href{http://adsabs.harvard.edu/abs/2000AJ....120.1579Y}{\color{blue}120},
  \href{http://adsabs.harvard.edu/abs/2000AJ....120.1579Y}{\color{blue}1579}

\bibitem[{{Zackay} {et~al.}(2016){Zackay}, {Ofek}, \& {Gal-Yam}}]{Zackay16}
{Zackay}, B., {Ofek}, E.~O., \& {Gal-Yam}, A. 2016,
  \href{http://dx.doi.org/10.3847/0004-637X/830/1/27}{\color{magenta}\apj},
  \href{https://ui.adsabs.harvard.edu/abs/2016ApJ...830...27Z}{\color{blue}830},
  \href{https://ui.adsabs.harvard.edu/abs/2016ApJ...830...27Z}{\color{blue}27}

\end{thebibliography}
\bibliographystyle{aas_arxiv}



\end{document}